\documentclass[aps,prd,nofootinbib,twocolumn,superscriptaddress,preprintnumbers]{revtex4-1}

\usepackage{amssymb}
\usepackage{amsmath}
\usepackage{epsfig}
\usepackage{hyperref}
\usepackage{breakurl}
\usepackage{bm}
\usepackage{color}
\usepackage{tikz}
\usepackage[utf8]{inputenc}
\usetikzlibrary{snakes}
\usetikzlibrary{decorations}
\usetikzlibrary{trees}
\usetikzlibrary{decorations.pathmorphing}
\usetikzlibrary{decorations.markings}
\usetikzlibrary{external}
\usetikzlibrary{intersections}
\usetikzlibrary{shapes,arrows}
\usetikzlibrary{arrows.meta}
\usetikzlibrary{calc}
\usetikzlibrary{shapes.misc}
\usetikzlibrary{decorations.text}
\usetikzlibrary{backgrounds}
\usetikzlibrary{fadings}
\usetikzlibrary{tikzmark,calc,arrows,shapes,decorations.pathreplacing}

\usepackage{subcaption}

\makeatletter
\long\def\@makecaption#1#2{%
  \par
  \vskip\abovecaptionskip
  \begingroup
    \small\rmfamily
    \sbox\@tempboxa{%
      \let\\\heading@cr
      \@make@capt@title{#1}{#2}%
    }%
    \@ifdim{\wd\@tempboxa >\hsize}{%
      \begingroup
        \samepage
        \flushing
        \let\footnote\@footnotemark@gobble
        \@make@capt@title{#1}{#2}\par
      \endgroup
    }{%
      \hb@xt@\hsize{\unhbox\@tempboxa\hfil}%
    }%
  \endgroup
  \vskip\belowcaptionskip
}%
\makeatother

\makeatletter
\def\simgt{\mathrel{\lower2.5pt\vbox{\lineskip=0pt\baselineskip=0pt
           \hbox{$>$}\hbox{$\sim$}}}}
\def\simlt{\mathrel{\lower2.5pt\vbox{\lineskip=0pt\baselineskip=0pt
           \hbox{$<$}\hbox{$\sim$}}}}

\def\fig#1{Fig.~\ref{#1}}

\makeatother

\def\eqn#1{Eq.~\eqref{#1}}
\def\spa#1.#2{\left\langle#1\,#2\right\rangle}
\def\spb#1.#2{\left[#1\,#2\right]}
\def\sand#1.#2.#3{%
\left\langle#1{\vphantom1}\right|{#2}\left|#3\right]}%
\def\sandmp#1.#2.#3{%
\left\langle#1{\vphantom1}\right|{#2}\left|#3\right]}%
\def\sandpm#1.#2.#3{%
\left[#1{\vphantom1}\right|{#2}\left|#3\right\rangle}%
\def\sandmm#1.#2.#3{%
\left\langle#1{\vphantom1}\right|{#2}\left|#3\right\rangle}%
\def\sandpp#1.#2.#3{%
\left[#1{\vphantom1}\right|{#2}\left|#3\right]}%

\def\Section#1{\vskip .15 cm 
\noindent {\it #1}}
\newcommand{\be}{\begin{equation}}
\newcommand{\ee}{\end{equation}}

\renewcommand{\imath}{\mathrm{i}}

\newcommand{\sigmabar}{\bar \sigma}
\newcommand{\vect}{\boldsymbol}
\renewcommand{\imath}{\mathrm{i}}
\newcommand{\eikSum}{\mathcal{E}}

\def\topbotatom#1{\hbox{\hbox to 0pt{$#1\bot$\hss}$#1\top$}}

\allowdisplaybreaks

\begin{document}

\title{Scattering Amplitudes and Conservative Binary Dynamics
at ${\cal O}(G^5)$ \\[3pt] without Self-Force Truncation}

\author{Zvi Bern}
\affiliation{
Mani L. Bhaumik Institute for Theoretical Physics,
University of California at Los Angeles,
Los Angeles, CA 90095, USA}

\author{Enrico Herrmann}
\affiliation{
	Mani L. Bhaumik Institute for Theoretical Physics,
	University of California at Los Angeles,
	Los Angeles, CA 90095, USA}

\author{Radu~Roiban}
\affiliation{Institute for Theoretical Studies, ETH Zurich, 8092 Zurich, Switzerland}
\affiliation{Institute for Gravitation and the Cosmos and 
Institute for Computational and Data Sciences\\
Pennsylvania State University,
University Park, PA 16802, USA}

\author{Michael~S.~Ruf}
\affiliation{
Stanford Linear Accelerator Center, Menlo Park, United States
}

\author{Alexander V. Smirnov}
\affiliation{Research Computing Center, Moscow State University, 119991 Moscow, Russia}
\affiliation{Moscow Center for Fundamental and Applied Mathematics, 119992 Moscow, Russia}

\author{Sid~Smith}
\affiliation{Dipartimento di Fisica e Astronomia, Universita di Padova, Via Marzolo 8, 35131 Padova, Italy}
\affiliation{INFN, Sezione di Padova,
Via Marzolo 8, I-35131 Padova, Italy.}
\affiliation{Higgs Centre for Theoretical Physics, University of Edinburgh, Edinburgh, EH9 3FD, United Kingdom}

\author{Mao Zeng}
\affiliation{Higgs Centre for Theoretical Physics, University of Edinburgh, Edinburgh, EH9 3FD, United Kingdom}

\begin{abstract}
We compute the complete potential-graviton contributions to the conservative radial action and scattering angle for two non-spinning bodies in general relativity, accurate through fifth order in Newton's constant and including second-order self-force (2SF) effects. The calculation is carried out in the scattering-amplitude framework, combining the double copy, effective field theory, and multi-loop integration techniques based on integration by parts and differential equations. To address a major computational bottleneck, we develop improved integration-by-parts algorithms that render calculations at this order tractable.   The post-Minkowskian amplitude is presented as a series expansion, following the strategy used earlier in maximal supergravity.   For the first self-force sector, which involves only polylogarithmic functions, we also provide a closed-form analytic expression.   For the second self-force sector, as in earlier supergravity work, we find nontrivial cancellations among contributions related to integrals supported on Calabi--Yau geometry.

\end{abstract}

\maketitle

%
\Section{Introduction---}
\label{sec:intro}
%
The detection of gravitational waves~\cite{LIGOScientific:2016aoc, LIGOScientific:2017vwq} marked a major milestone in physics, opening a new window into the universe.  Upcoming improvements in gravitational-wave detectors~\cite{Punturo:2010zz, LISA:2017pwj, Reitze:2019iox, LIGOasharp, Abac:2025saz} anticipate dramatic gains in both sensitivity and frequency coverage. Realizing the full scientific potential of these observations demands high-precision theoretical modeling across a large parameter space. Meeting this formidable challenge will require significant advances in multiple complementary theoretical frameworks, including numerical relativity~\cite{Pretorius:2005gq, Campanelli:2005dd, Baker:2005vv, Damour:2014afa},  the gravitational self-force (SF) program~\cite{Mino:1996nk, Quinn:1996am, Poisson:2011nh, Barack:2018yvs}, effective field theory (EFT)~\cite{Goldberger:2004jt, Cheung:2018wkq}, and both post-Newtonian (PN)~\cite{Droste:1916, Droste:1917, Einstein:1938yz, Ohta:1973je, Blanchet:2013haa} and post-Minkowskian (PM)~\cite{Bertotti:1956pxu, Kerr:1959zlt, Bertotti:1960wuq, Westpfahl:1979gu, Portilla:1980uz, Bel:1981be} methods. Advances in these various approaches must be synthesized into accurate and robust waveform models, such as those within the effective-one-body (EOB) framework~\cite{Buonanno:1998gg, Buonanno:2000ef}.  

In this paper, we compute the potential-mode contributions to the conservative scattering angle at fifth post-Minkowskian (5PM) order for two compact objects in Einstein gravity, including the technically challenging second self-force (2SF) terms. Our calculation closely follows the corresponding framework developed in maximal supergravity~\cite{Bern:2025zno}. While the computation is similar, the tensor ranks of the required integrals are generally higher, substantially increasing the complexity of the loop integrations, particularly in the 2SF sector, and necessitating further improvements to existing algorithms. 
This work follows earlier progress towards the complete 5PM dynamics in general relativity, which included the 5PM potential dynamics in electrodynamics~\cite{Bern:2023ccb}, the complete first self-force (1SF) result at 5PM order in Einstein gravity~\cite{Driesse:2024xad, Driesse:2024feo, Bini:2025vuk} and the complete 5PM potential dynamics in ${\cal N}=8$ supergravity~\cite{Bern:2024adl, Bern:2025zno}.

The post-Minkowskian framework is particularly natural for relativistic scattering: it aligns directly with the usual perturbative expansion in Newton's constant 
$G$, with the key difference that one evaluates the relevant observables in the appropriate classical limit. In the bound-state problem, this approximation is expected to be especially effective for highly eccentric orbits~\cite{Khalil:2022ylj}. Guided by standard effective-field-theory reasoning, compact astrophysical objects can be modeled as point particles whenever the separation is large compared to their size, and long-wavelength radiation is the relevant dynamical degree of freedom—both in unbound encounters and in bound binaries. Within this setting, scattering-amplitude methods offer a systematic and manifestly gauge-invariant route to pushing the post-Minkowskian (PM) expansion to higher orders.

Building on these tools, a variety of field-theoretic formulations have been proposed to address the classical two-body problem in gravity. These include approaches based on effective field theory matching~\cite{Cheung:2018wkq}, eikonal-based resummations~\cite{DiVecchia:2020ymx, DiVecchia:2021bdo}, exponential representations of the S-matrix~\cite{Damgaard:2021ipf, Damgaard:2023ttc}, observable-centered frameworks~\cite{Kosower:2018adc}, heavy-particle effective theories~\cite{Damgaard:2019lfh}, EFTs tailored to extreme mass-ratio systems~\cite{Cheung:2023lnj, Kosmopoulos:2023bwc}, and classical worldline constructions~\cite{Kalin:2020fhe, Mogull:2020sak, Kalin:2022hph, Jakobsen:2022psy}. These advances have enabled a series of high-order PM calculations for the gravitational two-body problem, particularly for non-spinning binaries~\cite{Bern:2019crd, Bern:2019nnu, Bern:2024adl, Bern:2021dqo, Bern:2021yeh, Driesse:2024xad, Driesse:2024feo, Jakobsen:2023hig, Jakobsen:2023ndj, Dlapa:2021vgp, Dlapa:2021npj, Bjerrum-Bohr:2021wwt}. (We note that effects due to spin, finite-size structure, tidal interactions, and the surrounding environment can also be incorporated in this framework; see, for example, Refs.~\cite{Blanchet:2013haa, Porto:2016pyg, Buonanno:2022pgc, Barausse:2014tra}.) The underlying amplitude computations have benefited from major theoretical and computational advances over the years, including generalized unitarity methods~\cite{Bern:1994zx, Bern:1994cg, Bern:1997sc, Britto:2004nc, Bern:2004cz, Bern:2007ct}, color-kinematics duality, and the double-copy construction~\cite{Kawai:1985xq, Bern:2008qj, Bern:2010ue, Bern:2019prr}, together with modern multi-loop integration techniques~\cite{Chetyrkin:1981qh, Tkachov:1981wb, Kotikov:1990kg, Bern:1993kr, Remiddi:1997ny, Gehrmann:1999as, Henn:2013pwa}.

The primary technical challenge lies in evaluating the loop integrals contributing to the 2SF sector. Reducing the resulting high-tensor-rank integrals to a comparatively small set of master integrals via integration-by-parts (IBP) identities~\cite{Chetyrkin:1981qh, Tkachov:1981wb} is particularly demanding. For this step, we employ a modified version of \texttt{FIRE7}~\cite{FIRE7}. Additional symmetry relations are found with \texttt{LiteRed}~\cite{Lee:2013mka}. The master integrals are then evaluated using the method of differential equations~\cite{Kotikov:1990kg, Bern:1993kr, Remiddi:1997ny, Gehrmann:1999as, Henn:2013pwa}. We benefit from substantial computational resources and from various advanced algorithms, including finite-field methods~\cite{vonManteuffel:2014ixa, Peraro:2016wsq}, spanning cuts~\cite{Larsen:2015ped}, presolving IBP identities~\cite{FIRE7}, improved seeding~\cite{Bern:2024adl, Driesse:2024xad, Guan:2024byi, Lange:2025fba}, and further optimization of seeding based on mass dimensions presented here.

\Section{Basic setup---}
We adopt the setup used to compute the potential-mode 2SF contributions in maximal supergravity~\cite{Bern:2025zno}, which in turn parallels the three-loop, amplitude-based framework developed in Refs.~\cite{Bern:2021dqo, Bern:2021yeh}. In Einstein gravity, the main qualitative difference is a substantial increase in the complexity of the integral reductions: in the absence of supersymmetry, the relevant integrands typically carry higher tensor rank, leading to significantly larger IBP systems. 

The extraction of classical observables from quantum scattering amplitudes is by now well established and extensively validated; accordingly, we provide only a brief summary here, referring the reader to Refs.~\cite{Cheung:2018wkq, Bern:2019crd, Kosower:2018adc} for further details.
In the post-Minkowskian regime, the separation of scales allows compact objects to be modeled as widely separated massive point particles interacting through long-range gravitational fields. This effective description, first developed in the worldline EFT approach~\cite{Goldberger:2004jt, Goldberger:2007hy, Porto:2016pyg} and later incorporated into relativistic scattering amplitudes~\cite{Cheung:2018wkq, Bern:2019crd, Kosower:2018adc, DiVecchia:2020ymx, DiVecchia:2021bdo, Bern:2021dqo, Damgaard:2023ttc}, enables the use of powerful techniques from perturbative quantum field theory.

The classical limit corresponds to large angular momentum, $J \gg \hbar$, or equivalently to small momentum transfer $q=p_1+p_4$ (with all momenta taken incoming). At loop level, this selects regions where loop momenta $\ell$ scale as $\ell \sim |q| \ll m_i\sim\sqrt{s}$, so that, within the EFT-based method of regions~\cite{Beneke:1997zp}, the classical contributions arise from the soft region. At order ${\cal O}(G^5)$ this yields a classical amplitude ${\cal M}^{\text{cl}}_5 \propto q^2\ln(-q^2)$, corresponding in position space to a conservative $1/r^5$ potential. To assist with the expansion in the soft region and the subsequent evaluation of integrals, we follow Refs.~\cite{Landshoff:1969yyn, Parra-Martinez:2020dzs} and employ special variables where all mass and momentum-transfer dependence factorizes from the integrals, leaving nontrivial functions of the single dimensionless boost parameter $\sigma = {p_1\cdot p_2}/{(m_1\, m_2)}={1}/{\sqrt{1-v^2}}$, or equivalently the relative velocity $v$.

In this Letter, we focus on the potential region, i.e., the subset of the soft region describing instantaneous interactions between massive particles, where loop momenta scale as $\ell = (\ell^0,\boldsymbol{\ell}) \sim |q|(v,\mathbf{1})$ with $v \ll 1$.   Classical observables are then extracted from the \emph{radial action}, which is related to the massive-scalar $2 \to 2$ scattering amplitude by~\cite{Bern:2021dqo},
\begin{equation}
\label{aarelation}
i \mathcal M^\text{cl}(\bm q) = \int_J \big(e^{i I_r(J)} - 1\big)\,, 
\end{equation}
where the integration measure is given in \eqn{eq:FT_J} below. The classical radial action~\cite{Landau:1975pou}, $I_r(J)=\int_\gamma p_r\, \mathrm{d}r$, is an integral over the radial momentum $p_r$ along the trajectory $\gamma$. It corresponds to the classical action with fixed total energy $E$ and angular momentum $J$ and determines the scattering angle
\begin{equation}
\label{eq:angle_Irad}
\chi= - \frac{\partial I_r(J)}{\partial J} 
         = \pi - 2 J \int\limits_{r_{\mathrm{min}}}^\infty \frac{\mathrm{d}r}{r^2 \sqrt{p^2_r(r)}}\,,
\end{equation}
where, in the second equality, we assume the isotropic gauge for the two-body Hamiltonian. In the PM expansion, the amplitude, the radial action, and the scattering angle are series in Newton's constant,
\begin{equation}
\label{eq:angle_PM}
    \mathcal{M}=\sum_{k=1}\mathcal{M}_k\,,\hskip.4cm \tilde{I}_r=\sum_{k=1}\tilde{I}_{r,k}\,,\hskip.4cm 
    \chi=\sum_{k=1}\chi_k\, ,
 \end{equation}
where $\mathcal{M}_k=\mathcal{O}(G^k)\,,\ \tilde{I}_{r,k}=\mathcal{O}(G^k)\,,$ and $\chi_k=\mathcal{O}(G^k)$ truncate exactly at the classical order, and we leave implicit their kinematic dependence.  For $k=5$, 
 \begin{align}
 {\cal M}_5(\bm q) ={}& 
            \tilde{I}_{r,5}(\bm q) 
          + \mathcal{M}_5^{{\rm it.}}(\bm q) \,.
 \label{eq:amp_action_5}
\end{align} 
$\mathcal{M}_5^{{\rm it.}}$ contains all the classically-singular terms, see e.g.\ Ref.~\cite{Bern:2024adl} for details. As noted there, such terms are separated by the boundary conditions in the differential equations for the master integrals. The classical part of the two-to-two  amplitude, i.e.~the radial action, is
\begin{equation}
\tilde I_r (\bm q) \, {=} 
\int_{J}\! I_r(J) :=
4E\, |\vect{p}|\,  \mu^{-2\epsilon}\!\!\int {\mathrm{d}^{D-2}\bm b}\, e^{\imath\bm q\cdot \bm b}\, I_r(J) \, .
\label{eq:FT_J}
\end{equation}
In the center-of-mass frame, the asymptotic impact parameter $\vect{b}$ and the momentum transfer~$\vect{q}$ are Fourier conjugate; $\vect{p}$ is the spatial momentum.   
The radial action is similar to the eikonal phase~\cite{DiVecchia:2020ymx, DiVecchia:2021bdo}, but more directly related to other concepts such as the exponential representation~\cite{Damgaard:2021ipf} of the S-matrix and its Magnus expansion~\cite{Kim:2025gis, Brandhuber:2025igz}. 

We employ dimensional regularization, working in $D=4-2\epsilon$ space-time dimensions, with renormalization scale $\mu$. The center-of-mass energy $E$, $|\vect{p}|$, and $\sigma$ are related by $E^2=m_1^2+m_2^2+2m_1m_2\,\sigma$ and 
$E |\vect{p}|={m_1m_2}\sqrt{\sigma^2-1}$. We organize the 5PM radial action according to the independent mass structures that can appear at this order
\begin{align}
    \frac{\tilde{I}_{r,5}}{m^4_1 m^4_2} ={}& 
        \frac{(m_1^4{+}m_2^4)}{m_1^2 m_2^2} \tilde{I}_{r,5}^{0 {\rm SF}} 
    + \frac{(m_1^2{+}m_2^2)}{m_1 m_2}    
        \tilde{I}_{r,5}^{1 {\rm SF}}
    + \tilde{I}_{r,5}^{2 {\rm SF}} \hskip .5 cm  \nonumber
   \\[3pt]  
    = {}&
       \frac{M^8\, \nu^2}{m_1^4 m_2^4}
            \left[ 
                      \hat{I}_{r,5}^{0 {\rm SF}} 
                +\nu  \hat{I}_{r,5}^{1 {\rm SF}}  +\nu^2\hat{I}_{r,5}^{2 {\rm SF}}
            \right] , 
   \label{eq:5PM_amp_SF_org}
\end{align}
where we define $M{=}m_1{+}m_2$, $\nu {=} \frac{m_1m_2}{M^2}$,  
$\hat{I}_{r,5}^{0 {\rm SF}} {=} \tilde{I}_{r,5}^{0 {\rm SF}} $, 
$\hat{I}_{r,5}^{1 {\rm SF}} {=} \tilde{I}_{r,5}^{1 {\rm SF}}{-}4\tilde{I}_{r,5}^{0 {\rm SF}}$ and 
$\hat{I}_{r,5}^{2 {\rm SF}} {=} \tilde{I}_{r,5}^{2 {\rm SF}}{-}2\tilde{I}_{r,5}^{1 {\rm SF}}{+}2\tilde{I}_{r,5}^{0 {\rm SF}}$.
While the first form in \eqn{eq:5PM_amp_SF_org} is natural from an amplitudes perspective, the second $\nu$-dependent form is more standard in the context of the self-force expansion.

\Section{Computational Setup---}
%
We extract classical conservative 5PM observables from the classical scattering amplitude of massive particles at $\mathcal{O}(G^5)$ in general relativity, i.e.~at four-loop order. 
We construct the integrand by first obtaining the requisite tree amplitudes via the double-copy, 
sewing them into generalized cuts, and then expanding in the soft region through ${\cal O}(|\vect{q}|^2)$. The resulting expressions are then mapped to a cut-integrand basis~\cite{Bern:2024vqs} to remove redundant labeling and subsequently merged into a soft-expanded integrand by picking each term exactly once while verifying cut consistency.

\begin{figure}[tb]
$
\vcenter{\hbox{
\includegraphics[scale=0.9]{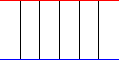}}} 
 \hskip .2 cm 
\vcenter{\hbox{\includegraphics[scale=0.9]{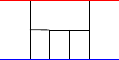}}} \hskip .2 cm 
\vcenter{\hbox{\includegraphics[scale=0.9]{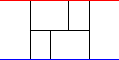}}} \hskip .2 cm 
\vcenter{\hbox{\includegraphics[scale=0.9]{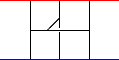}}}
$
\caption{\small Four-loop sample diagrams, including 2SF terms.}
\label{fig:egFeynmanDiags}
\end{figure}

The complete $|\vect{q}|$ and $m_i$ dependence in the resulting integrand is such that all integrals are functions of the boost parameter $\sigma$ only.
The even and odd powers of $|\vect{q}|$ decouple, so, without affecting the classical contributions, we restrict to terms that contain only even powers of $|\vect{q}|$~\cite{Bern:2023ccb}.
The result admits a convenient diagrammatic organization into 394 integral families built entirely from cubic vertices; representative topologies are shown in Fig.~\ref{fig:egFeynmanDiags}. Notably, the third and fourth diagrams arise for the first time at 2SF order, whereas the first two already appeared in earlier 1SF computations. Throughout, the matter lines denote linearized (eikonal) propagators.

Loop integration is the dominant computational bottleneck. Although eikonal propagators are present, the resulting integrals remain compatible with standard integration-by-parts (IBP) reduction. In practice, we reduce the $\mathcal{O}(10^{6})$ integrals appearing in the classical amplitude to $\mathcal{O}(10^{3})$ master integrals. We choose a master basis that avoids nonfactorized dependence on kinematic variables and the spacetime dimension in denominators~\cite{Smirnov:2020quc, Usovitsch:2020jrk}, which would otherwise introduce spurious poles and require nontrivial cancellations.

The IBP reduction is challenging due to the proliferation of irreducible numerators, doubled propagators, intricate topologies, and the large set of master integrals. We carry out the reductions with \textsc{FIRE7}~\cite{FIRE7}, leveraging improved finite-field performance~\cite{vonManteuffel:2014ixa, Peraro:2016wsq}. In addition, we refine the choice of ``seed integrals,'' extending our previous strategy~\cite{Bern:2024adl} and related developments~\cite{Driesse:2024xad, Guan:2024byi}. (see \texttt{Kira~3}~\cite{Lange:2025fba} for a public implementation, Ref.~\cite{Brunello:2025gpf} for a recent application in the post-Newtonian expansion, and Refs.~\cite{Larsen:2015ped, Wu:2023upw, Abreu:2017xsl, Abreu:2017hqn} for analogous seeding schemes in earlier non-Laporta approaches.)

A further key improvement is to enforce a weighted seed cutoff based on mass dimension \cite{massDimPaper}, assigning $+1$ ($-1$) to each linear numerator (denominator) and $+2$ ($-2$) to each quadratic numerator (denominator). Concretely, for the most difficult nonplanar family (the final diagram in Fig.~\ref{fig:egFeynmanDiags}), we impose a complexity cutoff of 8 on top-sector seeds and decrease this cutoff by one for each descent to a subsector with one fewer propagator. We also restrict additional propagator powers (``dots'') by allowing none in the top sector and at most one dot in sectors with two or more canceled propagators.  We impose a uniform mass-dimension upper bound of $-11$ (excluding the integration measure) across all sectors. These constraints shrink the seed set and lead to a significant speedup compared to our earlier setup.
The use of spanning cuts~\cite{Larsen:2015ped} further decomposes this family into 19 smaller IBP systems, which are solved independently and subsequently merged. After caching the non-redundant equations and pivot choices from Gaussian elimination at a single numerical probe point, this information is reused to substantially improve CPU and memory efficiency at additional points~\cite{Peraro:2016wsq, Magerya:2022hvj}. The result of the IBP reduction is verified through an independent calculation, without use of spanning cuts, at a distinct numerical point. 

After completing the IBP reduction of the amplitude in the classical limit, the result is expressed in terms of approximately $4{\small,}000$ master integrals, which depend on the spacetime dimension $D=4-2\epsilon$ and on the kinematic variable $\sigma$. These master integrals are evaluated via the method of differential equations~\cite{Kotikov:1990kg, Bern:1993kr, Remiddi:1997ny, Gehrmann:1999as, Henn:2013pwa}. The full set of master integrals is assembled into a single vector $\vec{I}$, whose dependence on $\sigma$ is governed by a first-order system of differential equations derived from IBP identities,
\begin{align}
\label{eq:diff_eq}
\frac{\mathrm{d}}{\mathrm{d}\sigma} \vec{I}(\sigma,\epsilon)
= \hat{M}(\sigma,\epsilon)\,\vec{I}(\sigma,\epsilon)\, .
\end{align}
In the present calculation, it is convenient to fix the required boundary conditions in the static limit $\sigma = 1$.
The form of the matrix $\hat{M}(\sigma,\epsilon)$ depends on the initial choice of basis $\vec{I}(\sigma,\epsilon)$, but is guaranteed to be in a block-triangular form inherited from the sector structure of Feynman integrals; each sector is a subset of Feynman integrals where the same propagators appear with a positive power in the denominator.

As in Ref.~\cite{Bern:2025zno}, the differential equations determine the complete PM amplitude. Here, instead of closed-form expressions in terms of special functions, in the potential region, we exploit that the master integrals are meromorphic in $\sigmabar \equiv\sigma-1$ near $\sigmabar=0$ and thus admit a local Laurent expansion about the static limit. Solving the differential equations by the Frobenius method yields a series solution in principle to any desired precision,
\begin{align}
\label{eq:diff_eq_sol}
    \vec{I}(\sigma,\epsilon)
    = \sum_{n=n_0}^{n_{\rm max}} \sigmabar^n\,\vec{c}_n(\epsilon)
    + \mathcal{O}(\sigmabar^{n_{\rm max}+1}) \, ,
\end{align}
with coefficients $\vec{c}_n(\epsilon)$ that are rational functions of the dimensional regulator $\epsilon$. The lower bound $n_0$ is determined by the leading velocity scaling in the static limit, while $n_{\max}$ is chosen to reach the target accuracy, here $\mathcal{O}(\bar{\sigma}^{32})$. Expanding $\hat{M}(\sigma,\epsilon)$ around $\bar{\sigma}=0$, we solve the system recursively in $\bar{\sigma}$: at each order, the unknowns follow from linear systems whose $\epsilon$ dependence is reconstructed~\cite{Wang:1981, Wang:1982} from numerical evaluations over prime fields. This approach builds on earlier applications in collider physics~\cite{Mistlberger:2018etf, Moriello:2019yhu, Pozzorini:2005ff, Hidding:2020ytt}.  Finally, after planarizing the most challenging 2SF classical contributions~\cite{Bern:2024adl, Bern:2025zno} using the support of cut matter lines~\cite{Saotome:2012vy, Akhoury:2013yua}, we obtain 371 boundary conditions, fixed from the static limit of scalar integrals that contain both classically singular and regular terms. Of these, only 95 enter the genuinely classical observables.

\begin{figure*}[tb]
  \begin{subfigure}[b]{0.11\textwidth}
    \centering
    \includegraphics[scale=0.75]{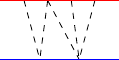}
    \caption{}
  \end{subfigure}
  \begin{subfigure}[b]{0.11\textwidth}
    \centering
    \includegraphics[scale=0.75]{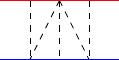}
    \caption{}
  \end{subfigure}
    \begin{subfigure}[b]{0.11\textwidth}
    \centering
    \includegraphics[scale=0.75]{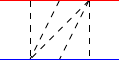}
    \caption{}
  \end{subfigure}
  \begin{subfigure}[b]{0.11\textwidth}
    \centering
    \includegraphics[scale=0.75]{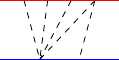}
    \caption{}
  \end{subfigure}
  \begin{subfigure}[b]{0.11\textwidth}
    \centering
    \includegraphics[scale=0.75]{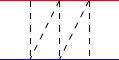}
    \caption{}
  \end{subfigure}
  \begin{subfigure}[b]{0.11\textwidth}
    \centering
    \includegraphics[scale=0.75]{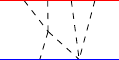}
    \caption{}
  \end{subfigure}
    \begin{subfigure}[b]{0.11\textwidth}
    \centering
    \includegraphics[scale=0.75]{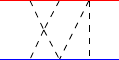}
    \caption{}
  \end{subfigure}
  \begin{subfigure}[b]{0.11\textwidth}
    \centering
    \includegraphics[scale=0.75]{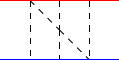}
    \caption{}
  \end{subfigure} 
  \\[5pt]
    \begin{subfigure}[b]{0.11\textwidth}
    \centering
    \includegraphics[scale=0.75]{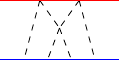}
    \caption{}
  \end{subfigure}
  \begin{subfigure}[b]{0.11\textwidth}
    \centering
    \includegraphics[scale=0.75]{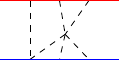}
    \caption{}
  \end{subfigure}
    \begin{subfigure}[b]{0.11\textwidth}
    \centering
    \includegraphics[scale=0.75]{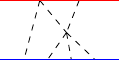}
    \caption{}
  \end{subfigure}
  \begin{subfigure}[b]{0.11\textwidth}
    \centering
    \includegraphics[scale=0.75]{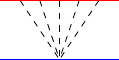}
    \caption{}
  \end{subfigure}
  \begin{subfigure}[b]{0.11\textwidth}
    \centering
    \includegraphics[scale=0.75]{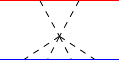}
    \caption{}
  \end{subfigure}
  \begin{subfigure}[b]{0.11\textwidth}
    \centering
    \includegraphics[scale=0.75]{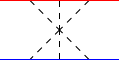}
    \caption{}
  \end{subfigure}
    \begin{subfigure}[b]{0.11\textwidth}
    \centering
    \includegraphics[scale=0.75]{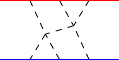}
    \caption{}
  \end{subfigure}
  \begin{subfigure}[b]{0.11\textwidth}
    \centering
    \includegraphics[scale=0.75]{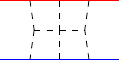}
    \caption{}
  \end{subfigure}
\vskip -.2 cm 
\caption{\small Representative planar diagrams defining the eikonal sums $\eikSum_{2,j}$ appearing in the classical amplitude, where $j$ corresponds to the diagram label.  The label $j = 12$ corresponds to the 0SF contribution, and the labels $j = \{1, 4, 6, 9, 11, 13\}$ correspond to the 1SF contributions, with the remaining contributions being 2SF.  The diagrams in each eikonal sum follow from permuting the attachment points of the dashed graviton lines.}  
 \label{fig:EikonalSumDiags}
\end{figure*}

\medskip
\Section{5PM Results---}
%
Once expressed in terms of boundary integrals, the results organize themselves into well-defined eikonal sums~\cite{Saotome:2012vy, Akhoury:2013yua}.  The diagrams within a given eikonal sum are related by permuting vertices on matter lines, resulting in a sum over 
eikonal propagators that gives an on-shell $\delta$-function, forcing them to be cut. \fig{fig:EikonalSumDiags} gives the planar representatives defining the eikonal sums relevant for the classical amplitude, with the remaining ones given by permuting the attachment point of the potential-region graviton lines, represented by dashed lines, on the matter lines.  See Refs.~\cite{Bern:2024adl, Bern:2025zno} for further details on eikonal sums in the 5PM amplitudes. 

In terms of eikonal sums, the $|\bm q|$-expanded potential-region amplitude through classical order has the form
\begin{equation}
\label{eq:amp_sterman_organized}
    \mathcal{M}_5=\left(\frac{|{\vect{q}}|^2}{\mu^2}\right)^{-4\epsilon}
    \sum_{k=-2}^2|{\vect{q}}|^{k}
    \sum_{n=1}^{N_k}
    f_{k,n}(\sigmabar,\epsilon) \, \eikSum_{k,n}(\epsilon)
    \,,
\end{equation}
where $N_k$ counts the independent eikonal sums and depends on the power, $k$, of $|{\vect{q}}|$. As mentioned above, we focus on the even-in-$|{\vect{q}}|$ terms. The functions $f_{k,n}(\sigmabar,\epsilon)$ are given as series in the velocity variable $\sigmabar$, and we suppress the mass dependence in the arguments. Retaining the computationally much simpler classical iterations serves as a non-trivial check of our setup; we have verified that the $k=-2,0$ terms have the correct values for them to be iteration terms.

We focus on the classical contributions $\tilde I_{r,5}$, cf.~Eq.~\eqref{eq:amp_action_5},
obtained from $\mathcal{M}_5$ by dropping iteration terms as identified by boundary integrals~\cite{Bern:2024adl} and corresponding to the $k=2$ term in Eq.~\eqref{eq:amp_sterman_organized}. Classical contributions arise only when the combination $f_{2,n}(\bar\sigma,\epsilon)\,\mathcal{E}_{2,n}(\epsilon)$ develops poles in $\epsilon$.  There are sixteen eikonal sums contributing to the classical order: one in the 0SF sector, six in the 1SF sector, and nine in the 2SF sector. The 0SF and 1SF cases were discussed in  Ref.~\cite{Bern:2024adl}. 

We find that the coefficient of one classical 2SF eikonal sum, $\mathcal{E}_{2,15}$, associated with a Calabi--Yau threefold integral~\cite{Frellesvig:2023bbf, Klemm:2024wtd, Brammer:2025rqo, Frellesvig:2024rea, Duhr:2025lbz}, vanishes to sufficiently high order in $\epsilon$ so that it does not contribute to the scattering angle. 
Similar cancellations have been observed at 5PM in maximal supergravity~\cite{Bern:2024adl, Bern:2025zno}, 4PN $\mathcal{O}(G^5)$~\cite{Foffa:2016rgu} and at 5PM 1SF in Einstein gravity~\cite{Driesse:2024xad}. This cancellation will be further discussed elsewhere~\cite{CYExplanation}.
In contrast to the maximal supergravity case~\cite{Bern:2025zno}, the coefficient of the $\mathcal{E}_{2,8}$ eikonal sum, associated with a Heun function~\cite{GSJoyceCubicLatticeGF, GSJoyce_2004, Ronveaux:1995Heun, Ablinger:2017bjx}, or equivalently symmetric-square K3, Picard--Fuchs equation~\cite{MRAmplitudes2023, Klemm:2024wtd, Duhr:2025lbz, Brammer:2025rqo}, does not vanish.

After evaluating the eikonal sums and assembling all contributions, we organize the $n$th SF term in the radial action as
\begin{align}
 \label{eq:resultnSF}
 \hspace{-.4cm}
    \tilde{I}_{r,5}^{n {\rm SF}}{=}
    G^5 \pi  \, 
     |{\vect{q}}|^2 \! 
     \left[
        \frac{|{\vect{q}}|^2}{\bar{\mu}^2}
     \right]^{\!-4\epsilon}\!\!
    \left[
        \frac{\tilde{I}_{r,5}^{{\rm nSF,div.\!\!\!\!}}}{\epsilon^2} 
        {+} \frac{\tilde{I}_{r,5}^{{\rm nSF,fin.\!\!\!\!}}}{\epsilon} 
        {+} \mathcal{O}(\epsilon^0)
    \right]\!, \,
  \hspace{-.4cm}  
\end{align}
where $\bar{\mu}^2 = 4\pi e^{-\gamma_{\rm E}}\mu^2$ denotes the $\overline{\rm MS}$ scale.  The double pole $1/\epsilon^2$ originates from the overlap of potential and radiation regions~\cite{Manohar:2006nz, Porto:2017dgs} and is expected to cancel against tail contributions, leaving a logarithmic dependence on $\sigmabar$ and a single pole in $\epsilon$. Such cancellations first appear at three loops~\cite{Galley:2015kus, Bernard:2017bvn, Porto:2017dgs, Bern:2021yeh} and are a generic feature of the method of regions~\cite{Beneke:1997zp}.  Expanding Eq.~(\ref{eq:resultnSF}) in $\epsilon$ yields the nonanalytic $|{\vect{q}}|^2\ln|{\vect{q}}|$ term, corresponding to a $1/r^5$ potential, encoding  classical 5PM dynamics, with coefficient $\tilde{I}_{r,5}^{{\rm nSF,fin.}}$.

We solved the system of differential equations through $\mathcal{O}(\sigmabar^{32})$; however, spurious singularities limit the result to the first 22 orders in the velocity expansion. Including the overall factor of $G^5$, this corresponds to terms through 26PN order.

The key new result of this work is the 2SF potential-graviton contribution to the classical radial action, which determines the corresponding potential-region scattering angle.  Interestingly, eight eikonal sums contribute $1/\epsilon^3$ singularities, but these are spurious and cancel in the full sum, leaving the expected $1/\epsilon^2$ divergence.  The values of the eikonal sums are essential for this cancellation, providing nontrivial consistency checks of our calculation.
The 1SF and 2SF divergent terms are (there is no divergence in the 0SF sector)
%
%
\begin{align}
\tilde{I}_{r,5}^{1 {\rm SF,div.}} &{=}
     \frac{196}{45}{+}\frac{4652 \sigmabar}{525}{+}\frac{31804 \sigmabar^2}{2205}{-}\frac{154772 \sigmabar^3}{121275}{+}\dots 
\\
%
%
\tilde{I}_{r,5}^{2 {\rm SF,div.}} &{=}
\frac{392}{45}{+}\frac{24832 \sigmabar}{1575}{+}\frac{349792 \sigmabar^2}{11025}{+}\frac{219232 \sigmabar^3}{72765}{+}\dots 
\, ,
\nonumber
\end{align}
where for the sake of brevity we wrote explicitly only terms through 7PN order; further terms through $\mathcal{O}(\sigmabar^{22})$ (26PN order) are included in the ancillary file \texttt{GR\_potential\_radial\_action.m}. 
Such divergent terms in the potential-graviton contribution to the radial action are expected to be related to the odd-in-velocity part of the energy loss at lower (in this case 4PM) order~\cite{Bini:2017wfr, Bini:2020hmy, Blanchet:2019rjs},
\begin{align}
\frac{8\nu^2 M^6}{b^4 \, E \sqrt{\sigma^2{-}1}} \left(\hat{I}_{r,5}^{1 {\rm SF,div.}} + \nu \hat{I}_{r,5}^{2 {\rm SF,div.}}\right) 
{=} E_{\rm rad}^{(4)}\Big|_{\rm v-odd} \, ,
\end{align}
with ${\hat I}$ defined below Eq.~\eqref{eq:5PM_amp_SF_org}. We have verified that this relation indeed holds, with the ${\cal O}(G^4)$ energy loss $E^{(4)}_{\rm rad}$ of Refs.~\cite{Dlapa:2022lmu, Jakobsen:2023hig}. The 1SF component of this relation is equivalent to Eq.~(17) of Ref.~\cite{Driesse:2024xad}.

The finite parts of the 0SF, 1SF radial actions, and our key new result, the 2SF radial action, are given through 7PN order:
\begin{widetext}
%
%
\begin{align}
\hspace{-.4cm}
    \tilde{I}_{r,5}^{0 {\rm SF,fin.}} {=} &
    -\frac{1}{640 \sigmabar^4}+\frac{11}{320 \sigmabar^3}-\frac{269}{256 \sigmabar^2}-\frac{4851}{256 \sigmabar}-\frac{144983}{2048}-\frac{120407 \sigmabar}{1280} -\frac{436583 \sigmabar^2}{10240}-\frac{2145 \sigmabar^3}{2048}{+}\cdots\!
 ,
   \label{eq:series_0SF}
\hspace{-.4cm}   
\\[4pt]
%
%
\tilde{I}_{r,5}^{1 {\rm SF,fin.}} = &
- \frac{1}{160 \sigmabar^4} + \frac{21}{160 \sigmabar^3} - \frac{1221}{320 \sigmabar^2} - \frac{671}{12 \sigmabar} -\frac{1460919}{12800} + \frac{3253136293 \sigmabar}{28224000} 
+ \frac{87034402519 \sigmabar^2}{1778112000} - \frac{1563634911847 \sigmabar^3}{470448000} +\dots
\nonumber\\[2pt]
&
-\pi^2 \left( \frac{41}{128 \sigmabar} +\frac{9769}{2304} + \frac{222713 \sigmabar}{7680} - \frac{668191 \sigmabar^2}{107520} 
- \frac{556863481 \sigmabar^3}{1935360} +\dots \right),
\\[4pt]
%
%
\tilde{I}_{r,5}^{2 {\rm SF,fin.}} = &
- \frac{3}{320 \sigmabar^4} + \frac{31}{160 \sigmabar^3} - \frac{3543}{640 \sigmabar^2} 
  - \frac{142651}{1920 \sigmabar} -\frac{22066709}{230400} - \frac{6359713 \sigmabar}{1568000} - \frac{1046640004657 \sigmabar^2}{711244800} 
  - \frac{322774326213709 \sigmabar^3}{36883123200} +\dots
  \nonumber\\[2pt]
 &-\pi^2\left( \frac{41}{64 \sigmabar} + \frac{9523}{1152} + \frac{44005 \sigmabar}{2304} - \frac{95445 \sigmabar^2}{512} - \frac{790497031 \sigmabar^3}{967680}+\dots \right) .
\end{align}
\end{widetext}
Terms through $\mathcal{O}(\sigmabar^{22})$ are included in the ancillary file \texttt{GR\textunderscore potential\textunderscore radial\textunderscore action.m}.  The ancillary file also contains the exact analytic expressions for the 0SF and 1SF contributions.
In these expressions, elliptic contributions cancel, as observed earlier in Ref.~\cite{Driesse:2024xad}. The remaining functions are expressible in terms of manifestly real cyclotomic harmonic polylogarithms (CPL) \cite{Ablinger:2011te, Bern:2024adl} (c.f. also Ref.~\cite{Ablinger:2021fnc} for real representations of functions beyond CPLs).
We note here the presence of terms proportional to $\pi^2$, which are absent from the analogous results in ${\cal N}=8$ supergravity \cite{Bern:2025zno}. 
In our exact 1SF potential radial action, as well as in the 1SF conservative results of Ref.~\cite{Driesse:2024xad}, such terms originate in (multiple) 
polylogarithmic functions. This is unlike the 4PM case, where $\pi^2$ terms tag only elliptic integrals.
It will be interesting to determine the special functions that come with $\pi^2$ factors at 2SF order.  

To extract the scattering angle from the momentum-space radial actions $\tilde{I}^{{\rm nSF}}_{r,5}$, we Fourier transform the $\vect{q}$ dependence to impact parameter space (i.e. invert Eq.~\eqref{eq:FT_J}), relate $b\equiv|\vect{b}|$ to the angular momentum via $b=J/|\vect{p}|$, and take a derivative with respect to $J$, see Eq.~\eqref{eq:angle_Irad}. Following the same steps that led to Eq.~(5.44) in Ref.~\cite{Bern:2024adl}, we can write the SF-split 5PM scattering angle as
\begin{align}
\frac{\chi_5}{ m^2_1 m^2_2 } 
    =\frac{(m^4_1{+}m^4_2)}{m^2_1 m^2_2}\chi^{{\rm 0SF}}_{5}
        {+} \frac{(m^2_1{+}m^2_2)}{m_1 m_2}\chi^{{\rm 1SF}}_{5}
        {+}\chi^{{\rm 2SF}}_{5}\,.
\end{align}
In impact parameter space, the $\chi^{{\rm nSF}}_{5}$ are related to the momentum-space radial action coefficients via
\begin{align}
\label{eq:angle5PM}
 \frac{\chi^{{\rm nSF}}_{5}}{m^2_1 m^2_2} = \frac{G^5 \left(\mu^2 \tilde{b}^2\right)^{5 \epsilon}}{b^5} \frac{1}{|\vect{p}|^2 E}
 \bigg[
 & -\frac{16}{\epsilon} \tilde{I}^{{\rm nSF,div.}}_{r,5} \\
 & \hspace{-2.5cm}+ \left(
     184 \tilde{I}^{{\rm nSF,div.}}_{r,5} 
    - 16 \tilde{I}^{{\rm nSF,fin.}}_{r,5}
   \right)
 \bigg] + \mathcal{O}(\epsilon)\,, \nonumber
\end{align}
where $\tilde{b}^2 = b^2 \pi e^{\gamma_{\rm E}}$.  As in the radial action, the unwanted singular term should cancel once the (conservative) radiation-reaction contributions to the scattering angle are included.

%
\begin{figure*}[tb]
\centering
\includegraphics[scale=.85]{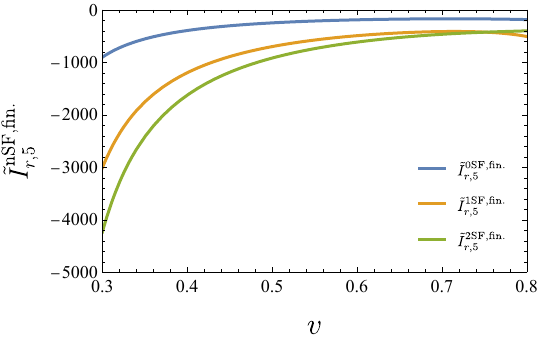}
\hskip 1.5 cm
\includegraphics[scale=.85]{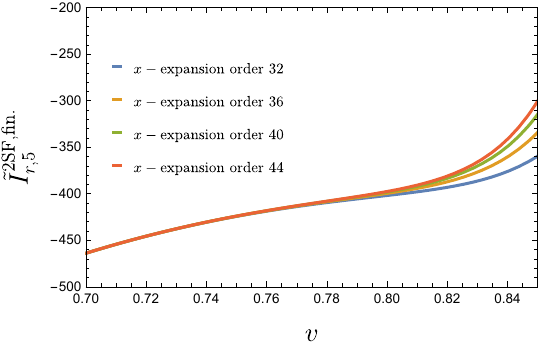}
\vspace{-.3cm}
\caption{\label{fig:RadialAction}
Potential-region momentum-space radial actions in Einstein gravity at 5PM as a function of the velocity $v$. The left panel compares the contribution of the different SF orders. The right panel shows the convergence of the series expansions of the finite part of the 2SF radial action.
}
\end{figure*}

\Section{Checks---}
Our result passes several nontrivial checks. First, as pointed out above, certain spurious higher-order poles in the dimensional regularization parameter $\epsilon$ cancel in the radial action and the remaining divergence matches the energy loss at 4PM order. 
At 0SF, we reproduce the exact probe limit result presented in e.g.~Ref.~\cite{Kol:2021jjc}.
Following the discussion in Ref.~\cite{Bern:2019crd}, the leading in velocity terms of the scattering angle are predicted by lower-order iterations. These velocity iterations arise from the integral representation of the scattering angle in terms of the radial momentum, see Eq.~(\ref{eq:angle_Irad}), and they determine the angle at orders $\sigmabar^{-j}$, $j=2,3,4,5$,
\begin{align}
 \hspace{-.3cm}
\chi_5 &= 
\frac{16 \chi_1 \chi_4[1+\epsilon A]}{3 \pi} -\frac{9 \chi_1^5}{80} - \frac{3 \chi_1^2 \chi_3}{2} 
\nonumber\\
& - \frac{12 \chi_1 \chi_2^2}{\pi^2}  + \frac{3 \chi_1^3 \chi_2}{\pi} + \frac{4 \chi_2 \chi_3}{\pi} 
    +\mathcal{O}\left(\frac{1}{\sigmabar},\epsilon\right),
    \hspace{-.3cm}
   \label{SmallVelPredict}
 \end{align}
where $A=\frac{55}{6}{-}16 \ln 2$. We checked using the 4PM angle $\chi_4$~\cite{Bern:2021dqo} that our results in Eq.~\eqref{eq:angle5PM} agree with this prediction. Note that the ${1}/{\epsilon}$ divergence of the 2SF angle is partially checked by the velocity iteration consistency at order $\sigmabar^{-2}$ and the presence of a similar divergence of the potential-region 4PM angle $\chi_4$. Notably, a combination of all these checks targets all classical eikonal sums, reinforcing the correctness of our result. We have compared with the scattering angle that follows from the PN Hamiltonian of Refs.~\cite{Blumlein:2020pog, Blumlein:2020pyo, Blumlein:2021txe} and found agreement.

\Section{Discussion---}
We include in Fig.~\ref{fig:RadialAction} plots of the finite part of the radial action that illustrate some of their features.
The left panel 
illustrates the relative importance of the different self-force components in the finite part of the radial action as a function of velocity. The right panel indicates that for $v \gtrsim .8$, higher-order terms and resummations will improve convergence. Once the radiation-mode contributions are incorporated, it will be essential to obtain accurate representations of the series solutions across the entire physical region.
Several complementary strategies are available.  As emphasized in Ref.~\cite{Bern:2025zno}, the choice of expansion variables plays a critical role in convergence. The conventional PN expansion in the velocity $v$ is not optimal, as it can introduce limitations from singularities outside the physical domain. A more suitable choice of expansion variable is $x$, related to the boost variable by $\sigma=(1+x^2)/(2x)$, which leads to better convergence properties in the physical region $x\in[0,1]$. This strategy is incorporated in \fig{fig:RadialAction}. See Ref.~\cite{Ruf:Amplitudes23} for a discussion of spurious singularities of the solution to the differential equations.

An obvious further improvement is to substantially extend the series to higher orders, which is not too challenging when using finite-field techniques and parallel computation.  Pad\'e approximants provide further improvements. An initial exploration of such approximants was presented in Ref.~\cite{Bern:2025zno}. Pad\'e approximants are widely used to enhance the accuracy of power-series expansions, both in waveform modeling~\cite{Damour:1997ub, Nagar:2016ayt, Nagar:2024oyk} and in earlier particle-physics applications~\cite{Bultheel:1984aa, Fleischer:1994ef, Fleischer:1996ju}. If complementary information on the asymptotic behavior in the ultra-relativistic limit $v\to1$~\cite{Rothstein:2024nlq, Barcaro:2025ifi, Alessio:2025isu} becomes available, these approximants could be anchored across the full kinematic range, further improving their reliability. 

\Section{Conclusions---} In this Letter, we computed the complete potential-mode contributions to the amplitude and conservative scattering angle at 5PM order.   Our calculations show that the computationally demanding 2SF component can be brought under systematic control.  We employed state-of-the-art Feynman integration tools, as published by two of the authors \cite{FIRE7}, along with various private enhancements.  Our framework closely parallels that employed in the corresponding analysis of maximal supergravity~\cite{Bern:2025zno}, augmented by substantial improvements in IBP seeding strategies,  which should likewise transfer to multi-loop collider-physics applications.  

After obtaining the four-scalar integrand and performing the IBP reduction, the differential equations and boundary integrals completely fix the PM amplitude and scattering angle. We explicitly solved the differential equations for all master integrals as a series expansion about the static limit $\sigmabar\to0$. The resulting solution determines the classical amplitude (radial action) at $\mathcal{O}(G^5)$ through 26PN order. Extending the series to substantially higher orders is, of course, doable using the differential equation Eq.~(\ref{eq:diff_eq}) and might be useful once radiation contributions are included. It would be interesting to promote our series solution to an exact analytic result\footnote{After this work was completed, such a solution was given in Ref.~\cite{Driesse:2026qiz}, which confirmed our potential region results. That paper also included conservative radiation effects.}, with full control over generalized polylogarithms, elliptic integrals, and Calabi--Yau integrals.

To complete the 5PM dynamics, radiation modes must also be included. While observables are independent of the precise split between conservative and dissipative dynamics, a definite scheme is required for comparison with other approaches~\cite{Wheeler:1949hn, Damour:1995kt, Damour:2016gwp, Barack:2018yvs, Driesse:2024xad}. For waveform modeling, the PM results can be embedded in the EOB framework~\cite{Khalil:2022ylj, Buonanno:2024byg}, potentially supplemented by resummations exploiting the singularity structure of observables~\cite{Damour:2022ybd, Long:2024ltn} or information from the high energy limit~\cite{Alessio:2025isu}. Finally, extending the results to bound orbits requires an analytic continuation using appropriate prescriptions, as discussed in Refs.~\cite{Khalil:2022ylj, Bini:2024tft, Dlapa:2024cje, Dlapa:2025biy}. Such continuations are nontrivial at 4PM and beyond due to the nonlocal-in-time contributions from the tail effect. Once the complete 5PM results are available, it would be very interesting to compare them with high-precision numerical relativity~\cite{Long:2025tvk} or self-force studies~\cite{Barack:2023oqp} to assess the magnitude of the effects of time nonlocality (which may be small cf. e.g. Ref.~\cite{Buonanno:2024byg}) and the size of the still-to-be-calculated 6PM contributions as well as to identify improved resummations of the PM series~\cite{Long:2024ltn}.

We look forward to incorporating radiative effects at fifth post-Minkowskian order and extending the post-Minkowskian framework to even higher orders to help support the precision goals of forthcoming gravitational-wave observatories.

%
\Section{Acknowledgements---}
%
%
We thank Alessandra Buonanno, Thibault Damour, Ira Rothstein, Michael Saavedra, Matteo Sergola, Lorenzo Tancredi, Johann Usovitsch, and Stefan Weinzierl for helpful discussions.   We especially thank Volodya Smirnov for his essential contributions and collaboration in the earlier stages of this project. We are grateful to Johannes Bl\"umlein, Andreas Maier, and Peter Marquard, as well as Raj Patil, for assistance in carrying out the comparison to PN results.
We also thank Alex Edison for contributing patches to the FIRE code used in this work.
Z.B. and E.H.~are supported in part by the U.S. Department of Energy (DOE) under award number DE-SC0009937, and by the European Research Council (ERC) Horizon Synergy Grant “Making Sense of the Unexpected in the Gravitational-Wave Sky” grant agreement no. GWSky-101167314.
R.R.~is supported by the U.S.  Department of Energy (DOE) under award number~DE-SC00019066 and by a Senior Fellowship at the Institute for Theoretical Studies, ETH Z\"urich.
M.R. is supported by the Department of Energy (DOE), Contract DE-AC02-76SF00515. 
The work of  A.S. was supported by the Moscow Center for Fundamental and Applied Mathematics of Lomonosov Moscow State University under Agreement No.~075-15-2025-345.
S.S. is supported in part by the \textit{Amplitudes} INFN scientific initiative.
M.Z.'s work is supported in part by the U.K.\ Royal Society through Grant URF\textbackslash R1\textbackslash 20109. For the purpose of open access, the author has applied a Creative Commons Attribution (CC BY) license to any Author Accepted Manuscript version arising from this submission.
This research used resources of the National Energy Research Scientific Computing Center (NERSC), a Department of Energy User Facility using NERSC awards ERCAP 0034824 and 0037814; we especially thank Wahid Bhimji at NERSC for his enthusiastic and critical assistance. The authors also acknowledge the Texas Advanced Computing Center (TACC) at the University of Texas at Austin for providing high-performance computing resources that have contributed to the research results reported within this paper. 
This work also made use of the resources provided by the Edinburgh Compute and Data Facility (ECDF) (http://www.ecdf.ed.ac.uk/).
In addition, this work used computational and storage services associated with the Hoffman2 Cluster, which is operated by the UCLA Office of Advanced Research Computing’s Research Technology Group.
We used \texttt{GNU parallel}~\cite{tange2018gnu} and \texttt{Julia}~\cite{bezanson2017julia} for some of the computations.
We are grateful to the Mani L. Bhaumik Institute for Theoretical Physics for support.


\bibliographystyle{apsrev4-1}
\bibliography{5PM_GR}

\begin{thebibliography}{159}%
\makeatletter
\providecommand \@ifxundefined [1]{%
 \@ifx{#1\undefined}
}%
\providecommand \@ifnum [1]{%
 \ifnum #1\expandafter \@firstoftwo
 \else \expandafter \@secondoftwo
 \fi
}%
\providecommand \@ifx [1]{%
 \ifx #1\expandafter \@firstoftwo
 \else \expandafter \@secondoftwo
 \fi
}%
\providecommand \natexlab [1]{#1}%
\providecommand \enquote  [1]{``#1''}%
\providecommand \bibnamefont  [1]{#1}%
\providecommand \bibfnamefont [1]{#1}%
\providecommand \citenamefont [1]{#1}%
\providecommand \href@noop [0]{\@secondoftwo}%
\providecommand \href [0]{\begingroup \@sanitize@url \@href}%
\providecommand \@href[1]{\@@startlink{#1}\@@href}%
\providecommand \@@href[1]{\endgroup#1\@@endlink}%
\providecommand \@sanitize@url [0]{\catcode `\\12\catcode `\$12\catcode
  `\&12\catcode `\#12\catcode `\^12\catcode `\_12\catcode `\%12\relax}%
\providecommand \@@startlink[1]{}%
\providecommand \@@endlink[0]{}%
\providecommand \url  [0]{\begingroup\@sanitize@url \@url }%
\providecommand \@url [1]{\endgroup\@href {#1}{\urlprefix }}%
\providecommand \urlprefix  [0]{URL }%
\providecommand \Eprint [0]{\href }%
\providecommand \doibase [0]{http://dx.doi.org/}%
\providecommand \selectlanguage [0]{\@gobble}%
\providecommand \bibinfo  [0]{\@secondoftwo}%
\providecommand \bibfield  [0]{\@secondoftwo}%
\providecommand \translation [1]{[#1]}%
\providecommand \BibitemOpen [0]{}%
\providecommand \bibitemStop [0]{}%
\providecommand \bibitemNoStop [0]{.\EOS\space}%
\providecommand \EOS [0]{\spacefactor3000\relax}%
\providecommand \BibitemShut  [1]{\csname bibitem#1\endcsname}%
\let\auto@bib@innerbib\@empty
\bibitem [{\citenamefont {Abbott}\ \emph {et~al.}(2016)\citenamefont {Abbott}
  \emph {et~al.}}]{LIGOScientific:2016aoc}%
  \BibitemOpen
  \bibfield  {author} {\bibinfo {author} {\bibfnamefont {B.~P.}\ \bibnamefont
  {Abbott}} \emph {et~al.} (\bibinfo {collaboration} {LIGO Scientific,
  Virgo}),\ }\href {\doibase 10.1103/PhysRevLett.116.061102} {\bibfield
  {journal} {\bibinfo  {journal} {Phys. Rev. Lett.}\ }\textbf {\bibinfo
  {volume} {116}},\ \bibinfo {pages} {061102} (\bibinfo {year} {2016})},\
  \Eprint {http://arxiv.org/abs/1602.03837} {arXiv:1602.03837 [gr-qc]}
  \BibitemShut {NoStop}%
\bibitem [{\citenamefont {Abbott}\ \emph {et~al.}(2017)\citenamefont {Abbott}
  \emph {et~al.}}]{LIGOScientific:2017vwq}%
  \BibitemOpen
  \bibfield  {author} {\bibinfo {author} {\bibfnamefont {B.~P.}\ \bibnamefont
  {Abbott}} \emph {et~al.} (\bibinfo {collaboration} {LIGO Scientific,
  Virgo}),\ }\href {\doibase 10.1103/PhysRevLett.119.161101} {\bibfield
  {journal} {\bibinfo  {journal} {Phys. Rev. Lett.}\ }\textbf {\bibinfo
  {volume} {119}},\ \bibinfo {pages} {161101} (\bibinfo {year} {2017})},\
  \Eprint {http://arxiv.org/abs/1710.05832} {arXiv:1710.05832 [gr-qc]}
  \BibitemShut {NoStop}%
\bibitem [{\citenamefont {Punturo}\ \emph {et~al.}(2010)\citenamefont {Punturo}
  \emph {et~al.}}]{Punturo:2010zz}%
  \BibitemOpen
  \bibfield  {author} {\bibinfo {author} {\bibfnamefont {M.}~\bibnamefont
  {Punturo}} \emph {et~al.},\ }\href {\doibase 10.1088/0264-9381/27/19/194002}
  {\bibfield  {journal} {\bibinfo  {journal} {Class. Quant. Grav.}\ }\textbf
  {\bibinfo {volume} {27}},\ \bibinfo {pages} {194002} (\bibinfo {year}
  {2010})}\BibitemShut {NoStop}%
\bibitem [{\citenamefont {Amaro-Seoane}\ \emph {et~al.}(2017)\citenamefont
  {Amaro-Seoane} \emph {et~al.}}]{LISA:2017pwj}%
  \BibitemOpen
  \bibfield  {author} {\bibinfo {author} {\bibfnamefont {P.}~\bibnamefont
  {Amaro-Seoane}} \emph {et~al.} (\bibinfo {collaboration} {LISA}),\
  }\href@noop {} {\  (\bibinfo {year} {2017})},\ \Eprint
  {http://arxiv.org/abs/1702.00786} {arXiv:1702.00786 [astro-ph.IM]}
  \BibitemShut {NoStop}%
\bibitem [{\citenamefont {Reitze}\ \emph {et~al.}(2019)\citenamefont {Reitze}
  \emph {et~al.}}]{Reitze:2019iox}%
  \BibitemOpen
  \bibfield  {author} {\bibinfo {author} {\bibfnamefont {D.}~\bibnamefont
  {Reitze}} \emph {et~al.},\ }\href@noop {} {\bibfield  {journal} {\bibinfo
  {journal} {Bull. Am. Astron. Soc.}\ }\textbf {\bibinfo {volume} {51}},\
  \bibinfo {pages} {035} (\bibinfo {year} {2019})},\ \Eprint
  {http://arxiv.org/abs/1907.04833} {arXiv:1907.04833 [astro-ph.IM]}
  \BibitemShut {NoStop}%
\bibitem [{\citenamefont {Fritschel}\ \emph {et~al.}()\citenamefont {Fritschel}
  \emph {et~al.}}]{LIGOasharp}%
  \BibitemOpen
  \bibfield  {author} {\bibinfo {author} {\bibfnamefont {P.}~\bibnamefont
  {Fritschel}} \emph {et~al.},\ }\href@noop {} {\enquote {\bibinfo {title}
  {{Report from the LSC Post-O5 Study Group}},}\ }\bibinfo {note} {Tech. Rep.
  T2200287 (LIGO, 2022)}\BibitemShut {NoStop}%
\bibitem [{\citenamefont {Abac}\ \emph {et~al.}(2025)\citenamefont {Abac} \emph
  {et~al.}}]{Abac:2025saz}%
  \BibitemOpen
  \bibfield  {author} {\bibinfo {author} {\bibfnamefont {A.}~\bibnamefont
  {Abac}} \emph {et~al.},\ }\href@noop {} {\  (\bibinfo {year} {2025})},\
  \Eprint {http://arxiv.org/abs/2503.12263} {arXiv:2503.12263 [gr-qc]}
  \BibitemShut {NoStop}%
\bibitem [{\citenamefont {Pretorius}(2005)}]{Pretorius:2005gq}%
  \BibitemOpen
  \bibfield  {author} {\bibinfo {author} {\bibfnamefont {F.}~\bibnamefont
  {Pretorius}},\ }\href {\doibase 10.1103/PhysRevLett.95.121101} {\bibfield
  {journal} {\bibinfo  {journal} {Phys. Rev. Lett.}\ }\textbf {\bibinfo
  {volume} {95}},\ \bibinfo {pages} {121101} (\bibinfo {year} {2005})},\
  \Eprint {http://arxiv.org/abs/gr-qc/0507014} {arXiv:gr-qc/0507014}
  \BibitemShut {NoStop}%
\bibitem [{\citenamefont {Campanelli}\ \emph {et~al.}(2006)\citenamefont
  {Campanelli}, \citenamefont {Lousto}, \citenamefont {Marronetti},\ and\
  \citenamefont {Zlochower}}]{Campanelli:2005dd}%
  \BibitemOpen
  \bibfield  {author} {\bibinfo {author} {\bibfnamefont {M.}~\bibnamefont
  {Campanelli}}, \bibinfo {author} {\bibfnamefont {C.~O.}\ \bibnamefont
  {Lousto}}, \bibinfo {author} {\bibfnamefont {P.}~\bibnamefont {Marronetti}},
  \ and\ \bibinfo {author} {\bibfnamefont {Y.}~\bibnamefont {Zlochower}},\
  }\href {\doibase 10.1103/PhysRevLett.96.111101} {\bibfield  {journal}
  {\bibinfo  {journal} {Phys. Rev. Lett.}\ }\textbf {\bibinfo {volume} {96}},\
  \bibinfo {pages} {111101} (\bibinfo {year} {2006})},\ \Eprint
  {http://arxiv.org/abs/gr-qc/0511048} {arXiv:gr-qc/0511048} \BibitemShut
  {NoStop}%
\bibitem [{\citenamefont {Baker}\ \emph {et~al.}(2006)\citenamefont {Baker},
  \citenamefont {Centrella}, \citenamefont {Choi}, \citenamefont {Koppitz},\
  and\ \citenamefont {van Meter}}]{Baker:2005vv}%
  \BibitemOpen
  \bibfield  {author} {\bibinfo {author} {\bibfnamefont {J.~G.}\ \bibnamefont
  {Baker}}, \bibinfo {author} {\bibfnamefont {J.}~\bibnamefont {Centrella}},
  \bibinfo {author} {\bibfnamefont {D.-I.}\ \bibnamefont {Choi}}, \bibinfo
  {author} {\bibfnamefont {M.}~\bibnamefont {Koppitz}}, \ and\ \bibinfo
  {author} {\bibfnamefont {J.}~\bibnamefont {van Meter}},\ }\href {\doibase
  10.1103/PhysRevLett.96.111102} {\bibfield  {journal} {\bibinfo  {journal}
  {Phys. Rev. Lett.}\ }\textbf {\bibinfo {volume} {96}},\ \bibinfo {pages}
  {111102} (\bibinfo {year} {2006})},\ \Eprint
  {http://arxiv.org/abs/gr-qc/0511103} {arXiv:gr-qc/0511103} \BibitemShut
  {NoStop}%
\bibitem [{\citenamefont {Damour}\ \emph {et~al.}(2014)\citenamefont {Damour},
  \citenamefont {Guercilena}, \citenamefont {Hinder}, \citenamefont {Hopper},
  \citenamefont {Nagar},\ and\ \citenamefont {Rezzolla}}]{Damour:2014afa}%
  \BibitemOpen
  \bibfield  {author} {\bibinfo {author} {\bibfnamefont {T.}~\bibnamefont
  {Damour}}, \bibinfo {author} {\bibfnamefont {F.}~\bibnamefont {Guercilena}},
  \bibinfo {author} {\bibfnamefont {I.}~\bibnamefont {Hinder}}, \bibinfo
  {author} {\bibfnamefont {S.}~\bibnamefont {Hopper}}, \bibinfo {author}
  {\bibfnamefont {A.}~\bibnamefont {Nagar}}, \ and\ \bibinfo {author}
  {\bibfnamefont {L.}~\bibnamefont {Rezzolla}},\ }\href {\doibase
  10.1103/PhysRevD.89.081503} {\bibfield  {journal} {\bibinfo  {journal} {Phys.
  Rev. D}\ }\textbf {\bibinfo {volume} {89}},\ \bibinfo {pages} {081503}
  (\bibinfo {year} {2014})},\ \Eprint {http://arxiv.org/abs/1402.7307}
  {arXiv:1402.7307 [gr-qc]} \BibitemShut {NoStop}%
\bibitem [{\citenamefont {Mino}\ \emph {et~al.}(1997)\citenamefont {Mino},
  \citenamefont {Sasaki},\ and\ \citenamefont {Tanaka}}]{Mino:1996nk}%
  \BibitemOpen
  \bibfield  {author} {\bibinfo {author} {\bibfnamefont {Y.}~\bibnamefont
  {Mino}}, \bibinfo {author} {\bibfnamefont {M.}~\bibnamefont {Sasaki}}, \ and\
  \bibinfo {author} {\bibfnamefont {T.}~\bibnamefont {Tanaka}},\ }\href
  {\doibase 10.1103/PhysRevD.55.3457} {\bibfield  {journal} {\bibinfo
  {journal} {Phys. Rev. D}\ }\textbf {\bibinfo {volume} {55}},\ \bibinfo
  {pages} {3457} (\bibinfo {year} {1997})},\ \Eprint
  {http://arxiv.org/abs/gr-qc/9606018} {arXiv:gr-qc/9606018} \BibitemShut
  {NoStop}%
\bibitem [{\citenamefont {Quinn}\ and\ \citenamefont
  {Wald}(1997)}]{Quinn:1996am}%
  \BibitemOpen
  \bibfield  {author} {\bibinfo {author} {\bibfnamefont {T.~C.}\ \bibnamefont
  {Quinn}}\ and\ \bibinfo {author} {\bibfnamefont {R.~M.}\ \bibnamefont
  {Wald}},\ }\href {\doibase 10.1103/PhysRevD.56.3381} {\bibfield  {journal}
  {\bibinfo  {journal} {Phys. Rev. D}\ }\textbf {\bibinfo {volume} {56}},\
  \bibinfo {pages} {3381} (\bibinfo {year} {1997})},\ \Eprint
  {http://arxiv.org/abs/gr-qc/9610053} {arXiv:gr-qc/9610053} \BibitemShut
  {NoStop}%
\bibitem [{\citenamefont {Poisson}\ \emph {et~al.}(2011)\citenamefont
  {Poisson}, \citenamefont {Pound},\ and\ \citenamefont
  {Vega}}]{Poisson:2011nh}%
  \BibitemOpen
  \bibfield  {author} {\bibinfo {author} {\bibfnamefont {E.}~\bibnamefont
  {Poisson}}, \bibinfo {author} {\bibfnamefont {A.}~\bibnamefont {Pound}}, \
  and\ \bibinfo {author} {\bibfnamefont {I.}~\bibnamefont {Vega}},\ }\href
  {\doibase 10.12942/lrr-2011-7} {\bibfield  {journal} {\bibinfo  {journal}
  {Living Rev. Rel.}\ }\textbf {\bibinfo {volume} {14}},\ \bibinfo {pages} {7}
  (\bibinfo {year} {2011})},\ \Eprint {http://arxiv.org/abs/1102.0529}
  {arXiv:1102.0529 [gr-qc]} \BibitemShut {NoStop}%
\bibitem [{\citenamefont {Barack}\ and\ \citenamefont
  {Pound}(2019)}]{Barack:2018yvs}%
  \BibitemOpen
  \bibfield  {author} {\bibinfo {author} {\bibfnamefont {L.}~\bibnamefont
  {Barack}}\ and\ \bibinfo {author} {\bibfnamefont {A.}~\bibnamefont {Pound}},\
  }\href {\doibase 10.1088/1361-6633/aae552} {\bibfield  {journal} {\bibinfo
  {journal} {Rept. Prog. Phys.}\ }\textbf {\bibinfo {volume} {82}},\ \bibinfo
  {pages} {016904} (\bibinfo {year} {2019})},\ \Eprint
  {http://arxiv.org/abs/1805.10385} {arXiv:1805.10385 [gr-qc]} \BibitemShut
  {NoStop}%
\bibitem [{\citenamefont {Goldberger}\ and\ \citenamefont
  {Rothstein}(2006)}]{Goldberger:2004jt}%
  \BibitemOpen
  \bibfield  {author} {\bibinfo {author} {\bibfnamefont {W.~D.}\ \bibnamefont
  {Goldberger}}\ and\ \bibinfo {author} {\bibfnamefont {I.~Z.}\ \bibnamefont
  {Rothstein}},\ }\href {\doibase 10.1103/PhysRevD.73.104029} {\bibfield
  {journal} {\bibinfo  {journal} {Phys. Rev. D}\ }\textbf {\bibinfo {volume}
  {73}},\ \bibinfo {pages} {104029} (\bibinfo {year} {2006})},\ \Eprint
  {http://arxiv.org/abs/hep-th/0409156} {arXiv:hep-th/0409156} \BibitemShut
  {NoStop}%
\bibitem [{\citenamefont {Cheung}\ \emph {et~al.}(2018)\citenamefont {Cheung},
  \citenamefont {Rothstein},\ and\ \citenamefont {Solon}}]{Cheung:2018wkq}%
  \BibitemOpen
  \bibfield  {author} {\bibinfo {author} {\bibfnamefont {C.}~\bibnamefont
  {Cheung}}, \bibinfo {author} {\bibfnamefont {I.~Z.}\ \bibnamefont
  {Rothstein}}, \ and\ \bibinfo {author} {\bibfnamefont {M.~P.}\ \bibnamefont
  {Solon}},\ }\href {\doibase 10.1103/PhysRevLett.121.251101} {\bibfield
  {journal} {\bibinfo  {journal} {Phys. Rev. Lett.}\ }\textbf {\bibinfo
  {volume} {121}},\ \bibinfo {pages} {251101} (\bibinfo {year} {2018})},\
  \Eprint {http://arxiv.org/abs/1808.02489} {arXiv:1808.02489 [hep-th]}
  \BibitemShut {NoStop}%
\bibitem [{\citenamefont {Droste}(1916)}]{Droste:1916}%
  \BibitemOpen
  \bibfield  {author} {\bibinfo {author} {\bibfnamefont {J.}~\bibnamefont
  {Droste}},\ }\href@noop {} {\bibfield  {journal} {\bibinfo  {journal} {Proc.
  Acad. Sci. Amst.}\ }\textbf {\bibinfo {volume} {19}},\ \bibinfo {pages} {447}
  (\bibinfo {year} {1916})}\BibitemShut {NoStop}%
\bibitem [{\citenamefont {Lorentz}\ and\ \citenamefont
  {Droste}(1917)}]{Droste:1917}%
  \BibitemOpen
  \bibfield  {author} {\bibinfo {author} {\bibfnamefont {H.}~\bibnamefont
  {Lorentz}}\ and\ \bibinfo {author} {\bibfnamefont {J.}~\bibnamefont
  {Droste}},\ }\href@noop {} {\bibfield  {journal} {\bibinfo  {journal}
  {Verslagen der Afdeeling Natuurkunde van de Koninklijke Akademie van
  Wetenschappen}\ }\textbf {\bibinfo {volume} {26}},\ \bibinfo {pages} {392}
  (\bibinfo {year} {1917})}\BibitemShut {NoStop}%
\bibitem [{\citenamefont {Einstein}\ \emph {et~al.}(1938)\citenamefont
  {Einstein}, \citenamefont {Infeld},\ and\ \citenamefont
  {Hoffmann}}]{Einstein:1938yz}%
  \BibitemOpen
  \bibfield  {author} {\bibinfo {author} {\bibfnamefont {A.}~\bibnamefont
  {Einstein}}, \bibinfo {author} {\bibfnamefont {L.}~\bibnamefont {Infeld}}, \
  and\ \bibinfo {author} {\bibfnamefont {B.}~\bibnamefont {Hoffmann}},\ }\href
  {\doibase 10.2307/1968714} {\bibfield  {journal} {\bibinfo  {journal} {Annals
  Math.}\ }\textbf {\bibinfo {volume} {39}},\ \bibinfo {pages} {65} (\bibinfo
  {year} {1938})}\BibitemShut {NoStop}%
\bibitem [{\citenamefont {Ohta}\ \emph {et~al.}(1973)\citenamefont {Ohta},
  \citenamefont {Okamura}, \citenamefont {Kimura},\ and\ \citenamefont
  {Hiida}}]{Ohta:1973je}%
  \BibitemOpen
  \bibfield  {author} {\bibinfo {author} {\bibfnamefont {T.}~\bibnamefont
  {Ohta}}, \bibinfo {author} {\bibfnamefont {H.}~\bibnamefont {Okamura}},
  \bibinfo {author} {\bibfnamefont {T.}~\bibnamefont {Kimura}}, \ and\ \bibinfo
  {author} {\bibfnamefont {K.}~\bibnamefont {Hiida}},\ }\href {\doibase
  10.1143/PTP.50.492} {\bibfield  {journal} {\bibinfo  {journal} {Prog. Theor.
  Phys.}\ }\textbf {\bibinfo {volume} {50}},\ \bibinfo {pages} {492} (\bibinfo
  {year} {1973})}\BibitemShut {NoStop}%
\bibitem [{\citenamefont {Blanchet}(2014)}]{Blanchet:2013haa}%
  \BibitemOpen
  \bibfield  {author} {\bibinfo {author} {\bibfnamefont {L.}~\bibnamefont
  {Blanchet}},\ }\href {\doibase 10.12942/lrr-2014-2} {\bibfield  {journal}
  {\bibinfo  {journal} {Living Rev. Rel.}\ }\textbf {\bibinfo {volume} {17}},\
  \bibinfo {pages} {2} (\bibinfo {year} {2014})},\ \Eprint
  {http://arxiv.org/abs/1310.1528} {arXiv:1310.1528 [gr-qc]} \BibitemShut
  {NoStop}%
\bibitem [{\citenamefont {Bertotti}(1956)}]{Bertotti:1956pxu}%
  \BibitemOpen
  \bibfield  {author} {\bibinfo {author} {\bibfnamefont {B.}~\bibnamefont
  {Bertotti}},\ }\href {\doibase 10.1007/bf02746175} {\bibfield  {journal}
  {\bibinfo  {journal} {Nuovo Cim.}\ }\textbf {\bibinfo {volume} {4}},\
  \bibinfo {pages} {898} (\bibinfo {year} {1956})}\BibitemShut {NoStop}%
\bibitem [{\citenamefont {Kerr}(1959)}]{Kerr:1959zlt}%
  \BibitemOpen
  \bibfield  {author} {\bibinfo {author} {\bibfnamefont {R.~P.}\ \bibnamefont
  {Kerr}},\ }\href {\doibase 10.1007/bf02732767} {\bibfield  {journal}
  {\bibinfo  {journal} {Nuovo Cim.}\ }\textbf {\bibinfo {volume} {13}},\
  \bibinfo {pages} {469} (\bibinfo {year} {1959})}\BibitemShut {NoStop}%
\bibitem [{\citenamefont {Bertotti}\ and\ \citenamefont
  {Plebanski}(1960)}]{Bertotti:1960wuq}%
  \BibitemOpen
  \bibfield  {author} {\bibinfo {author} {\bibfnamefont {B.}~\bibnamefont
  {Bertotti}}\ and\ \bibinfo {author} {\bibfnamefont {J.}~\bibnamefont
  {Plebanski}},\ }\href {\doibase 10.1016/0003-4916(60)90132-9} {\bibfield
  {journal} {\bibinfo  {journal} {Annals Phys.}\ }\textbf {\bibinfo {volume}
  {11}},\ \bibinfo {pages} {169} (\bibinfo {year} {1960})}\BibitemShut
  {NoStop}%
\bibitem [{\citenamefont {Westpfahl}\ and\ \citenamefont
  {Goller}(1979)}]{Westpfahl:1979gu}%
  \BibitemOpen
  \bibfield  {author} {\bibinfo {author} {\bibfnamefont {K.}~\bibnamefont
  {Westpfahl}}\ and\ \bibinfo {author} {\bibfnamefont {M.}~\bibnamefont
  {Goller}},\ }\href {\doibase 10.1007/BF02817047} {\bibfield  {journal}
  {\bibinfo  {journal} {Lett. Nuovo Cim.}\ }\textbf {\bibinfo {volume} {26}},\
  \bibinfo {pages} {573} (\bibinfo {year} {1979})}\BibitemShut {NoStop}%
\bibitem [{\citenamefont {Portilla}(1980)}]{Portilla:1980uz}%
  \BibitemOpen
  \bibfield  {author} {\bibinfo {author} {\bibfnamefont {M.}~\bibnamefont
  {Portilla}},\ }\href {\doibase 10.1088/0305-4470/13/12/017} {\bibfield
  {journal} {\bibinfo  {journal} {J. Phys. A}\ }\textbf {\bibinfo {volume}
  {13}},\ \bibinfo {pages} {3677} (\bibinfo {year} {1980})}\BibitemShut
  {NoStop}%
\bibitem [{\citenamefont {Bel}\ \emph {et~al.}(1981)\citenamefont {Bel},
  \citenamefont {Damour}, \citenamefont {Deruelle}, \citenamefont {Ibanez},\
  and\ \citenamefont {Martin}}]{Bel:1981be}%
  \BibitemOpen
  \bibfield  {author} {\bibinfo {author} {\bibfnamefont {L.}~\bibnamefont
  {Bel}}, \bibinfo {author} {\bibfnamefont {T.}~\bibnamefont {Damour}},
  \bibinfo {author} {\bibfnamefont {N.}~\bibnamefont {Deruelle}}, \bibinfo
  {author} {\bibfnamefont {J.}~\bibnamefont {Ibanez}}, \ and\ \bibinfo {author}
  {\bibfnamefont {J.}~\bibnamefont {Martin}},\ }\href {\doibase
  10.1007/BF00756073} {\bibfield  {journal} {\bibinfo  {journal} {Gen. Rel.
  Grav.}\ }\textbf {\bibinfo {volume} {13}},\ \bibinfo {pages} {963} (\bibinfo
  {year} {1981})}\BibitemShut {NoStop}%
\bibitem [{\citenamefont {Buonanno}\ and\ \citenamefont
  {Damour}(1999)}]{Buonanno:1998gg}%
  \BibitemOpen
  \bibfield  {author} {\bibinfo {author} {\bibfnamefont {A.}~\bibnamefont
  {Buonanno}}\ and\ \bibinfo {author} {\bibfnamefont {T.}~\bibnamefont
  {Damour}},\ }\href {\doibase 10.1103/PhysRevD.59.084006} {\bibfield
  {journal} {\bibinfo  {journal} {Phys. Rev. D}\ }\textbf {\bibinfo {volume}
  {59}},\ \bibinfo {pages} {084006} (\bibinfo {year} {1999})},\ \Eprint
  {http://arxiv.org/abs/gr-qc/9811091} {arXiv:gr-qc/9811091} \BibitemShut
  {NoStop}%
\bibitem [{\citenamefont {Buonanno}\ and\ \citenamefont
  {Damour}(2000)}]{Buonanno:2000ef}%
  \BibitemOpen
  \bibfield  {author} {\bibinfo {author} {\bibfnamefont {A.}~\bibnamefont
  {Buonanno}}\ and\ \bibinfo {author} {\bibfnamefont {T.}~\bibnamefont
  {Damour}},\ }\href {\doibase 10.1103/PhysRevD.62.064015} {\bibfield
  {journal} {\bibinfo  {journal} {Phys. Rev. D}\ }\textbf {\bibinfo {volume}
  {62}},\ \bibinfo {pages} {064015} (\bibinfo {year} {2000})},\ \Eprint
  {http://arxiv.org/abs/gr-qc/0001013} {arXiv:gr-qc/0001013} \BibitemShut
  {NoStop}%
\bibitem [{\citenamefont {Bern}\ \emph
  {et~al.}(2025{\natexlab{a}})\citenamefont {Bern}, \citenamefont {Herrmann},
  \citenamefont {Roiban}, \citenamefont {Ruf}, \citenamefont {Smirnov},
  \citenamefont {Smirnov},\ and\ \citenamefont {Zeng}}]{Bern:2025zno}%
  \BibitemOpen
  \bibfield  {author} {\bibinfo {author} {\bibfnamefont {Z.}~\bibnamefont
  {Bern}}, \bibinfo {author} {\bibfnamefont {E.}~\bibnamefont {Herrmann}},
  \bibinfo {author} {\bibfnamefont {R.}~\bibnamefont {Roiban}}, \bibinfo
  {author} {\bibfnamefont {M.~S.}\ \bibnamefont {Ruf}}, \bibinfo {author}
  {\bibfnamefont {A.~V.}\ \bibnamefont {Smirnov}}, \bibinfo {author}
  {\bibfnamefont {V.~A.}\ \bibnamefont {Smirnov}}, \ and\ \bibinfo {author}
  {\bibfnamefont {M.}~\bibnamefont {Zeng}},\ }\href@noop {} {\  (\bibinfo
  {year} {2025}{\natexlab{a}})},\ \Eprint {http://arxiv.org/abs/2509.17412}
  {arXiv:2509.17412 [hep-th]} \BibitemShut {NoStop}%
\bibitem [{\citenamefont {Bern}\ \emph
  {et~al.}(2024{\natexlab{a}})\citenamefont {Bern}, \citenamefont {Herrmann},
  \citenamefont {Roiban}, \citenamefont {Ruf}, \citenamefont {Smirnov},
  \citenamefont {Smirnov},\ and\ \citenamefont {Zeng}}]{Bern:2023ccb}%
  \BibitemOpen
  \bibfield  {author} {\bibinfo {author} {\bibfnamefont {Z.}~\bibnamefont
  {Bern}}, \bibinfo {author} {\bibfnamefont {E.}~\bibnamefont {Herrmann}},
  \bibinfo {author} {\bibfnamefont {R.}~\bibnamefont {Roiban}}, \bibinfo
  {author} {\bibfnamefont {M.~S.}\ \bibnamefont {Ruf}}, \bibinfo {author}
  {\bibfnamefont {A.~V.}\ \bibnamefont {Smirnov}}, \bibinfo {author}
  {\bibfnamefont {V.~A.}\ \bibnamefont {Smirnov}}, \ and\ \bibinfo {author}
  {\bibfnamefont {M.}~\bibnamefont {Zeng}},\ }\href {\doibase
  10.1103/PhysRevLett.132.251601} {\bibfield  {journal} {\bibinfo  {journal}
  {Phys. Rev. Lett.}\ }\textbf {\bibinfo {volume} {132}},\ \bibinfo {pages}
  {251601} (\bibinfo {year} {2024}{\natexlab{a}})},\ \Eprint
  {http://arxiv.org/abs/2305.08981} {arXiv:2305.08981 [hep-th]} \BibitemShut
  {NoStop}%
\bibitem [{\citenamefont {Driesse}\ \emph {et~al.}(2024)\citenamefont
  {Driesse}, \citenamefont {Jakobsen}, \citenamefont {Mogull}, \citenamefont
  {Plefka}, \citenamefont {Sauer},\ and\ \citenamefont
  {Usovitsch}}]{Driesse:2024xad}%
  \BibitemOpen
  \bibfield  {author} {\bibinfo {author} {\bibfnamefont {M.}~\bibnamefont
  {Driesse}}, \bibinfo {author} {\bibfnamefont {G.~U.}\ \bibnamefont
  {Jakobsen}}, \bibinfo {author} {\bibfnamefont {G.}~\bibnamefont {Mogull}},
  \bibinfo {author} {\bibfnamefont {J.}~\bibnamefont {Plefka}}, \bibinfo
  {author} {\bibfnamefont {B.}~\bibnamefont {Sauer}}, \ and\ \bibinfo {author}
  {\bibfnamefont {J.}~\bibnamefont {Usovitsch}},\ }\href {\doibase
  10.1103/PhysRevLett.132.241402} {\bibfield  {journal} {\bibinfo  {journal}
  {Phys. Rev. Lett.}\ }\textbf {\bibinfo {volume} {132}},\ \bibinfo {pages}
  {241402} (\bibinfo {year} {2024})},\ \Eprint
  {http://arxiv.org/abs/2403.07781} {arXiv:2403.07781 [hep-th]} \BibitemShut
  {NoStop}%
\bibitem [{\citenamefont {Driesse}\ \emph {et~al.}(2025)\citenamefont
  {Driesse}, \citenamefont {Jakobsen}, \citenamefont {Klemm}, \citenamefont
  {Mogull}, \citenamefont {Nega}, \citenamefont {Plefka}, \citenamefont
  {Sauer},\ and\ \citenamefont {Usovitsch}}]{Driesse:2024feo}%
  \BibitemOpen
  \bibfield  {author} {\bibinfo {author} {\bibfnamefont {M.}~\bibnamefont
  {Driesse}}, \bibinfo {author} {\bibfnamefont {G.~U.}\ \bibnamefont
  {Jakobsen}}, \bibinfo {author} {\bibfnamefont {A.}~\bibnamefont {Klemm}},
  \bibinfo {author} {\bibfnamefont {G.}~\bibnamefont {Mogull}}, \bibinfo
  {author} {\bibfnamefont {C.}~\bibnamefont {Nega}}, \bibinfo {author}
  {\bibfnamefont {J.}~\bibnamefont {Plefka}}, \bibinfo {author} {\bibfnamefont
  {B.}~\bibnamefont {Sauer}}, \ and\ \bibinfo {author} {\bibfnamefont
  {J.}~\bibnamefont {Usovitsch}},\ }\href {\doibase 10.1038/s41586-025-08984-2}
  {\bibfield  {journal} {\bibinfo  {journal} {Nature}\ }\textbf {\bibinfo
  {volume} {641}},\ \bibinfo {pages} {603} (\bibinfo {year} {2025})},\ \Eprint
  {http://arxiv.org/abs/2411.11846} {arXiv:2411.11846 [hep-th]} \BibitemShut
  {NoStop}%
\bibitem [{\citenamefont {Bini}\ and\ \citenamefont
  {Damour}(2025)}]{Bini:2025vuk}%
  \BibitemOpen
  \bibfield  {author} {\bibinfo {author} {\bibfnamefont {D.}~\bibnamefont
  {Bini}}\ and\ \bibinfo {author} {\bibfnamefont {T.}~\bibnamefont {Damour}},\
  }\href {\doibase 10.1103/8ks7-2blq} {\bibfield  {journal} {\bibinfo
  {journal} {Phys. Rev. D}\ }\textbf {\bibinfo {volume} {112}},\ \bibinfo
  {pages} {044002} (\bibinfo {year} {2025})},\ \Eprint
  {http://arxiv.org/abs/2504.20204} {arXiv:2504.20204 [hep-th]} \BibitemShut
  {NoStop}%
\bibitem [{\citenamefont {Bern}\ \emph
  {et~al.}(2024{\natexlab{b}})\citenamefont {Bern}, \citenamefont {Herrmann},
  \citenamefont {Roiban}, \citenamefont {Ruf}, \citenamefont {Smirnov},
  \citenamefont {Smirnov},\ and\ \citenamefont {Zeng}}]{Bern:2024adl}%
  \BibitemOpen
  \bibfield  {author} {\bibinfo {author} {\bibfnamefont {Z.}~\bibnamefont
  {Bern}}, \bibinfo {author} {\bibfnamefont {E.}~\bibnamefont {Herrmann}},
  \bibinfo {author} {\bibfnamefont {R.}~\bibnamefont {Roiban}}, \bibinfo
  {author} {\bibfnamefont {M.~S.}\ \bibnamefont {Ruf}}, \bibinfo {author}
  {\bibfnamefont {A.~V.}\ \bibnamefont {Smirnov}}, \bibinfo {author}
  {\bibfnamefont {V.~A.}\ \bibnamefont {Smirnov}}, \ and\ \bibinfo {author}
  {\bibfnamefont {M.}~\bibnamefont {Zeng}},\ }\href {\doibase
  10.1007/JHEP10(2024)023} {\bibfield  {journal} {\bibinfo  {journal} {JHEP}\
  }\textbf {\bibinfo {volume} {10}},\ \bibinfo {pages} {023} (\bibinfo {year}
  {2024}{\natexlab{b}})},\ \Eprint {http://arxiv.org/abs/2406.01554}
  {arXiv:2406.01554 [hep-th]} \BibitemShut {NoStop}%
\bibitem [{\citenamefont {Khalil}\ \emph {et~al.}(2022)\citenamefont {Khalil},
  \citenamefont {Buonanno}, \citenamefont {Steinhoff},\ and\ \citenamefont
  {Vines}}]{Khalil:2022ylj}%
  \BibitemOpen
  \bibfield  {author} {\bibinfo {author} {\bibfnamefont {M.}~\bibnamefont
  {Khalil}}, \bibinfo {author} {\bibfnamefont {A.}~\bibnamefont {Buonanno}},
  \bibinfo {author} {\bibfnamefont {J.}~\bibnamefont {Steinhoff}}, \ and\
  \bibinfo {author} {\bibfnamefont {J.}~\bibnamefont {Vines}},\ }\href
  {\doibase 10.1103/PhysRevD.106.024042} {\bibfield  {journal} {\bibinfo
  {journal} {Phys. Rev. D}\ }\textbf {\bibinfo {volume} {106}},\ \bibinfo
  {pages} {024042} (\bibinfo {year} {2022})},\ \Eprint
  {http://arxiv.org/abs/2204.05047} {arXiv:2204.05047 [gr-qc]} \BibitemShut
  {NoStop}%
\bibitem [{\citenamefont {Di~Vecchia}\ \emph {et~al.}(2020)\citenamefont
  {Di~Vecchia}, \citenamefont {Heissenberg}, \citenamefont {Russo},\ and\
  \citenamefont {Veneziano}}]{DiVecchia:2020ymx}%
  \BibitemOpen
  \bibfield  {author} {\bibinfo {author} {\bibfnamefont {P.}~\bibnamefont
  {Di~Vecchia}}, \bibinfo {author} {\bibfnamefont {C.}~\bibnamefont
  {Heissenberg}}, \bibinfo {author} {\bibfnamefont {R.}~\bibnamefont {Russo}},
  \ and\ \bibinfo {author} {\bibfnamefont {G.}~\bibnamefont {Veneziano}},\
  }\href {\doibase 10.1016/j.physletb.2020.135924} {\bibfield  {journal}
  {\bibinfo  {journal} {Phys. Lett. B}\ }\textbf {\bibinfo {volume} {811}},\
  \bibinfo {pages} {135924} (\bibinfo {year} {2020})},\ \Eprint
  {http://arxiv.org/abs/2008.12743} {arXiv:2008.12743 [hep-th]} \BibitemShut
  {NoStop}%
\bibitem [{\citenamefont {Di~Vecchia}\ \emph {et~al.}(2021)\citenamefont
  {Di~Vecchia}, \citenamefont {Heissenberg}, \citenamefont {Russo},\ and\
  \citenamefont {Veneziano}}]{DiVecchia:2021bdo}%
  \BibitemOpen
  \bibfield  {author} {\bibinfo {author} {\bibfnamefont {P.}~\bibnamefont
  {Di~Vecchia}}, \bibinfo {author} {\bibfnamefont {C.}~\bibnamefont
  {Heissenberg}}, \bibinfo {author} {\bibfnamefont {R.}~\bibnamefont {Russo}},
  \ and\ \bibinfo {author} {\bibfnamefont {G.}~\bibnamefont {Veneziano}},\
  }\href {\doibase 10.1007/JHEP07(2021)169} {\bibfield  {journal} {\bibinfo
  {journal} {JHEP}\ }\textbf {\bibinfo {volume} {07}},\ \bibinfo {pages} {169}
  (\bibinfo {year} {2021})},\ \Eprint {http://arxiv.org/abs/2104.03256}
  {arXiv:2104.03256 [hep-th]} \BibitemShut {NoStop}%
\bibitem [{\citenamefont {Damgaard}\ \emph {et~al.}(2021)\citenamefont
  {Damgaard}, \citenamefont {Plante},\ and\ \citenamefont
  {Vanhove}}]{Damgaard:2021ipf}%
  \BibitemOpen
  \bibfield  {author} {\bibinfo {author} {\bibfnamefont {P.~H.}\ \bibnamefont
  {Damgaard}}, \bibinfo {author} {\bibfnamefont {L.}~\bibnamefont {Plante}}, \
  and\ \bibinfo {author} {\bibfnamefont {P.}~\bibnamefont {Vanhove}},\ }\href
  {\doibase 10.1007/JHEP11(2021)213} {\bibfield  {journal} {\bibinfo  {journal}
  {JHEP}\ }\textbf {\bibinfo {volume} {11}},\ \bibinfo {pages} {213} (\bibinfo
  {year} {2021})},\ \Eprint {http://arxiv.org/abs/2107.12891} {arXiv:2107.12891
  [hep-th]} \BibitemShut {NoStop}%
\bibitem [{\citenamefont {Damgaard}\ \emph {et~al.}(2023)\citenamefont
  {Damgaard}, \citenamefont {Hansen}, \citenamefont {Plant\'e},\ and\
  \citenamefont {Vanhove}}]{Damgaard:2023ttc}%
  \BibitemOpen
  \bibfield  {author} {\bibinfo {author} {\bibfnamefont {P.~H.}\ \bibnamefont
  {Damgaard}}, \bibinfo {author} {\bibfnamefont {E.~R.}\ \bibnamefont
  {Hansen}}, \bibinfo {author} {\bibfnamefont {L.}~\bibnamefont {Plant\'e}}, \
  and\ \bibinfo {author} {\bibfnamefont {P.}~\bibnamefont {Vanhove}},\ }\href
  {\doibase 10.1007/JHEP09(2023)183} {\bibfield  {journal} {\bibinfo  {journal}
  {JHEP}\ }\textbf {\bibinfo {volume} {09}},\ \bibinfo {pages} {183} (\bibinfo
  {year} {2023})},\ \Eprint {http://arxiv.org/abs/2307.04746} {arXiv:2307.04746
  [hep-th]} \BibitemShut {NoStop}%
\bibitem [{\citenamefont {Kosower}\ \emph {et~al.}(2019)\citenamefont
  {Kosower}, \citenamefont {Maybee},\ and\ \citenamefont
  {O'Connell}}]{Kosower:2018adc}%
  \BibitemOpen
  \bibfield  {author} {\bibinfo {author} {\bibfnamefont {D.~A.}\ \bibnamefont
  {Kosower}}, \bibinfo {author} {\bibfnamefont {B.}~\bibnamefont {Maybee}}, \
  and\ \bibinfo {author} {\bibfnamefont {D.}~\bibnamefont {O'Connell}},\ }\href
  {\doibase 10.1007/JHEP02(2019)137} {\bibfield  {journal} {\bibinfo  {journal}
  {JHEP}\ }\textbf {\bibinfo {volume} {02}},\ \bibinfo {pages} {137} (\bibinfo
  {year} {2019})},\ \Eprint {http://arxiv.org/abs/1811.10950} {arXiv:1811.10950
  [hep-th]} \BibitemShut {NoStop}%
\bibitem [{\citenamefont {Damgaard}\ \emph {et~al.}(2019)\citenamefont
  {Damgaard}, \citenamefont {Haddad},\ and\ \citenamefont
  {Helset}}]{Damgaard:2019lfh}%
  \BibitemOpen
  \bibfield  {author} {\bibinfo {author} {\bibfnamefont {P.~H.}\ \bibnamefont
  {Damgaard}}, \bibinfo {author} {\bibfnamefont {K.}~\bibnamefont {Haddad}}, \
  and\ \bibinfo {author} {\bibfnamefont {A.}~\bibnamefont {Helset}},\ }\href
  {\doibase 10.1007/JHEP11(2019)070} {\bibfield  {journal} {\bibinfo  {journal}
  {JHEP}\ }\textbf {\bibinfo {volume} {11}},\ \bibinfo {pages} {070} (\bibinfo
  {year} {2019})},\ \Eprint {http://arxiv.org/abs/1908.10308} {arXiv:1908.10308
  [hep-ph]} \BibitemShut {NoStop}%
\bibitem [{\citenamefont {Cheung}\ \emph {et~al.}(2024)\citenamefont {Cheung},
  \citenamefont {Parra-Martinez}, \citenamefont {Rothstein}, \citenamefont
  {Shah},\ and\ \citenamefont {Wilson-Gerow}}]{Cheung:2023lnj}%
  \BibitemOpen
  \bibfield  {author} {\bibinfo {author} {\bibfnamefont {C.}~\bibnamefont
  {Cheung}}, \bibinfo {author} {\bibfnamefont {J.}~\bibnamefont
  {Parra-Martinez}}, \bibinfo {author} {\bibfnamefont {I.~Z.}\ \bibnamefont
  {Rothstein}}, \bibinfo {author} {\bibfnamefont {N.}~\bibnamefont {Shah}}, \
  and\ \bibinfo {author} {\bibfnamefont {J.}~\bibnamefont {Wilson-Gerow}},\
  }\href {\doibase 10.1103/PhysRevLett.132.091402} {\bibfield  {journal}
  {\bibinfo  {journal} {Phys. Rev. Lett.}\ }\textbf {\bibinfo {volume} {132}},\
  \bibinfo {pages} {091402} (\bibinfo {year} {2024})},\ \Eprint
  {http://arxiv.org/abs/2308.14832} {arXiv:2308.14832 [hep-th]} \BibitemShut
  {NoStop}%
\bibitem [{\citenamefont {Kosmopoulos}\ and\ \citenamefont
  {Solon}(2024)}]{Kosmopoulos:2023bwc}%
  \BibitemOpen
  \bibfield  {author} {\bibinfo {author} {\bibfnamefont {D.}~\bibnamefont
  {Kosmopoulos}}\ and\ \bibinfo {author} {\bibfnamefont {M.~P.}\ \bibnamefont
  {Solon}},\ }\href {\doibase 10.1007/JHEP03(2024)125} {\bibfield  {journal}
  {\bibinfo  {journal} {JHEP}\ }\textbf {\bibinfo {volume} {03}},\ \bibinfo
  {pages} {125} (\bibinfo {year} {2024})},\ \Eprint
  {http://arxiv.org/abs/2308.15304} {arXiv:2308.15304 [hep-th]} \BibitemShut
  {NoStop}%
\bibitem [{\citenamefont {K\"alin}\ \emph {et~al.}(2020)\citenamefont
  {K\"alin}, \citenamefont {Liu},\ and\ \citenamefont {Porto}}]{Kalin:2020fhe}%
  \BibitemOpen
  \bibfield  {author} {\bibinfo {author} {\bibfnamefont {G.}~\bibnamefont
  {K\"alin}}, \bibinfo {author} {\bibfnamefont {Z.}~\bibnamefont {Liu}}, \ and\
  \bibinfo {author} {\bibfnamefont {R.~A.}\ \bibnamefont {Porto}},\ }\href
  {\doibase 10.1103/PhysRevLett.125.261103} {\bibfield  {journal} {\bibinfo
  {journal} {Phys. Rev. Lett.}\ }\textbf {\bibinfo {volume} {125}},\ \bibinfo
  {pages} {261103} (\bibinfo {year} {2020})},\ \Eprint
  {http://arxiv.org/abs/2007.04977} {arXiv:2007.04977 [hep-th]} \BibitemShut
  {NoStop}%
\bibitem [{\citenamefont {Mogull}\ \emph {et~al.}(2021)\citenamefont {Mogull},
  \citenamefont {Plefka},\ and\ \citenamefont {Steinhoff}}]{Mogull:2020sak}%
  \BibitemOpen
  \bibfield  {author} {\bibinfo {author} {\bibfnamefont {G.}~\bibnamefont
  {Mogull}}, \bibinfo {author} {\bibfnamefont {J.}~\bibnamefont {Plefka}}, \
  and\ \bibinfo {author} {\bibfnamefont {J.}~\bibnamefont {Steinhoff}},\ }\href
  {\doibase 10.1007/JHEP02(2021)048} {\bibfield  {journal} {\bibinfo  {journal}
  {JHEP}\ }\textbf {\bibinfo {volume} {02}},\ \bibinfo {pages} {048} (\bibinfo
  {year} {2021})},\ \Eprint {http://arxiv.org/abs/2010.02865} {arXiv:2010.02865
  [hep-th]} \BibitemShut {NoStop}%
\bibitem [{\citenamefont {K{\"a}lin}\ \emph {et~al.}(2023)\citenamefont
  {K{\"a}lin}, \citenamefont {Neef},\ and\ \citenamefont
  {Porto}}]{Kalin:2022hph}%
  \BibitemOpen
  \bibfield  {author} {\bibinfo {author} {\bibfnamefont {G.}~\bibnamefont
  {K{\"a}lin}}, \bibinfo {author} {\bibfnamefont {J.}~\bibnamefont {Neef}}, \
  and\ \bibinfo {author} {\bibfnamefont {R.~A.}\ \bibnamefont {Porto}},\ }\href
  {\doibase 10.1007/JHEP01(2023)140} {\bibfield  {journal} {\bibinfo  {journal}
  {JHEP}\ }\textbf {\bibinfo {volume} {01}},\ \bibinfo {pages} {140} (\bibinfo
  {year} {2023})},\ \Eprint {http://arxiv.org/abs/2207.00580} {arXiv:2207.00580
  [hep-th]} \BibitemShut {NoStop}%
\bibitem [{\citenamefont {Jakobsen}\ \emph {et~al.}(2022)\citenamefont
  {Jakobsen}, \citenamefont {Mogull}, \citenamefont {Plefka},\ and\
  \citenamefont {Sauer}}]{Jakobsen:2022psy}%
  \BibitemOpen
  \bibfield  {author} {\bibinfo {author} {\bibfnamefont {G.~U.}\ \bibnamefont
  {Jakobsen}}, \bibinfo {author} {\bibfnamefont {G.}~\bibnamefont {Mogull}},
  \bibinfo {author} {\bibfnamefont {J.}~\bibnamefont {Plefka}}, \ and\ \bibinfo
  {author} {\bibfnamefont {B.}~\bibnamefont {Sauer}},\ }\href {\doibase
  10.1007/JHEP10(2022)128} {\bibfield  {journal} {\bibinfo  {journal} {JHEP}\
  }\textbf {\bibinfo {volume} {10}},\ \bibinfo {pages} {128} (\bibinfo {year}
  {2022})},\ \Eprint {http://arxiv.org/abs/2207.00569} {arXiv:2207.00569
  [hep-th]} \BibitemShut {NoStop}%
\bibitem [{\citenamefont {Bern}\ \emph
  {et~al.}(2019{\natexlab{a}})\citenamefont {Bern}, \citenamefont {Cheung},
  \citenamefont {Roiban}, \citenamefont {Shen}, \citenamefont {Solon},\ and\
  \citenamefont {Zeng}}]{Bern:2019crd}%
  \BibitemOpen
  \bibfield  {author} {\bibinfo {author} {\bibfnamefont {Z.}~\bibnamefont
  {Bern}}, \bibinfo {author} {\bibfnamefont {C.}~\bibnamefont {Cheung}},
  \bibinfo {author} {\bibfnamefont {R.}~\bibnamefont {Roiban}}, \bibinfo
  {author} {\bibfnamefont {C.-H.}\ \bibnamefont {Shen}}, \bibinfo {author}
  {\bibfnamefont {M.~P.}\ \bibnamefont {Solon}}, \ and\ \bibinfo {author}
  {\bibfnamefont {M.}~\bibnamefont {Zeng}},\ }\href {\doibase
  10.1007/JHEP10(2019)206} {\bibfield  {journal} {\bibinfo  {journal} {JHEP}\
  }\textbf {\bibinfo {volume} {10}},\ \bibinfo {pages} {206} (\bibinfo {year}
  {2019}{\natexlab{a}})},\ \Eprint {http://arxiv.org/abs/1908.01493}
  {arXiv:1908.01493 [hep-th]} \BibitemShut {NoStop}%
\bibitem [{\citenamefont {Bern}\ \emph
  {et~al.}(2019{\natexlab{b}})\citenamefont {Bern}, \citenamefont {Cheung},
  \citenamefont {Roiban}, \citenamefont {Shen}, \citenamefont {Solon},\ and\
  \citenamefont {Zeng}}]{Bern:2019nnu}%
  \BibitemOpen
  \bibfield  {author} {\bibinfo {author} {\bibfnamefont {Z.}~\bibnamefont
  {Bern}}, \bibinfo {author} {\bibfnamefont {C.}~\bibnamefont {Cheung}},
  \bibinfo {author} {\bibfnamefont {R.}~\bibnamefont {Roiban}}, \bibinfo
  {author} {\bibfnamefont {C.-H.}\ \bibnamefont {Shen}}, \bibinfo {author}
  {\bibfnamefont {M.~P.}\ \bibnamefont {Solon}}, \ and\ \bibinfo {author}
  {\bibfnamefont {M.}~\bibnamefont {Zeng}},\ }\href {\doibase
  10.1103/PhysRevLett.122.201603} {\bibfield  {journal} {\bibinfo  {journal}
  {Phys. Rev. Lett.}\ }\textbf {\bibinfo {volume} {122}},\ \bibinfo {pages}
  {201603} (\bibinfo {year} {2019}{\natexlab{b}})},\ \Eprint
  {http://arxiv.org/abs/1901.04424} {arXiv:1901.04424 [hep-th]} \BibitemShut
  {NoStop}%
\bibitem [{\citenamefont {Bern}\ \emph {et~al.}(2021)\citenamefont {Bern},
  \citenamefont {Parra-Martinez}, \citenamefont {Roiban}, \citenamefont {Ruf},
  \citenamefont {Shen}, \citenamefont {Solon},\ and\ \citenamefont
  {Zeng}}]{Bern:2021dqo}%
  \BibitemOpen
  \bibfield  {author} {\bibinfo {author} {\bibfnamefont {Z.}~\bibnamefont
  {Bern}}, \bibinfo {author} {\bibfnamefont {J.}~\bibnamefont
  {Parra-Martinez}}, \bibinfo {author} {\bibfnamefont {R.}~\bibnamefont
  {Roiban}}, \bibinfo {author} {\bibfnamefont {M.~S.}\ \bibnamefont {Ruf}},
  \bibinfo {author} {\bibfnamefont {C.-H.}\ \bibnamefont {Shen}}, \bibinfo
  {author} {\bibfnamefont {M.~P.}\ \bibnamefont {Solon}}, \ and\ \bibinfo
  {author} {\bibfnamefont {M.}~\bibnamefont {Zeng}},\ }\href {\doibase
  10.1103/PhysRevLett.126.171601} {\bibfield  {journal} {\bibinfo  {journal}
  {Phys. Rev. Lett.}\ }\textbf {\bibinfo {volume} {126}},\ \bibinfo {pages}
  {171601} (\bibinfo {year} {2021})},\ \Eprint
  {http://arxiv.org/abs/2101.07254} {arXiv:2101.07254 [hep-th]} \BibitemShut
  {NoStop}%
\bibitem [{\citenamefont {Bern}\ \emph {et~al.}(2022)\citenamefont {Bern},
  \citenamefont {Parra-Martinez}, \citenamefont {Roiban}, \citenamefont {Ruf},
  \citenamefont {Shen}, \citenamefont {Solon},\ and\ \citenamefont
  {Zeng}}]{Bern:2021yeh}%
  \BibitemOpen
  \bibfield  {author} {\bibinfo {author} {\bibfnamefont {Z.}~\bibnamefont
  {Bern}}, \bibinfo {author} {\bibfnamefont {J.}~\bibnamefont
  {Parra-Martinez}}, \bibinfo {author} {\bibfnamefont {R.}~\bibnamefont
  {Roiban}}, \bibinfo {author} {\bibfnamefont {M.~S.}\ \bibnamefont {Ruf}},
  \bibinfo {author} {\bibfnamefont {C.-H.}\ \bibnamefont {Shen}}, \bibinfo
  {author} {\bibfnamefont {M.~P.}\ \bibnamefont {Solon}}, \ and\ \bibinfo
  {author} {\bibfnamefont {M.}~\bibnamefont {Zeng}},\ }\href {\doibase
  10.1103/PhysRevLett.128.161103} {\bibfield  {journal} {\bibinfo  {journal}
  {Phys. Rev. Lett.}\ }\textbf {\bibinfo {volume} {128}},\ \bibinfo {pages}
  {161103} (\bibinfo {year} {2022})},\ \Eprint
  {http://arxiv.org/abs/2112.10750} {arXiv:2112.10750 [hep-th]} \BibitemShut
  {NoStop}%
\bibitem [{\citenamefont {Jakobsen}\ \emph
  {et~al.}(2023{\natexlab{a}})\citenamefont {Jakobsen}, \citenamefont {Mogull},
  \citenamefont {Plefka},\ and\ \citenamefont {Sauer}}]{Jakobsen:2023hig}%
  \BibitemOpen
  \bibfield  {author} {\bibinfo {author} {\bibfnamefont {G.~U.}\ \bibnamefont
  {Jakobsen}}, \bibinfo {author} {\bibfnamefont {G.}~\bibnamefont {Mogull}},
  \bibinfo {author} {\bibfnamefont {J.}~\bibnamefont {Plefka}}, \ and\ \bibinfo
  {author} {\bibfnamefont {B.}~\bibnamefont {Sauer}},\ }\href {\doibase
  10.1103/PhysRevLett.131.241402} {\bibfield  {journal} {\bibinfo  {journal}
  {Phys. Rev. Lett.}\ }\textbf {\bibinfo {volume} {131}},\ \bibinfo {pages}
  {241402} (\bibinfo {year} {2023}{\natexlab{a}})},\ \Eprint
  {http://arxiv.org/abs/2308.11514} {arXiv:2308.11514 [hep-th]} \BibitemShut
  {NoStop}%
\bibitem [{\citenamefont {Jakobsen}\ \emph
  {et~al.}(2023{\natexlab{b}})\citenamefont {Jakobsen}, \citenamefont {Mogull},
  \citenamefont {Plefka}, \citenamefont {Sauer},\ and\ \citenamefont
  {Xu}}]{Jakobsen:2023ndj}%
  \BibitemOpen
  \bibfield  {author} {\bibinfo {author} {\bibfnamefont {G.~U.}\ \bibnamefont
  {Jakobsen}}, \bibinfo {author} {\bibfnamefont {G.}~\bibnamefont {Mogull}},
  \bibinfo {author} {\bibfnamefont {J.}~\bibnamefont {Plefka}}, \bibinfo
  {author} {\bibfnamefont {B.}~\bibnamefont {Sauer}}, \ and\ \bibinfo {author}
  {\bibfnamefont {Y.}~\bibnamefont {Xu}},\ }\href {\doibase
  10.1103/PhysRevLett.131.151401} {\bibfield  {journal} {\bibinfo  {journal}
  {Phys. Rev. Lett.}\ }\textbf {\bibinfo {volume} {131}},\ \bibinfo {pages}
  {151401} (\bibinfo {year} {2023}{\natexlab{b}})},\ \Eprint
  {http://arxiv.org/abs/2306.01714} {arXiv:2306.01714 [hep-th]} \BibitemShut
  {NoStop}%
\bibitem [{\citenamefont {Dlapa}\ \emph
  {et~al.}(2022{\natexlab{a}})\citenamefont {Dlapa}, \citenamefont {K\"alin},
  \citenamefont {Liu},\ and\ \citenamefont {Porto}}]{Dlapa:2021vgp}%
  \BibitemOpen
  \bibfield  {author} {\bibinfo {author} {\bibfnamefont {C.}~\bibnamefont
  {Dlapa}}, \bibinfo {author} {\bibfnamefont {G.}~\bibnamefont {K\"alin}},
  \bibinfo {author} {\bibfnamefont {Z.}~\bibnamefont {Liu}}, \ and\ \bibinfo
  {author} {\bibfnamefont {R.~A.}\ \bibnamefont {Porto}},\ }\href {\doibase
  10.1103/PhysRevLett.128.161104} {\bibfield  {journal} {\bibinfo  {journal}
  {Phys. Rev. Lett.}\ }\textbf {\bibinfo {volume} {128}},\ \bibinfo {pages}
  {161104} (\bibinfo {year} {2022}{\natexlab{a}})},\ \Eprint
  {http://arxiv.org/abs/2112.11296} {arXiv:2112.11296 [hep-th]} \BibitemShut
  {NoStop}%
\bibitem [{\citenamefont {Dlapa}\ \emph
  {et~al.}(2022{\natexlab{b}})\citenamefont {Dlapa}, \citenamefont {K\"alin},
  \citenamefont {Liu},\ and\ \citenamefont {Porto}}]{Dlapa:2021npj}%
  \BibitemOpen
  \bibfield  {author} {\bibinfo {author} {\bibfnamefont {C.}~\bibnamefont
  {Dlapa}}, \bibinfo {author} {\bibfnamefont {G.}~\bibnamefont {K\"alin}},
  \bibinfo {author} {\bibfnamefont {Z.}~\bibnamefont {Liu}}, \ and\ \bibinfo
  {author} {\bibfnamefont {R.~A.}\ \bibnamefont {Porto}},\ }\href {\doibase
  10.1016/j.physletb.2022.137203} {\bibfield  {journal} {\bibinfo  {journal}
  {Phys. Lett. B}\ }\textbf {\bibinfo {volume} {831}},\ \bibinfo {pages}
  {137203} (\bibinfo {year} {2022}{\natexlab{b}})},\ \Eprint
  {http://arxiv.org/abs/2106.08276} {arXiv:2106.08276 [hep-th]} \BibitemShut
  {NoStop}%
\bibitem [{\citenamefont {Bjerrum-Bohr}\ \emph {et~al.}(2022)\citenamefont
  {Bjerrum-Bohr}, \citenamefont {Plant\'e},\ and\ \citenamefont
  {Vanhove}}]{Bjerrum-Bohr:2021wwt}%
  \BibitemOpen
  \bibfield  {author} {\bibinfo {author} {\bibfnamefont {N.~E.~J.}\
  \bibnamefont {Bjerrum-Bohr}}, \bibinfo {author} {\bibfnamefont
  {L.}~\bibnamefont {Plant\'e}}, \ and\ \bibinfo {author} {\bibfnamefont
  {P.}~\bibnamefont {Vanhove}},\ }\href {\doibase 10.1007/JHEP03(2022)071}
  {\bibfield  {journal} {\bibinfo  {journal} {JHEP}\ }\textbf {\bibinfo
  {volume} {03}},\ \bibinfo {pages} {071} (\bibinfo {year} {2022})},\ \Eprint
  {http://arxiv.org/abs/2111.02976} {arXiv:2111.02976 [hep-th]} \BibitemShut
  {NoStop}%
\bibitem [{\citenamefont {Porto}(2016)}]{Porto:2016pyg}%
  \BibitemOpen
  \bibfield  {author} {\bibinfo {author} {\bibfnamefont {R.~A.}\ \bibnamefont
  {Porto}},\ }\href {\doibase 10.1016/j.physrep.2016.04.003} {\bibfield
  {journal} {\bibinfo  {journal} {Phys. Rept.}\ }\textbf {\bibinfo {volume}
  {633}},\ \bibinfo {pages} {1} (\bibinfo {year} {2016})},\ \Eprint
  {http://arxiv.org/abs/1601.04914} {arXiv:1601.04914 [hep-th]} \BibitemShut
  {NoStop}%
\bibitem [{\citenamefont {Buonanno}\ \emph {et~al.}(2022)\citenamefont
  {Buonanno}, \citenamefont {Khalil}, \citenamefont {O'Connell}, \citenamefont
  {Roiban}, \citenamefont {Solon},\ and\ \citenamefont
  {Zeng}}]{Buonanno:2022pgc}%
  \BibitemOpen
  \bibfield  {author} {\bibinfo {author} {\bibfnamefont {A.}~\bibnamefont
  {Buonanno}}, \bibinfo {author} {\bibfnamefont {M.}~\bibnamefont {Khalil}},
  \bibinfo {author} {\bibfnamefont {D.}~\bibnamefont {O'Connell}}, \bibinfo
  {author} {\bibfnamefont {R.}~\bibnamefont {Roiban}}, \bibinfo {author}
  {\bibfnamefont {M.~P.}\ \bibnamefont {Solon}}, \ and\ \bibinfo {author}
  {\bibfnamefont {M.}~\bibnamefont {Zeng}},\ }in\ \href@noop {} {\emph
  {\bibinfo {booktitle} {{Snowmass 2021}}}}\ (\bibinfo {year} {2022})\ \Eprint
  {http://arxiv.org/abs/2204.05194} {arXiv:2204.05194 [hep-th]} \BibitemShut
  {NoStop}%
\bibitem [{\citenamefont {Barausse}\ \emph {et~al.}(2014)\citenamefont
  {Barausse}, \citenamefont {Cardoso},\ and\ \citenamefont
  {Pani}}]{Barausse:2014tra}%
  \BibitemOpen
  \bibfield  {author} {\bibinfo {author} {\bibfnamefont {E.}~\bibnamefont
  {Barausse}}, \bibinfo {author} {\bibfnamefont {V.}~\bibnamefont {Cardoso}}, \
  and\ \bibinfo {author} {\bibfnamefont {P.}~\bibnamefont {Pani}},\ }\href
  {\doibase 10.1103/PhysRevD.89.104059} {\bibfield  {journal} {\bibinfo
  {journal} {Phys. Rev. D}\ }\textbf {\bibinfo {volume} {89}},\ \bibinfo
  {pages} {104059} (\bibinfo {year} {2014})},\ \Eprint
  {http://arxiv.org/abs/1404.7149} {arXiv:1404.7149 [gr-qc]} \BibitemShut
  {NoStop}%
\bibitem [{\citenamefont {Bern}\ \emph
  {et~al.}(1994{\natexlab{a}})\citenamefont {Bern}, \citenamefont {Dixon},
  \citenamefont {Dunbar},\ and\ \citenamefont {Kosower}}]{Bern:1994zx}%
  \BibitemOpen
  \bibfield  {author} {\bibinfo {author} {\bibfnamefont {Z.}~\bibnamefont
  {Bern}}, \bibinfo {author} {\bibfnamefont {L.~J.}\ \bibnamefont {Dixon}},
  \bibinfo {author} {\bibfnamefont {D.~C.}\ \bibnamefont {Dunbar}}, \ and\
  \bibinfo {author} {\bibfnamefont {D.~A.}\ \bibnamefont {Kosower}},\ }\href
  {\doibase 10.1016/0550-3213(94)90179-1} {\bibfield  {journal} {\bibinfo
  {journal} {Nucl. Phys. B}\ }\textbf {\bibinfo {volume} {425}},\ \bibinfo
  {pages} {217} (\bibinfo {year} {1994}{\natexlab{a}})},\ \Eprint
  {http://arxiv.org/abs/hep-ph/9403226} {arXiv:hep-ph/9403226} \BibitemShut
  {NoStop}%
\bibitem [{\citenamefont {Bern}\ \emph {et~al.}(1995)\citenamefont {Bern},
  \citenamefont {Dixon}, \citenamefont {Dunbar},\ and\ \citenamefont
  {Kosower}}]{Bern:1994cg}%
  \BibitemOpen
  \bibfield  {author} {\bibinfo {author} {\bibfnamefont {Z.}~\bibnamefont
  {Bern}}, \bibinfo {author} {\bibfnamefont {L.~J.}\ \bibnamefont {Dixon}},
  \bibinfo {author} {\bibfnamefont {D.~C.}\ \bibnamefont {Dunbar}}, \ and\
  \bibinfo {author} {\bibfnamefont {D.~A.}\ \bibnamefont {Kosower}},\ }\href
  {\doibase 10.1016/0550-3213(94)00488-Z} {\bibfield  {journal} {\bibinfo
  {journal} {Nucl. Phys. B}\ }\textbf {\bibinfo {volume} {435}},\ \bibinfo
  {pages} {59} (\bibinfo {year} {1995})},\ \Eprint
  {http://arxiv.org/abs/hep-ph/9409265} {arXiv:hep-ph/9409265} \BibitemShut
  {NoStop}%
\bibitem [{\citenamefont {Bern}\ \emph {et~al.}(1998)\citenamefont {Bern},
  \citenamefont {Dixon},\ and\ \citenamefont {Kosower}}]{Bern:1997sc}%
  \BibitemOpen
  \bibfield  {author} {\bibinfo {author} {\bibfnamefont {Z.}~\bibnamefont
  {Bern}}, \bibinfo {author} {\bibfnamefont {L.~J.}\ \bibnamefont {Dixon}}, \
  and\ \bibinfo {author} {\bibfnamefont {D.~A.}\ \bibnamefont {Kosower}},\
  }\href {\doibase 10.1016/S0550-3213(97)00703-7} {\bibfield  {journal}
  {\bibinfo  {journal} {Nucl. Phys. B}\ }\textbf {\bibinfo {volume} {513}},\
  \bibinfo {pages} {3} (\bibinfo {year} {1998})},\ \Eprint
  {http://arxiv.org/abs/hep-ph/9708239} {arXiv:hep-ph/9708239} \BibitemShut
  {NoStop}%
\bibitem [{\citenamefont {Britto}\ \emph {et~al.}(2005)\citenamefont {Britto},
  \citenamefont {Cachazo},\ and\ \citenamefont {Feng}}]{Britto:2004nc}%
  \BibitemOpen
  \bibfield  {author} {\bibinfo {author} {\bibfnamefont {R.}~\bibnamefont
  {Britto}}, \bibinfo {author} {\bibfnamefont {F.}~\bibnamefont {Cachazo}}, \
  and\ \bibinfo {author} {\bibfnamefont {B.}~\bibnamefont {Feng}},\ }\href
  {\doibase 10.1016/j.nuclphysb.2005.07.014} {\bibfield  {journal} {\bibinfo
  {journal} {Nucl. Phys. B}\ }\textbf {\bibinfo {volume} {725}},\ \bibinfo
  {pages} {275} (\bibinfo {year} {2005})},\ \Eprint
  {http://arxiv.org/abs/hep-th/0412103} {arXiv:hep-th/0412103} \BibitemShut
  {NoStop}%
\bibitem [{\citenamefont {Bern}\ \emph {et~al.}(2004)\citenamefont {Bern},
  \citenamefont {Dixon},\ and\ \citenamefont {Kosower}}]{Bern:2004cz}%
  \BibitemOpen
  \bibfield  {author} {\bibinfo {author} {\bibfnamefont {Z.}~\bibnamefont
  {Bern}}, \bibinfo {author} {\bibfnamefont {L.~J.}\ \bibnamefont {Dixon}}, \
  and\ \bibinfo {author} {\bibfnamefont {D.~A.}\ \bibnamefont {Kosower}},\
  }\href {\doibase 10.1088/1126-6708/2004/08/012} {\bibfield  {journal}
  {\bibinfo  {journal} {JHEP}\ }\textbf {\bibinfo {volume} {08}},\ \bibinfo
  {pages} {012} (\bibinfo {year} {2004})},\ \Eprint
  {http://arxiv.org/abs/hep-ph/0404293} {arXiv:hep-ph/0404293} \BibitemShut
  {NoStop}%
\bibitem [{\citenamefont {Bern}\ \emph {et~al.}(2007)\citenamefont {Bern},
  \citenamefont {Carrasco}, \citenamefont {Johansson},\ and\ \citenamefont
  {Kosower}}]{Bern:2007ct}%
  \BibitemOpen
  \bibfield  {author} {\bibinfo {author} {\bibfnamefont {Z.}~\bibnamefont
  {Bern}}, \bibinfo {author} {\bibfnamefont {J.~J.~M.}\ \bibnamefont
  {Carrasco}}, \bibinfo {author} {\bibfnamefont {H.}~\bibnamefont {Johansson}},
  \ and\ \bibinfo {author} {\bibfnamefont {D.~A.}\ \bibnamefont {Kosower}},\
  }\href {\doibase 10.1103/PhysRevD.76.125020} {\bibfield  {journal} {\bibinfo
  {journal} {Phys. Rev. D}\ }\textbf {\bibinfo {volume} {76}},\ \bibinfo
  {pages} {125020} (\bibinfo {year} {2007})},\ \Eprint
  {http://arxiv.org/abs/0705.1864} {arXiv:0705.1864 [hep-th]} \BibitemShut
  {NoStop}%
\bibitem [{\citenamefont {Kawai}\ \emph {et~al.}(1986)\citenamefont {Kawai},
  \citenamefont {Lewellen},\ and\ \citenamefont {Tye}}]{Kawai:1985xq}%
  \BibitemOpen
  \bibfield  {author} {\bibinfo {author} {\bibfnamefont {H.}~\bibnamefont
  {Kawai}}, \bibinfo {author} {\bibfnamefont {D.~C.}\ \bibnamefont {Lewellen}},
  \ and\ \bibinfo {author} {\bibfnamefont {S.~H.~H.}\ \bibnamefont {Tye}},\
  }\href {\doibase 10.1016/0550-3213(86)90362-7} {\bibfield  {journal}
  {\bibinfo  {journal} {Nucl. Phys. B}\ }\textbf {\bibinfo {volume} {269}},\
  \bibinfo {pages} {1} (\bibinfo {year} {1986})}\BibitemShut {NoStop}%
\bibitem [{\citenamefont {Bern}\ \emph {et~al.}(2008)\citenamefont {Bern},
  \citenamefont {Carrasco},\ and\ \citenamefont {Johansson}}]{Bern:2008qj}%
  \BibitemOpen
  \bibfield  {author} {\bibinfo {author} {\bibfnamefont {Z.}~\bibnamefont
  {Bern}}, \bibinfo {author} {\bibfnamefont {J.~J.~M.}\ \bibnamefont
  {Carrasco}}, \ and\ \bibinfo {author} {\bibfnamefont {H.}~\bibnamefont
  {Johansson}},\ }\href {\doibase 10.1103/PhysRevD.78.085011} {\bibfield
  {journal} {\bibinfo  {journal} {Phys. Rev. D}\ }\textbf {\bibinfo {volume}
  {78}},\ \bibinfo {pages} {085011} (\bibinfo {year} {2008})},\ \Eprint
  {http://arxiv.org/abs/0805.3993} {arXiv:0805.3993 [hep-ph]} \BibitemShut
  {NoStop}%
\bibitem [{\citenamefont {Bern}\ \emph {et~al.}(2010)\citenamefont {Bern},
  \citenamefont {Carrasco},\ and\ \citenamefont {Johansson}}]{Bern:2010ue}%
  \BibitemOpen
  \bibfield  {author} {\bibinfo {author} {\bibfnamefont {Z.}~\bibnamefont
  {Bern}}, \bibinfo {author} {\bibfnamefont {J.~J.~M.}\ \bibnamefont
  {Carrasco}}, \ and\ \bibinfo {author} {\bibfnamefont {H.}~\bibnamefont
  {Johansson}},\ }\href {\doibase 10.1103/PhysRevLett.105.061602} {\bibfield
  {journal} {\bibinfo  {journal} {Phys. Rev. Lett.}\ }\textbf {\bibinfo
  {volume} {105}},\ \bibinfo {pages} {061602} (\bibinfo {year} {2010})},\
  \Eprint {http://arxiv.org/abs/1004.0476} {arXiv:1004.0476 [hep-th]}
  \BibitemShut {NoStop}%
\bibitem [{\citenamefont {Bern}\ \emph
  {et~al.}(2024{\natexlab{c}})\citenamefont {Bern}, \citenamefont {Carrasco},
  \citenamefont {Chiodaroli}, \citenamefont {Johansson},\ and\ \citenamefont
  {Roiban}}]{Bern:2019prr}%
  \BibitemOpen
  \bibfield  {author} {\bibinfo {author} {\bibfnamefont {Z.}~\bibnamefont
  {Bern}}, \bibinfo {author} {\bibfnamefont {J.~J.}\ \bibnamefont {Carrasco}},
  \bibinfo {author} {\bibfnamefont {M.}~\bibnamefont {Chiodaroli}}, \bibinfo
  {author} {\bibfnamefont {H.}~\bibnamefont {Johansson}}, \ and\ \bibinfo
  {author} {\bibfnamefont {R.}~\bibnamefont {Roiban}},\ }\href {\doibase
  10.1088/1751-8121/ad5fd0} {\bibfield  {journal} {\bibinfo  {journal} {J.
  Phys. A}\ }\textbf {\bibinfo {volume} {57}},\ \bibinfo {pages} {333002}
  (\bibinfo {year} {2024}{\natexlab{c}})},\ \Eprint
  {http://arxiv.org/abs/1909.01358} {arXiv:1909.01358 [hep-th]} \BibitemShut
  {NoStop}%
\bibitem [{\citenamefont {Chetyrkin}\ and\ \citenamefont
  {Tkachov}(1981)}]{Chetyrkin:1981qh}%
  \BibitemOpen
  \bibfield  {author} {\bibinfo {author} {\bibfnamefont {K.~G.}\ \bibnamefont
  {Chetyrkin}}\ and\ \bibinfo {author} {\bibfnamefont {F.~V.}\ \bibnamefont
  {Tkachov}},\ }\href {\doibase 10.1016/0550-3213(81)90199-1} {\bibfield
  {journal} {\bibinfo  {journal} {Nucl. Phys. B}\ }\textbf {\bibinfo {volume}
  {192}},\ \bibinfo {pages} {159} (\bibinfo {year} {1981})}\BibitemShut
  {NoStop}%
\bibitem [{\citenamefont {Tkachov}(1981)}]{Tkachov:1981wb}%
  \BibitemOpen
  \bibfield  {author} {\bibinfo {author} {\bibfnamefont {F.~V.}\ \bibnamefont
  {Tkachov}},\ }\href {\doibase 10.1016/0370-2693(81)90288-4} {\bibfield
  {journal} {\bibinfo  {journal} {Phys. Lett. B}\ }\textbf {\bibinfo {volume}
  {100}},\ \bibinfo {pages} {65} (\bibinfo {year} {1981})}\BibitemShut
  {NoStop}%
\bibitem [{\citenamefont {Kotikov}(1991)}]{Kotikov:1990kg}%
  \BibitemOpen
  \bibfield  {author} {\bibinfo {author} {\bibfnamefont {A.~V.}\ \bibnamefont
  {Kotikov}},\ }\href {\doibase 10.1016/0370-2693(91)90413-K} {\bibfield
  {journal} {\bibinfo  {journal} {Phys. Lett. B}\ }\textbf {\bibinfo {volume}
  {254}},\ \bibinfo {pages} {158} (\bibinfo {year} {1991})}\BibitemShut
  {NoStop}%
\bibitem [{\citenamefont {Bern}\ \emph
  {et~al.}(1994{\natexlab{b}})\citenamefont {Bern}, \citenamefont {Dixon},\
  and\ \citenamefont {Kosower}}]{Bern:1993kr}%
  \BibitemOpen
  \bibfield  {author} {\bibinfo {author} {\bibfnamefont {Z.}~\bibnamefont
  {Bern}}, \bibinfo {author} {\bibfnamefont {L.~J.}\ \bibnamefont {Dixon}}, \
  and\ \bibinfo {author} {\bibfnamefont {D.~A.}\ \bibnamefont {Kosower}},\
  }\href {\doibase 10.1016/0550-3213(94)90398-0} {\bibfield  {journal}
  {\bibinfo  {journal} {Nucl. Phys. B}\ }\textbf {\bibinfo {volume} {412}},\
  \bibinfo {pages} {751} (\bibinfo {year} {1994}{\natexlab{b}})},\ \Eprint
  {http://arxiv.org/abs/hep-ph/9306240} {arXiv:hep-ph/9306240} \BibitemShut
  {NoStop}%
\bibitem [{\citenamefont {Remiddi}(1997)}]{Remiddi:1997ny}%
  \BibitemOpen
  \bibfield  {author} {\bibinfo {author} {\bibfnamefont {E.}~\bibnamefont
  {Remiddi}},\ }\href {\doibase 10.1007/BF03185566} {\bibfield  {journal}
  {\bibinfo  {journal} {Nuovo Cim. A}\ }\textbf {\bibinfo {volume} {110}},\
  \bibinfo {pages} {1435} (\bibinfo {year} {1997})},\ \Eprint
  {http://arxiv.org/abs/hep-th/9711188} {arXiv:hep-th/9711188} \BibitemShut
  {NoStop}%
\bibitem [{\citenamefont {Gehrmann}\ and\ \citenamefont
  {Remiddi}(2000)}]{Gehrmann:1999as}%
  \BibitemOpen
  \bibfield  {author} {\bibinfo {author} {\bibfnamefont {T.}~\bibnamefont
  {Gehrmann}}\ and\ \bibinfo {author} {\bibfnamefont {E.}~\bibnamefont
  {Remiddi}},\ }\href {\doibase 10.1016/S0550-3213(00)00223-6} {\bibfield
  {journal} {\bibinfo  {journal} {Nucl. Phys. B}\ }\textbf {\bibinfo {volume}
  {580}},\ \bibinfo {pages} {485} (\bibinfo {year} {2000})},\ \Eprint
  {http://arxiv.org/abs/hep-ph/9912329} {arXiv:hep-ph/9912329} \BibitemShut
  {NoStop}%
\bibitem [{\citenamefont {Henn}(2013)}]{Henn:2013pwa}%
  \BibitemOpen
  \bibfield  {author} {\bibinfo {author} {\bibfnamefont {J.~M.}\ \bibnamefont
  {Henn}},\ }\href {\doibase 10.1103/PhysRevLett.110.251601} {\bibfield
  {journal} {\bibinfo  {journal} {Phys. Rev. Lett.}\ }\textbf {\bibinfo
  {volume} {110}},\ \bibinfo {pages} {251601} (\bibinfo {year} {2013})},\
  \Eprint {http://arxiv.org/abs/1304.1806} {arXiv:1304.1806 [hep-th]}
  \BibitemShut {NoStop}%
\bibitem [{\citenamefont {Smirnov}\ and\ \citenamefont {Zeng}(2025)}]{FIRE7}%
  \BibitemOpen
  \bibfield  {author} {\bibinfo {author} {\bibfnamefont {A.~V.}\ \bibnamefont
  {Smirnov}}\ and\ \bibinfo {author} {\bibfnamefont {M.}~\bibnamefont {Zeng}},\
  }\href@noop {} {\  (\bibinfo {year} {2025})},\ \Eprint
  {http://arxiv.org/abs/2510.07150} {arXiv:2510.07150 [hep-ph]} \BibitemShut
  {NoStop}%
\bibitem [{\citenamefont {Lee}(2014)}]{Lee:2013mka}%
  \BibitemOpen
  \bibfield  {author} {\bibinfo {author} {\bibfnamefont {R.~N.}\ \bibnamefont
  {Lee}},\ }\href {\doibase 10.1088/1742-6596/523/1/012059} {\bibfield
  {journal} {\bibinfo  {journal} {J. Phys. Conf. Ser.}\ }\textbf {\bibinfo
  {volume} {523}},\ \bibinfo {pages} {012059} (\bibinfo {year} {2014})},\
  \Eprint {http://arxiv.org/abs/1310.1145} {arXiv:1310.1145 [hep-ph]}
  \BibitemShut {NoStop}%
\bibitem [{\citenamefont {von Manteuffel}\ and\ \citenamefont
  {Schabinger}(2015)}]{vonManteuffel:2014ixa}%
  \BibitemOpen
  \bibfield  {author} {\bibinfo {author} {\bibfnamefont {A.}~\bibnamefont {von
  Manteuffel}}\ and\ \bibinfo {author} {\bibfnamefont {R.~M.}\ \bibnamefont
  {Schabinger}},\ }\href {\doibase 10.1016/j.physletb.2015.03.029} {\bibfield
  {journal} {\bibinfo  {journal} {Phys. Lett. B}\ }\textbf {\bibinfo {volume}
  {744}},\ \bibinfo {pages} {101} (\bibinfo {year} {2015})},\ \Eprint
  {http://arxiv.org/abs/1406.4513} {arXiv:1406.4513 [hep-ph]} \BibitemShut
  {NoStop}%
\bibitem [{\citenamefont {Peraro}(2016)}]{Peraro:2016wsq}%
  \BibitemOpen
  \bibfield  {author} {\bibinfo {author} {\bibfnamefont {T.}~\bibnamefont
  {Peraro}},\ }\href {\doibase 10.1007/JHEP12(2016)030} {\bibfield  {journal}
  {\bibinfo  {journal} {JHEP}\ }\textbf {\bibinfo {volume} {12}},\ \bibinfo
  {pages} {030} (\bibinfo {year} {2016})},\ \Eprint
  {http://arxiv.org/abs/1608.01902} {arXiv:1608.01902 [hep-ph]} \BibitemShut
  {NoStop}%
\bibitem [{\citenamefont {Larsen}\ and\ \citenamefont
  {Zhang}(2016)}]{Larsen:2015ped}%
  \BibitemOpen
  \bibfield  {author} {\bibinfo {author} {\bibfnamefont {K.~J.}\ \bibnamefont
  {Larsen}}\ and\ \bibinfo {author} {\bibfnamefont {Y.}~\bibnamefont {Zhang}},\
  }\href {\doibase 10.1103/PhysRevD.93.041701} {\bibfield  {journal} {\bibinfo
  {journal} {Phys. Rev. D}\ }\textbf {\bibinfo {volume} {93}},\ \bibinfo
  {pages} {041701} (\bibinfo {year} {2016})},\ \Eprint
  {http://arxiv.org/abs/1511.01071} {arXiv:1511.01071 [hep-th]} \BibitemShut
  {NoStop}%
\bibitem [{\citenamefont {Guan}\ \emph {et~al.}(2025)\citenamefont {Guan},
  \citenamefont {Liu}, \citenamefont {Ma},\ and\ \citenamefont
  {Wu}}]{Guan:2024byi}%
  \BibitemOpen
  \bibfield  {author} {\bibinfo {author} {\bibfnamefont {X.}~\bibnamefont
  {Guan}}, \bibinfo {author} {\bibfnamefont {X.}~\bibnamefont {Liu}}, \bibinfo
  {author} {\bibfnamefont {Y.-Q.}\ \bibnamefont {Ma}}, \ and\ \bibinfo {author}
  {\bibfnamefont {W.-H.}\ \bibnamefont {Wu}},\ }\href {\doibase
  10.1016/j.cpc.2025.109538} {\bibfield  {journal} {\bibinfo  {journal}
  {Comput. Phys. Commun.}\ }\textbf {\bibinfo {volume} {310}},\ \bibinfo
  {pages} {109538} (\bibinfo {year} {2025})},\ \Eprint
  {http://arxiv.org/abs/2405.14621} {arXiv:2405.14621 [hep-ph]} \BibitemShut
  {NoStop}%
\bibitem [{\citenamefont {Lange}\ \emph {et~al.}(2025)\citenamefont {Lange},
  \citenamefont {Usovitsch},\ and\ \citenamefont {Wu}}]{Lange:2025fba}%
  \BibitemOpen
  \bibfield  {author} {\bibinfo {author} {\bibfnamefont {F.}~\bibnamefont
  {Lange}}, \bibinfo {author} {\bibfnamefont {J.}~\bibnamefont {Usovitsch}}, \
  and\ \bibinfo {author} {\bibfnamefont {Z.}~\bibnamefont {Wu}},\ }\href@noop
  {} {\  (\bibinfo {year} {2025})},\ \Eprint {http://arxiv.org/abs/2505.20197}
  {arXiv:2505.20197 [hep-ph]} \BibitemShut {NoStop}%
\bibitem [{\citenamefont {Goldberger}(2007)}]{Goldberger:2007hy}%
  \BibitemOpen
  \bibfield  {author} {\bibinfo {author} {\bibfnamefont {W.~D.}\ \bibnamefont
  {Goldberger}},\ }in\ \href@noop {} {\emph {\bibinfo {booktitle} {{Les Houches
  Summer School - Session 86: Particle Physics and Cosmology: The Fabric of
  Spacetime}}}}\ (\bibinfo {year} {2007})\ \Eprint
  {http://arxiv.org/abs/hep-ph/0701129} {arXiv:hep-ph/0701129} \BibitemShut
  {NoStop}%
\bibitem [{\citenamefont {Beneke}\ and\ \citenamefont
  {Smirnov}(1998)}]{Beneke:1997zp}%
  \BibitemOpen
  \bibfield  {author} {\bibinfo {author} {\bibfnamefont {M.}~\bibnamefont
  {Beneke}}\ and\ \bibinfo {author} {\bibfnamefont {V.~A.}\ \bibnamefont
  {Smirnov}},\ }\href {\doibase 10.1016/S0550-3213(98)00138-2} {\bibfield
  {journal} {\bibinfo  {journal} {Nucl. Phys. B}\ }\textbf {\bibinfo {volume}
  {522}},\ \bibinfo {pages} {321} (\bibinfo {year} {1998})},\ \Eprint
  {http://arxiv.org/abs/hep-ph/9711391} {arXiv:hep-ph/9711391} \BibitemShut
  {NoStop}%
\bibitem [{\citenamefont {Landshoff}\ and\ \citenamefont
  {Polkinghorne}(1969)}]{Landshoff:1969yyn}%
  \BibitemOpen
  \bibfield  {author} {\bibinfo {author} {\bibfnamefont {P.~V.}\ \bibnamefont
  {Landshoff}}\ and\ \bibinfo {author} {\bibfnamefont {J.~C.}\ \bibnamefont
  {Polkinghorne}},\ }\href {\doibase 10.1103/PhysRev.181.1989} {\bibfield
  {journal} {\bibinfo  {journal} {Phys. Rev.}\ }\textbf {\bibinfo {volume}
  {181}},\ \bibinfo {pages} {1989} (\bibinfo {year} {1969})}\BibitemShut
  {NoStop}%
\bibitem [{\citenamefont {Parra-Martinez}\ \emph {et~al.}(2020)\citenamefont
  {Parra-Martinez}, \citenamefont {Ruf},\ and\ \citenamefont
  {Zeng}}]{Parra-Martinez:2020dzs}%
  \BibitemOpen
  \bibfield  {author} {\bibinfo {author} {\bibfnamefont {J.}~\bibnamefont
  {Parra-Martinez}}, \bibinfo {author} {\bibfnamefont {M.~S.}\ \bibnamefont
  {Ruf}}, \ and\ \bibinfo {author} {\bibfnamefont {M.}~\bibnamefont {Zeng}},\
  }\href {\doibase 10.1007/JHEP11(2020)023} {\bibfield  {journal} {\bibinfo
  {journal} {JHEP}\ }\textbf {\bibinfo {volume} {11}},\ \bibinfo {pages} {023}
  (\bibinfo {year} {2020})},\ \Eprint {http://arxiv.org/abs/2005.04236}
  {arXiv:2005.04236 [hep-th]} \BibitemShut {NoStop}%
\bibitem [{\citenamefont {Landau}\ and\ \citenamefont
  {Lifschits}(1975)}]{Landau:1975pou}%
  \BibitemOpen
  \bibfield  {author} {\bibinfo {author} {\bibfnamefont {L.~D.}\ \bibnamefont
  {Landau}}\ and\ \bibinfo {author} {\bibfnamefont {E.~M.}\ \bibnamefont
  {Lifschits}},\ }\href@noop {} {\emph {\bibinfo {title} {{The Classical Theory
  of Fields}}}},\ \bibinfo {series} {Course of Theoretical Physics}, Vol.\
  \bibinfo {volume} {Volume 2}\ (\bibinfo  {publisher} {Pergamon Press},\
  \bibinfo {address} {Oxford},\ \bibinfo {year} {1975})\BibitemShut {NoStop}%
\bibitem [{\citenamefont {Kim}\ \emph {et~al.}(2025)\citenamefont {Kim},
  \citenamefont {Patil}, \citenamefont {Scheopner},\ and\ \citenamefont
  {Steinhoff}}]{Kim:2025gis}%
  \BibitemOpen
  \bibfield  {author} {\bibinfo {author} {\bibfnamefont {J.-W.}\ \bibnamefont
  {Kim}}, \bibinfo {author} {\bibfnamefont {R.}~\bibnamefont {Patil}}, \bibinfo
  {author} {\bibfnamefont {T.}~\bibnamefont {Scheopner}}, \ and\ \bibinfo
  {author} {\bibfnamefont {J.}~\bibnamefont {Steinhoff}},\ }\href@noop {} {\
  (\bibinfo {year} {2025})},\ \Eprint {http://arxiv.org/abs/2511.05649}
  {arXiv:2511.05649 [hep-th]} \BibitemShut {NoStop}%
\bibitem [{\citenamefont {Brandhuber}\ \emph {et~al.}(2025)\citenamefont
  {Brandhuber}, \citenamefont {Brown}, \citenamefont {Pichini}, \citenamefont
  {Travaglini},\ and\ \citenamefont {Vives~Matasan}}]{Brandhuber:2025igz}%
  \BibitemOpen
  \bibfield  {author} {\bibinfo {author} {\bibfnamefont {A.}~\bibnamefont
  {Brandhuber}}, \bibinfo {author} {\bibfnamefont {G.~R.}\ \bibnamefont
  {Brown}}, \bibinfo {author} {\bibfnamefont {P.}~\bibnamefont {Pichini}},
  \bibinfo {author} {\bibfnamefont {G.}~\bibnamefont {Travaglini}}, \ and\
  \bibinfo {author} {\bibfnamefont {P.}~\bibnamefont {Vives~Matasan}},\
  }\href@noop {} {\  (\bibinfo {year} {2025})},\ \Eprint
  {http://arxiv.org/abs/2512.05017} {arXiv:2512.05017 [hep-th]} \BibitemShut
  {NoStop}%
\bibitem [{\citenamefont {Bern}\ \emph
  {et~al.}(2025{\natexlab{b}})\citenamefont {Bern}, \citenamefont {Herrmann},
  \citenamefont {Roiban}, \citenamefont {Ruf},\ and\ \citenamefont
  {Zeng}}]{Bern:2024vqs}%
  \BibitemOpen
  \bibfield  {author} {\bibinfo {author} {\bibfnamefont {Z.}~\bibnamefont
  {Bern}}, \bibinfo {author} {\bibfnamefont {E.}~\bibnamefont {Herrmann}},
  \bibinfo {author} {\bibfnamefont {R.}~\bibnamefont {Roiban}}, \bibinfo
  {author} {\bibfnamefont {M.~S.}\ \bibnamefont {Ruf}}, \ and\ \bibinfo
  {author} {\bibfnamefont {M.}~\bibnamefont {Zeng}},\ }\href {\doibase
  10.1007/JHEP06(2025)115} {\bibfield  {journal} {\bibinfo  {journal} {JHEP}\
  }\textbf {\bibinfo {volume} {06}},\ \bibinfo {pages} {115} (\bibinfo {year}
  {2025}{\natexlab{b}})},\ \Eprint {http://arxiv.org/abs/2408.06686}
  {arXiv:2408.06686 [hep-th]} \BibitemShut {NoStop}%
\bibitem [{\citenamefont {Smirnov}\ and\ \citenamefont
  {Smirnov}(2020)}]{Smirnov:2020quc}%
  \BibitemOpen
  \bibfield  {author} {\bibinfo {author} {\bibfnamefont {A.~V.}\ \bibnamefont
  {Smirnov}}\ and\ \bibinfo {author} {\bibfnamefont {V.~A.}\ \bibnamefont
  {Smirnov}},\ }\href {\doibase 10.1016/j.nuclphysb.2020.115213} {\bibfield
  {journal} {\bibinfo  {journal} {Nucl. Phys. B}\ }\textbf {\bibinfo {volume}
  {960}},\ \bibinfo {pages} {115213} (\bibinfo {year} {2020})},\ \Eprint
  {http://arxiv.org/abs/2002.08042} {arXiv:2002.08042 [hep-ph]} \BibitemShut
  {NoStop}%
\bibitem [{\citenamefont {Usovitsch}(2020)}]{Usovitsch:2020jrk}%
  \BibitemOpen
  \bibfield  {author} {\bibinfo {author} {\bibfnamefont {J.}~\bibnamefont
  {Usovitsch}},\ }\href@noop {} {\  (\bibinfo {year} {2020})},\ \Eprint
  {http://arxiv.org/abs/2002.08173} {arXiv:2002.08173 [hep-ph]} \BibitemShut
  {NoStop}%
\bibitem [{\citenamefont {Brunello}\ \emph {et~al.}(2025)\citenamefont
  {Brunello}, \citenamefont {Mandal}, \citenamefont {Mastrolia}, \citenamefont
  {Patil}, \citenamefont {Pegorin}, \citenamefont {Ronca}, \citenamefont
  {Smith}, \citenamefont {Steinhoff},\ and\ \citenamefont
  {Torres~Bobadilla}}]{Brunello:2025gpf}%
  \BibitemOpen
  \bibfield  {author} {\bibinfo {author} {\bibfnamefont {G.}~\bibnamefont
  {Brunello}}, \bibinfo {author} {\bibfnamefont {M.~K.}\ \bibnamefont
  {Mandal}}, \bibinfo {author} {\bibfnamefont {P.}~\bibnamefont {Mastrolia}},
  \bibinfo {author} {\bibfnamefont {R.}~\bibnamefont {Patil}}, \bibinfo
  {author} {\bibfnamefont {M.}~\bibnamefont {Pegorin}}, \bibinfo {author}
  {\bibfnamefont {J.}~\bibnamefont {Ronca}}, \bibinfo {author} {\bibfnamefont
  {S.}~\bibnamefont {Smith}}, \bibinfo {author} {\bibfnamefont
  {J.}~\bibnamefont {Steinhoff}}, \ and\ \bibinfo {author} {\bibfnamefont
  {W.~J.}\ \bibnamefont {Torres~Bobadilla}},\ }\href@noop {} {\  (\bibinfo
  {year} {2025})},\ \Eprint {http://arxiv.org/abs/2512.19498} {arXiv:2512.19498
  [hep-th]} \BibitemShut {NoStop}%
\bibitem [{\citenamefont {Wu}\ \emph {et~al.}(2023)\citenamefont {Wu},
  \citenamefont {Boehm}, \citenamefont {Ma}, \citenamefont {Xu},\ and\
  \citenamefont {Zhang}}]{Wu:2023upw}%
  \BibitemOpen
  \bibfield  {author} {\bibinfo {author} {\bibfnamefont {Z.}~\bibnamefont
  {Wu}}, \bibinfo {author} {\bibfnamefont {J.}~\bibnamefont {Boehm}}, \bibinfo
  {author} {\bibfnamefont {R.}~\bibnamefont {Ma}}, \bibinfo {author}
  {\bibfnamefont {H.}~\bibnamefont {Xu}}, \ and\ \bibinfo {author}
  {\bibfnamefont {Y.}~\bibnamefont {Zhang}},\ }\href@noop {} {\  (\bibinfo
  {year} {2023})},\ \Eprint {http://arxiv.org/abs/2305.08783} {arXiv:2305.08783
  [hep-ph]} \BibitemShut {NoStop}%
\bibitem [{\citenamefont {Abreu}\ \emph {et~al.}(2017)\citenamefont {Abreu},
  \citenamefont {Febres~Cordero}, \citenamefont {Ita}, \citenamefont {Jaquier},
  \citenamefont {Page},\ and\ \citenamefont {Zeng}}]{Abreu:2017xsl}%
  \BibitemOpen
  \bibfield  {author} {\bibinfo {author} {\bibfnamefont {S.}~\bibnamefont
  {Abreu}}, \bibinfo {author} {\bibfnamefont {F.}~\bibnamefont
  {Febres~Cordero}}, \bibinfo {author} {\bibfnamefont {H.}~\bibnamefont {Ita}},
  \bibinfo {author} {\bibfnamefont {M.}~\bibnamefont {Jaquier}}, \bibinfo
  {author} {\bibfnamefont {B.}~\bibnamefont {Page}}, \ and\ \bibinfo {author}
  {\bibfnamefont {M.}~\bibnamefont {Zeng}},\ }\href {\doibase
  10.1103/PhysRevLett.119.142001} {\bibfield  {journal} {\bibinfo  {journal}
  {Phys. Rev. Lett.}\ }\textbf {\bibinfo {volume} {119}},\ \bibinfo {pages}
  {142001} (\bibinfo {year} {2017})},\ \Eprint
  {http://arxiv.org/abs/1703.05273} {arXiv:1703.05273 [hep-ph]} \BibitemShut
  {NoStop}%
\bibitem [{\citenamefont {Abreu}\ \emph {et~al.}(2018)\citenamefont {Abreu},
  \citenamefont {Febres~Cordero}, \citenamefont {Ita}, \citenamefont {Page},\
  and\ \citenamefont {Zeng}}]{Abreu:2017hqn}%
  \BibitemOpen
  \bibfield  {author} {\bibinfo {author} {\bibfnamefont {S.}~\bibnamefont
  {Abreu}}, \bibinfo {author} {\bibfnamefont {F.}~\bibnamefont
  {Febres~Cordero}}, \bibinfo {author} {\bibfnamefont {H.}~\bibnamefont {Ita}},
  \bibinfo {author} {\bibfnamefont {B.}~\bibnamefont {Page}}, \ and\ \bibinfo
  {author} {\bibfnamefont {M.}~\bibnamefont {Zeng}},\ }\href {\doibase
  10.1103/PhysRevD.97.116014} {\bibfield  {journal} {\bibinfo  {journal} {Phys.
  Rev. D}\ }\textbf {\bibinfo {volume} {97}},\ \bibinfo {pages} {116014}
  (\bibinfo {year} {2018})},\ \Eprint {http://arxiv.org/abs/1712.03946}
  {arXiv:1712.03946 [hep-ph]} \BibitemShut {NoStop}%
\bibitem [{\citenamefont {Zeng}\ and\ \citenamefont {Smith}()}]{massDimPaper}%
  \BibitemOpen
  \bibfield  {author} {\bibinfo {author} {\bibfnamefont {M.}~\bibnamefont
  {Zeng}}\ and\ \bibinfo {author} {\bibfnamefont {S.}~\bibnamefont {Smith}},\
  }\href@noop {} {\bibinfo  {journal} {in progress}\ }\BibitemShut {NoStop}%
\bibitem [{\citenamefont {Magerya}(2022)}]{Magerya:2022hvj}%
  \BibitemOpen
\bibfield  {journal} {  }\bibfield  {author} {\bibinfo {author} {\bibfnamefont
  {V.}~\bibnamefont {Magerya}},\ }\href@noop {} {\  (\bibinfo {year} {2022})},\
  \Eprint {http://arxiv.org/abs/2211.03572} {arXiv:2211.03572
  [physics.data-an]} \BibitemShut {NoStop}%
\bibitem [{\citenamefont {Wang}(1981)}]{Wang:1981}%
  \BibitemOpen
  \bibfield  {author} {\bibinfo {author} {\bibfnamefont {P.~S.}\ \bibnamefont
  {Wang}},\ }in\ \href {\doibase 10.1145/800206.806398} {\emph {\bibinfo
  {booktitle} {Proceedings of the Fourth ACM Symposium on Symbolic and
  Algebraic Computation}}},\ \bibinfo {series and number} {SYMSAC '81}\
  (\bibinfo  {publisher} {Association for Computing Machinery},\ \bibinfo
  {address} {New York, NY, USA},\ \bibinfo {year} {1981})\ p.\ \bibinfo {pages}
  {212–217}\BibitemShut {NoStop}%
\bibitem [{\citenamefont {Wang}\ \emph {et~al.}(1982)\citenamefont {Wang},
  \citenamefont {Guy},\ and\ \citenamefont {Davenport}}]{Wang:1982}%
  \BibitemOpen
  \bibfield  {author} {\bibinfo {author} {\bibfnamefont {P.~S.}\ \bibnamefont
  {Wang}}, \bibinfo {author} {\bibfnamefont {M.~J.~T.}\ \bibnamefont {Guy}}, \
  and\ \bibinfo {author} {\bibfnamefont {J.~H.}\ \bibnamefont {Davenport}},\
  }\href {\doibase 10.1145/1089292.1089293} {\bibfield  {journal} {\bibinfo
  {journal} {SIGSAM Bull.}\ }\textbf {\bibinfo {volume} {16}},\ \bibinfo
  {pages} {2–3} (\bibinfo {year} {1982})}\BibitemShut {NoStop}%
\bibitem [{\citenamefont {Mistlberger}(2018)}]{Mistlberger:2018etf}%
  \BibitemOpen
  \bibfield  {author} {\bibinfo {author} {\bibfnamefont {B.}~\bibnamefont
  {Mistlberger}},\ }\href {\doibase 10.1007/JHEP05(2018)028} {\bibfield
  {journal} {\bibinfo  {journal} {JHEP}\ }\textbf {\bibinfo {volume} {05}},\
  \bibinfo {pages} {028} (\bibinfo {year} {2018})},\ \Eprint
  {http://arxiv.org/abs/1802.00833} {arXiv:1802.00833 [hep-ph]} \BibitemShut
  {NoStop}%
\bibitem [{\citenamefont {Moriello}(2020)}]{Moriello:2019yhu}%
  \BibitemOpen
  \bibfield  {author} {\bibinfo {author} {\bibfnamefont {F.}~\bibnamefont
  {Moriello}},\ }\href {\doibase 10.1007/JHEP01(2020)150} {\bibfield  {journal}
  {\bibinfo  {journal} {JHEP}\ }\textbf {\bibinfo {volume} {01}},\ \bibinfo
  {pages} {150} (\bibinfo {year} {2020})},\ \Eprint
  {http://arxiv.org/abs/1907.13234} {arXiv:1907.13234 [hep-ph]} \BibitemShut
  {NoStop}%
\bibitem [{\citenamefont {Pozzorini}\ and\ \citenamefont
  {Remiddi}(2006)}]{Pozzorini:2005ff}%
  \BibitemOpen
  \bibfield  {author} {\bibinfo {author} {\bibfnamefont {S.}~\bibnamefont
  {Pozzorini}}\ and\ \bibinfo {author} {\bibfnamefont {E.}~\bibnamefont
  {Remiddi}},\ }\href {\doibase 10.1016/j.cpc.2006.05.005} {\bibfield
  {journal} {\bibinfo  {journal} {Comput. Phys. Commun.}\ }\textbf {\bibinfo
  {volume} {175}},\ \bibinfo {pages} {381} (\bibinfo {year} {2006})},\ \Eprint
  {http://arxiv.org/abs/hep-ph/0505041} {arXiv:hep-ph/0505041} \BibitemShut
  {NoStop}%
\bibitem [{\citenamefont {Hidding}(2021)}]{Hidding:2020ytt}%
  \BibitemOpen
  \bibfield  {author} {\bibinfo {author} {\bibfnamefont {M.}~\bibnamefont
  {Hidding}},\ }\href {\doibase 10.1016/j.cpc.2021.108125} {\bibfield
  {journal} {\bibinfo  {journal} {Comput. Phys. Commun.}\ }\textbf {\bibinfo
  {volume} {269}},\ \bibinfo {pages} {108125} (\bibinfo {year} {2021})},\
  \Eprint {http://arxiv.org/abs/2006.05510} {arXiv:2006.05510 [hep-ph]}
  \BibitemShut {NoStop}%
\bibitem [{\citenamefont {Saotome}\ and\ \citenamefont
  {Akhoury}(2013)}]{Saotome:2012vy}%
  \BibitemOpen
  \bibfield  {author} {\bibinfo {author} {\bibfnamefont {R.}~\bibnamefont
  {Saotome}}\ and\ \bibinfo {author} {\bibfnamefont {R.}~\bibnamefont
  {Akhoury}},\ }\href {\doibase 10.1007/JHEP01(2013)123} {\bibfield  {journal}
  {\bibinfo  {journal} {JHEP}\ }\textbf {\bibinfo {volume} {01}},\ \bibinfo
  {pages} {123} (\bibinfo {year} {2013})},\ \Eprint
  {http://arxiv.org/abs/1210.8111} {arXiv:1210.8111 [hep-th]} \BibitemShut
  {NoStop}%
\bibitem [{\citenamefont {Akhoury}\ \emph {et~al.}(2021)\citenamefont
  {Akhoury}, \citenamefont {Saotome},\ and\ \citenamefont
  {Sterman}}]{Akhoury:2013yua}%
  \BibitemOpen
  \bibfield  {author} {\bibinfo {author} {\bibfnamefont {R.}~\bibnamefont
  {Akhoury}}, \bibinfo {author} {\bibfnamefont {R.}~\bibnamefont {Saotome}}, \
  and\ \bibinfo {author} {\bibfnamefont {G.}~\bibnamefont {Sterman}},\ }\href
  {\doibase 10.1103/PhysRevD.103.064036} {\bibfield  {journal} {\bibinfo
  {journal} {Phys. Rev. D}\ }\textbf {\bibinfo {volume} {103}},\ \bibinfo
  {pages} {064036} (\bibinfo {year} {2021})},\ \Eprint
  {http://arxiv.org/abs/1308.5204} {arXiv:1308.5204 [hep-th]} \BibitemShut
  {NoStop}%
\bibitem [{\citenamefont {Frellesvig}\ \emph {et~al.}(2024)\citenamefont
  {Frellesvig}, \citenamefont {Morales},\ and\ \citenamefont
  {Wilhelm}}]{Frellesvig:2023bbf}%
  \BibitemOpen
  \bibfield  {author} {\bibinfo {author} {\bibfnamefont {H.}~\bibnamefont
  {Frellesvig}}, \bibinfo {author} {\bibfnamefont {R.}~\bibnamefont {Morales}},
  \ and\ \bibinfo {author} {\bibfnamefont {M.}~\bibnamefont {Wilhelm}},\ }\href
  {\doibase 10.1103/PhysRevLett.132.201602} {\bibfield  {journal} {\bibinfo
  {journal} {Phys. Rev. Lett.}\ }\textbf {\bibinfo {volume} {132}},\ \bibinfo
  {pages} {201602} (\bibinfo {year} {2024})},\ \Eprint
  {http://arxiv.org/abs/2312.11371} {arXiv:2312.11371 [hep-th]} \BibitemShut
  {NoStop}%
\bibitem [{\citenamefont {Klemm}\ \emph {et~al.}(2024)\citenamefont {Klemm},
  \citenamefont {Nega}, \citenamefont {Sauer},\ and\ \citenamefont
  {Plefka}}]{Klemm:2024wtd}%
  \BibitemOpen
  \bibfield  {author} {\bibinfo {author} {\bibfnamefont {A.}~\bibnamefont
  {Klemm}}, \bibinfo {author} {\bibfnamefont {C.}~\bibnamefont {Nega}},
  \bibinfo {author} {\bibfnamefont {B.}~\bibnamefont {Sauer}}, \ and\ \bibinfo
  {author} {\bibfnamefont {J.}~\bibnamefont {Plefka}},\ }\href {\doibase
  10.1103/PhysRevD.109.124046} {\bibfield  {journal} {\bibinfo  {journal}
  {Phys. Rev. D}\ }\textbf {\bibinfo {volume} {109}},\ \bibinfo {pages}
  {124046} (\bibinfo {year} {2024})},\ \Eprint
  {http://arxiv.org/abs/2401.07899} {arXiv:2401.07899 [hep-th]} \BibitemShut
  {NoStop}%
\bibitem [{\citenamefont {Brammer}\ \emph {et~al.}(2025)\citenamefont
  {Brammer}, \citenamefont {Frellesvig}, \citenamefont {Morales},\ and\
  \citenamefont {Wilhelm}}]{Brammer:2025rqo}%
  \BibitemOpen
  \bibfield  {author} {\bibinfo {author} {\bibfnamefont {D.}~\bibnamefont
  {Brammer}}, \bibinfo {author} {\bibfnamefont {H.}~\bibnamefont {Frellesvig}},
  \bibinfo {author} {\bibfnamefont {R.}~\bibnamefont {Morales}}, \ and\
  \bibinfo {author} {\bibfnamefont {M.}~\bibnamefont {Wilhelm}},\ }\href@noop
  {} {\  (\bibinfo {year} {2025})},\ \Eprint {http://arxiv.org/abs/2505.10274}
  {arXiv:2505.10274 [hep-th]} \BibitemShut {NoStop}%
\bibitem [{\citenamefont {Frellesvig}\ \emph {et~al.}(2025)\citenamefont
  {Frellesvig}, \citenamefont {Morales}, \citenamefont {P{\"o}gel},
  \citenamefont {Weinzierl},\ and\ \citenamefont
  {Wilhelm}}]{Frellesvig:2024rea}%
  \BibitemOpen
  \bibfield  {author} {\bibinfo {author} {\bibfnamefont {H.}~\bibnamefont
  {Frellesvig}}, \bibinfo {author} {\bibfnamefont {R.}~\bibnamefont {Morales}},
  \bibinfo {author} {\bibfnamefont {S.}~\bibnamefont {P{\"o}gel}}, \bibinfo
  {author} {\bibfnamefont {S.}~\bibnamefont {Weinzierl}}, \ and\ \bibinfo
  {author} {\bibfnamefont {M.}~\bibnamefont {Wilhelm}},\ }\href {\doibase
  10.1007/JHEP02(2025)209} {\bibfield  {journal} {\bibinfo  {journal} {JHEP}\
  }\textbf {\bibinfo {volume} {02}},\ \bibinfo {pages} {209} (\bibinfo {year}
  {2025})},\ \Eprint {http://arxiv.org/abs/2412.12057} {arXiv:2412.12057
  [hep-th]} \BibitemShut {NoStop}%
\bibitem [{\citenamefont {Duhr}\ \emph {et~al.}(2025)\citenamefont {Duhr},
  \citenamefont {Maggio}, \citenamefont {Nega}, \citenamefont {Sauer},
  \citenamefont {Tancredi},\ and\ \citenamefont {Wagner}}]{Duhr:2025lbz}%
  \BibitemOpen
  \bibfield  {author} {\bibinfo {author} {\bibfnamefont {C.}~\bibnamefont
  {Duhr}}, \bibinfo {author} {\bibfnamefont {S.}~\bibnamefont {Maggio}},
  \bibinfo {author} {\bibfnamefont {C.}~\bibnamefont {Nega}}, \bibinfo {author}
  {\bibfnamefont {B.}~\bibnamefont {Sauer}}, \bibinfo {author} {\bibfnamefont
  {L.}~\bibnamefont {Tancredi}}, \ and\ \bibinfo {author} {\bibfnamefont
  {F.~J.}\ \bibnamefont {Wagner}},\ }\href {\doibase 10.1007/JHEP06(2025)128}
  {\bibfield  {journal} {\bibinfo  {journal} {JHEP}\ }\textbf {\bibinfo
  {volume} {06}},\ \bibinfo {pages} {128} (\bibinfo {year} {2025})},\ \Eprint
  {http://arxiv.org/abs/2503.20655} {arXiv:2503.20655 [hep-th]} \BibitemShut
  {NoStop}%
\bibitem [{\citenamefont {Foffa}\ \emph {et~al.}(2017)\citenamefont {Foffa},
  \citenamefont {Mastrolia}, \citenamefont {Sturani},\ and\ \citenamefont
  {Sturm}}]{Foffa:2016rgu}%
  \BibitemOpen
  \bibfield  {author} {\bibinfo {author} {\bibfnamefont {S.}~\bibnamefont
  {Foffa}}, \bibinfo {author} {\bibfnamefont {P.}~\bibnamefont {Mastrolia}},
  \bibinfo {author} {\bibfnamefont {R.}~\bibnamefont {Sturani}}, \ and\
  \bibinfo {author} {\bibfnamefont {C.}~\bibnamefont {Sturm}},\ }\href
  {\doibase 10.1103/PhysRevD.95.104009} {\bibfield  {journal} {\bibinfo
  {journal} {Phys. Rev. D}\ }\textbf {\bibinfo {volume} {95}},\ \bibinfo
  {pages} {104009} (\bibinfo {year} {2017})},\ \Eprint
  {http://arxiv.org/abs/1612.00482} {arXiv:1612.00482 [gr-qc]} \BibitemShut
  {NoStop}%
\bibitem [{\citenamefont {Bern}\ \emph {et~al.}()\citenamefont {Bern},
  \citenamefont {Jackman}, \citenamefont {Mansfield},\ and\ \citenamefont
  {Ruf}}]{CYExplanation}%
  \BibitemOpen
  \bibfield  {author} {\bibinfo {author} {\bibfnamefont {Z.}~\bibnamefont
  {Bern}}, \bibinfo {author} {\bibfnamefont {A.}~\bibnamefont {Jackman}},
  \bibinfo {author} {\bibfnamefont {G.}~\bibnamefont {Mansfield}}, \ and\
  \bibinfo {author} {\bibfnamefont {M.}~\bibnamefont {Ruf}},\ }\href@noop {}
  {\bibinfo  {journal} {in progress}\ }\BibitemShut {NoStop}%
\bibitem [{\citenamefont {Joyce}(1973)}]{GSJoyceCubicLatticeGF}%
  \BibitemOpen
\bibfield  {journal} {  }\bibfield  {author} {\bibinfo {author} {\bibfnamefont
  {G.~S.}\ \bibnamefont {Joyce}},\ }\href {http://www.jstor.org/stable/74037}
  {\bibfield  {journal} {\bibinfo  {journal} {Philosophical Transactions of the
  Royal Society of London. Series A, Mathematical and Physical Sciences}\
  }\textbf {\bibinfo {volume} {273}},\ \bibinfo {pages} {583} (\bibinfo {year}
  {1973})}\BibitemShut {NoStop}%
\bibitem [{\citenamefont {Joyce}\ and\ \citenamefont
  {Delves}(2004)}]{GSJoyce_2004}%
  \BibitemOpen
  \bibfield  {author} {\bibinfo {author} {\bibfnamefont {G.~S.}\ \bibnamefont
  {Joyce}}\ and\ \bibinfo {author} {\bibfnamefont {R.~T.}\ \bibnamefont
  {Delves}},\ }\href {\doibase 10.1088/0305-4470/37/20/012} {\bibfield
  {journal} {\bibinfo  {journal} {Journal of Physics A: Mathematical and
  General}\ }\textbf {\bibinfo {volume} {37}},\ \bibinfo {pages} {5417}
  (\bibinfo {year} {2004})}\BibitemShut {NoStop}%
\bibitem [{\citenamefont {Ronveaux}(1995)}]{Ronveaux:1995Heun}%
  \BibitemOpen
  \bibinfo {editor} {\bibfnamefont {A.}~\bibnamefont {Ronveaux}},\ ed.,\
  \href@noop {} {\emph {\bibinfo {title} {Heun's Differential Equations}}}\
  (\bibinfo  {publisher} {Oxford University Press},\ \bibinfo {address}
  {Oxford},\ \bibinfo {year} {1995})\BibitemShut {NoStop}%
\bibitem [{\citenamefont {Ablinger}\ \emph {et~al.}(2018)\citenamefont
  {Ablinger}, \citenamefont {Bl{\"u}mlein}, \citenamefont {De~Freitas},
  \citenamefont {van Hoeij}, \citenamefont {Imamoglu}, \citenamefont {Raab},
  \citenamefont {Radu},\ and\ \citenamefont {Schneider}}]{Ablinger:2017bjx}%
  \BibitemOpen
  \bibfield  {author} {\bibinfo {author} {\bibfnamefont {J.}~\bibnamefont
  {Ablinger}}, \bibinfo {author} {\bibfnamefont {J.}~\bibnamefont
  {Bl{\"u}mlein}}, \bibinfo {author} {\bibfnamefont {A.}~\bibnamefont
  {De~Freitas}}, \bibinfo {author} {\bibfnamefont {M.}~\bibnamefont {van
  Hoeij}}, \bibinfo {author} {\bibfnamefont {E.}~\bibnamefont {Imamoglu}},
  \bibinfo {author} {\bibfnamefont {C.~G.}\ \bibnamefont {Raab}}, \bibinfo
  {author} {\bibfnamefont {C.~S.}\ \bibnamefont {Radu}}, \ and\ \bibinfo
  {author} {\bibfnamefont {C.}~\bibnamefont {Schneider}},\ }\href {\doibase
  10.1063/1.4986417} {\bibfield  {journal} {\bibinfo  {journal} {J. Math.
  Phys.}\ }\textbf {\bibinfo {volume} {59}},\ \bibinfo {pages} {062305}
  (\bibinfo {year} {2018})},\ \Eprint {http://arxiv.org/abs/1706.01299}
  {arXiv:1706.01299 [hep-th]} \BibitemShut {NoStop}%
\bibitem [{\citenamefont {Ruf}(2023)}]{MRAmplitudes2023}%
  \BibitemOpen
  \bibfield  {author} {\bibinfo {author} {\bibfnamefont {M.}~\bibnamefont
  {Ruf}},\ }\href {https://indico.cern.ch/event/1228963/contributions/5506364/}
  {\enquote {\bibinfo {title} {Towards gravitational scattering at the fifth
  order in $g$},}\ } (\bibinfo {year} {2023}),\ \bibinfo {note} {{Amplitudes
  2023 Conference}}\BibitemShut {NoStop}%
\bibitem [{\citenamefont {Manohar}\ and\ \citenamefont
  {Stewart}(2007)}]{Manohar:2006nz}%
  \BibitemOpen
  \bibfield  {author} {\bibinfo {author} {\bibfnamefont {A.~V.}\ \bibnamefont
  {Manohar}}\ and\ \bibinfo {author} {\bibfnamefont {I.~W.}\ \bibnamefont
  {Stewart}},\ }\href {\doibase 10.1103/PhysRevD.76.074002} {\bibfield
  {journal} {\bibinfo  {journal} {Phys. Rev. D}\ }\textbf {\bibinfo {volume}
  {76}},\ \bibinfo {pages} {074002} (\bibinfo {year} {2007})},\ \Eprint
  {http://arxiv.org/abs/hep-ph/0605001} {arXiv:hep-ph/0605001} \BibitemShut
  {NoStop}%
\bibitem [{\citenamefont {Porto}\ and\ \citenamefont
  {Rothstein}(2017)}]{Porto:2017dgs}%
  \BibitemOpen
  \bibfield  {author} {\bibinfo {author} {\bibfnamefont {R.~A.}\ \bibnamefont
  {Porto}}\ and\ \bibinfo {author} {\bibfnamefont {I.~Z.}\ \bibnamefont
  {Rothstein}},\ }\href {\doibase 10.1103/PhysRevD.96.024062} {\bibfield
  {journal} {\bibinfo  {journal} {Phys. Rev. D}\ }\textbf {\bibinfo {volume}
  {96}},\ \bibinfo {pages} {024062} (\bibinfo {year} {2017})},\ \Eprint
  {http://arxiv.org/abs/1703.06433} {arXiv:1703.06433 [gr-qc]} \BibitemShut
  {NoStop}%
\bibitem [{\citenamefont {Galley}\ \emph {et~al.}(2016)\citenamefont {Galley},
  \citenamefont {Leibovich}, \citenamefont {Porto},\ and\ \citenamefont
  {Ross}}]{Galley:2015kus}%
  \BibitemOpen
  \bibfield  {author} {\bibinfo {author} {\bibfnamefont {C.~R.}\ \bibnamefont
  {Galley}}, \bibinfo {author} {\bibfnamefont {A.~K.}\ \bibnamefont
  {Leibovich}}, \bibinfo {author} {\bibfnamefont {R.~A.}\ \bibnamefont
  {Porto}}, \ and\ \bibinfo {author} {\bibfnamefont {A.}~\bibnamefont {Ross}},\
  }\href {\doibase 10.1103/PhysRevD.93.124010} {\bibfield  {journal} {\bibinfo
  {journal} {Phys. Rev. D}\ }\textbf {\bibinfo {volume} {93}},\ \bibinfo
  {pages} {124010} (\bibinfo {year} {2016})},\ \Eprint
  {http://arxiv.org/abs/1511.07379} {arXiv:1511.07379 [gr-qc]} \BibitemShut
  {NoStop}%
\bibitem [{\citenamefont {Bernard}\ \emph {et~al.}(2017)\citenamefont
  {Bernard}, \citenamefont {Blanchet}, \citenamefont {Boh{\'e}}, \citenamefont
  {Faye},\ and\ \citenamefont {Marsat}}]{Bernard:2017bvn}%
  \BibitemOpen
  \bibfield  {author} {\bibinfo {author} {\bibfnamefont {L.}~\bibnamefont
  {Bernard}}, \bibinfo {author} {\bibfnamefont {L.}~\bibnamefont {Blanchet}},
  \bibinfo {author} {\bibfnamefont {A.}~\bibnamefont {Boh{\'e}}}, \bibinfo
  {author} {\bibfnamefont {G.}~\bibnamefont {Faye}}, \ and\ \bibinfo {author}
  {\bibfnamefont {S.}~\bibnamefont {Marsat}},\ }\href {\doibase
  10.1103/PhysRevD.96.104043} {\bibfield  {journal} {\bibinfo  {journal} {Phys.
  Rev. D}\ }\textbf {\bibinfo {volume} {96}},\ \bibinfo {pages} {104043}
  (\bibinfo {year} {2017})},\ \Eprint {http://arxiv.org/abs/1706.08480}
  {arXiv:1706.08480 [gr-qc]} \BibitemShut {NoStop}%
\bibitem [{\citenamefont {Bini}\ and\ \citenamefont
  {Damour}(2017)}]{Bini:2017wfr}%
  \BibitemOpen
  \bibfield  {author} {\bibinfo {author} {\bibfnamefont {D.}~\bibnamefont
  {Bini}}\ and\ \bibinfo {author} {\bibfnamefont {T.}~\bibnamefont {Damour}},\
  }\href {\doibase 10.1103/PhysRevD.96.064021} {\bibfield  {journal} {\bibinfo
  {journal} {Phys. Rev. D}\ }\textbf {\bibinfo {volume} {96}},\ \bibinfo
  {pages} {064021} (\bibinfo {year} {2017})},\ \Eprint
  {http://arxiv.org/abs/1706.06877} {arXiv:1706.06877 [gr-qc]} \BibitemShut
  {NoStop}%
\bibitem [{\citenamefont {Bini}\ \emph {et~al.}(2020)\citenamefont {Bini},
  \citenamefont {Damour},\ and\ \citenamefont {Geralico}}]{Bini:2020hmy}%
  \BibitemOpen
  \bibfield  {author} {\bibinfo {author} {\bibfnamefont {D.}~\bibnamefont
  {Bini}}, \bibinfo {author} {\bibfnamefont {T.}~\bibnamefont {Damour}}, \ and\
  \bibinfo {author} {\bibfnamefont {A.}~\bibnamefont {Geralico}},\ }\href
  {\doibase 10.1103/PhysRevD.102.084047} {\bibfield  {journal} {\bibinfo
  {journal} {Phys. Rev. D}\ }\textbf {\bibinfo {volume} {102}},\ \bibinfo
  {pages} {084047} (\bibinfo {year} {2020})},\ \Eprint
  {http://arxiv.org/abs/2007.11239} {arXiv:2007.11239 [gr-qc]} \BibitemShut
  {NoStop}%
\bibitem [{\citenamefont {Blanchet}\ \emph {et~al.}(2020)\citenamefont
  {Blanchet}, \citenamefont {Foffa}, \citenamefont {Larrouturou},\ and\
  \citenamefont {Sturani}}]{Blanchet:2019rjs}%
  \BibitemOpen
  \bibfield  {author} {\bibinfo {author} {\bibfnamefont {L.}~\bibnamefont
  {Blanchet}}, \bibinfo {author} {\bibfnamefont {S.}~\bibnamefont {Foffa}},
  \bibinfo {author} {\bibfnamefont {F.}~\bibnamefont {Larrouturou}}, \ and\
  \bibinfo {author} {\bibfnamefont {R.}~\bibnamefont {Sturani}},\ }\href
  {\doibase 10.1103/PhysRevD.101.084045} {\bibfield  {journal} {\bibinfo
  {journal} {Phys. Rev. D}\ }\textbf {\bibinfo {volume} {101}},\ \bibinfo
  {pages} {084045} (\bibinfo {year} {2020})},\ \Eprint
  {http://arxiv.org/abs/1912.12359} {arXiv:1912.12359 [gr-qc]} \BibitemShut
  {NoStop}%
\bibitem [{\citenamefont {Dlapa}\ \emph {et~al.}(2023)\citenamefont {Dlapa},
  \citenamefont {K\"alin}, \citenamefont {Liu}, \citenamefont {Neef},\ and\
  \citenamefont {Porto}}]{Dlapa:2022lmu}%
  \BibitemOpen
  \bibfield  {author} {\bibinfo {author} {\bibfnamefont {C.}~\bibnamefont
  {Dlapa}}, \bibinfo {author} {\bibfnamefont {G.}~\bibnamefont {K\"alin}},
  \bibinfo {author} {\bibfnamefont {Z.}~\bibnamefont {Liu}}, \bibinfo {author}
  {\bibfnamefont {J.}~\bibnamefont {Neef}}, \ and\ \bibinfo {author}
  {\bibfnamefont {R.~A.}\ \bibnamefont {Porto}},\ }\href {\doibase
  10.1103/PhysRevLett.130.101401} {\bibfield  {journal} {\bibinfo  {journal}
  {Phys. Rev. Lett.}\ }\textbf {\bibinfo {volume} {130}},\ \bibinfo {pages}
  {101401} (\bibinfo {year} {2023})},\ \Eprint
  {http://arxiv.org/abs/2210.05541} {arXiv:2210.05541 [hep-th]} \BibitemShut
  {NoStop}%
\bibitem [{\citenamefont {Ablinger}\ \emph {et~al.}(2011)\citenamefont
  {Ablinger}, \citenamefont {Bl{\"u}mlein},\ and\ \citenamefont
  {Schneider}}]{Ablinger:2011te}%
  \BibitemOpen
  \bibfield  {author} {\bibinfo {author} {\bibfnamefont {J.}~\bibnamefont
  {Ablinger}}, \bibinfo {author} {\bibfnamefont {J.}~\bibnamefont
  {Bl{\"u}mlein}}, \ and\ \bibinfo {author} {\bibfnamefont {C.}~\bibnamefont
  {Schneider}},\ }\href {\doibase 10.1063/1.3629472} {\bibfield  {journal}
  {\bibinfo  {journal} {J. Math. Phys.}\ }\textbf {\bibinfo {volume} {52}},\
  \bibinfo {pages} {102301} (\bibinfo {year} {2011})},\ \Eprint
  {http://arxiv.org/abs/1105.6063} {arXiv:1105.6063 [math-ph]} \BibitemShut
  {NoStop}%
\bibitem [{\citenamefont {Ablinger}\ \emph {et~al.}(2021)\citenamefont
  {Ablinger}, \citenamefont {Bl{\"u}mlein},\ and\ \citenamefont
  {Schneider}}]{Ablinger:2021fnc}%
  \BibitemOpen
  \bibfield  {author} {\bibinfo {author} {\bibfnamefont {J.}~\bibnamefont
  {Ablinger}}, \bibinfo {author} {\bibfnamefont {J.}~\bibnamefont
  {Bl{\"u}mlein}}, \ and\ \bibinfo {author} {\bibfnamefont {C.}~\bibnamefont
  {Schneider}},\ }\href {\doibase 10.1103/PhysRevD.103.096025} {\bibfield
  {journal} {\bibinfo  {journal} {Phys. Rev. D}\ }\textbf {\bibinfo {volume}
  {103}},\ \bibinfo {pages} {096025} (\bibinfo {year} {2021})},\ \Eprint
  {http://arxiv.org/abs/2103.08330} {arXiv:2103.08330 [hep-th]} \BibitemShut
  {NoStop}%
\bibitem [{\citenamefont {Kol}\ \emph {et~al.}(2022)\citenamefont {Kol},
  \citenamefont {O'connell},\ and\ \citenamefont {Telem}}]{Kol:2021jjc}%
  \BibitemOpen
  \bibfield  {author} {\bibinfo {author} {\bibfnamefont {U.}~\bibnamefont
  {Kol}}, \bibinfo {author} {\bibfnamefont {D.}~\bibnamefont {O'connell}}, \
  and\ \bibinfo {author} {\bibfnamefont {O.}~\bibnamefont {Telem}},\ }\href
  {\doibase 10.1007/JHEP03(2022)141} {\bibfield  {journal} {\bibinfo  {journal}
  {JHEP}\ }\textbf {\bibinfo {volume} {03}},\ \bibinfo {pages} {141} (\bibinfo
  {year} {2022})},\ \Eprint {http://arxiv.org/abs/2109.12092} {arXiv:2109.12092
  [hep-th]} \BibitemShut {NoStop}%
\bibitem [{\citenamefont {Bl{\"u}mlein}\ \emph {et~al.}(2020)\citenamefont
  {Bl{\"u}mlein}, \citenamefont {Maier}, \citenamefont {Marquard},\ and\
  \citenamefont {Sch{\"a}fer}}]{Blumlein:2020pog}%
  \BibitemOpen
  \bibfield  {author} {\bibinfo {author} {\bibfnamefont {J.}~\bibnamefont
  {Bl{\"u}mlein}}, \bibinfo {author} {\bibfnamefont {A.}~\bibnamefont {Maier}},
  \bibinfo {author} {\bibfnamefont {P.}~\bibnamefont {Marquard}}, \ and\
  \bibinfo {author} {\bibfnamefont {G.}~\bibnamefont {Sch{\"a}fer}},\ }\href
  {\doibase 10.1016/j.nuclphysb.2020.115041} {\bibfield  {journal} {\bibinfo
  {journal} {Nucl. Phys. B}\ }\textbf {\bibinfo {volume} {955}},\ \bibinfo
  {pages} {115041} (\bibinfo {year} {2020})},\ \Eprint
  {http://arxiv.org/abs/2003.01692} {arXiv:2003.01692 [gr-qc]} \BibitemShut
  {NoStop}%
\bibitem [{\citenamefont {Bl{\"u}mlein}\ \emph {et~al.}(2021)\citenamefont
  {Bl{\"u}mlein}, \citenamefont {Maier}, \citenamefont {Marquard},\ and\
  \citenamefont {Sch{\"a}fer}}]{Blumlein:2020pyo}%
  \BibitemOpen
  \bibfield  {author} {\bibinfo {author} {\bibfnamefont {J.}~\bibnamefont
  {Bl{\"u}mlein}}, \bibinfo {author} {\bibfnamefont {A.}~\bibnamefont {Maier}},
  \bibinfo {author} {\bibfnamefont {P.}~\bibnamefont {Marquard}}, \ and\
  \bibinfo {author} {\bibfnamefont {G.}~\bibnamefont {Sch{\"a}fer}},\ }\href
  {\doibase 10.1016/j.nuclphysb.2021.115352} {\bibfield  {journal} {\bibinfo
  {journal} {Nucl. Phys. B}\ }\textbf {\bibinfo {volume} {965}},\ \bibinfo
  {pages} {115352} (\bibinfo {year} {2021})},\ \Eprint
  {http://arxiv.org/abs/2010.13672} {arXiv:2010.13672 [gr-qc]} \BibitemShut
  {NoStop}%
\bibitem [{\citenamefont {Bl{\"u}mlein}\ \emph {et~al.}(2022)\citenamefont
  {Bl{\"u}mlein}, \citenamefont {Maier}, \citenamefont {Marquard},\ and\
  \citenamefont {Sch{\"a}fer}}]{Blumlein:2021txe}%
  \BibitemOpen
  \bibfield  {author} {\bibinfo {author} {\bibfnamefont {J.}~\bibnamefont
  {Bl{\"u}mlein}}, \bibinfo {author} {\bibfnamefont {A.}~\bibnamefont {Maier}},
  \bibinfo {author} {\bibfnamefont {P.}~\bibnamefont {Marquard}}, \ and\
  \bibinfo {author} {\bibfnamefont {G.}~\bibnamefont {Sch{\"a}fer}},\ }\href
  {\doibase 10.1016/j.nuclphysb.2022.115900} {\bibfield  {journal} {\bibinfo
  {journal} {Nucl. Phys. B}\ }\textbf {\bibinfo {volume} {983}},\ \bibinfo
  {pages} {115900} (\bibinfo {year} {2022})},\ \bibinfo {note} {[Erratum:
  Nucl.Phys.B 985, 115991 (2022)]},\ \Eprint {http://arxiv.org/abs/2110.13822}
  {arXiv:2110.13822 [gr-qc]} \BibitemShut {NoStop}%
\bibitem [{\citenamefont {Ruf}()}]{Ruf:Amplitudes23}%
  \BibitemOpen
  \bibfield  {author} {\bibinfo {author} {\bibfnamefont {M.}~\bibnamefont
  {Ruf}},\ }\href@noop {} {\ }\Eprint {http://arxiv.org/abs/Talk at Amplitudes
  23, CERN, August, 2023, Slide 22} {Talk at Amplitudes 23, CERN, August, 2023,
  Slide 22} \BibitemShut {NoStop}%
\bibitem [{\citenamefont {Damour}\ \emph {et~al.}(1998)\citenamefont {Damour},
  \citenamefont {Iyer},\ and\ \citenamefont {Sathyaprakash}}]{Damour:1997ub}%
  \BibitemOpen
  \bibfield  {author} {\bibinfo {author} {\bibfnamefont {T.}~\bibnamefont
  {Damour}}, \bibinfo {author} {\bibfnamefont {B.~R.}\ \bibnamefont {Iyer}}, \
  and\ \bibinfo {author} {\bibfnamefont {B.~S.}\ \bibnamefont
  {Sathyaprakash}},\ }\href {\doibase 10.1103/PhysRevD.57.885} {\bibfield
  {journal} {\bibinfo  {journal} {Phys. Rev. D}\ }\textbf {\bibinfo {volume}
  {57}},\ \bibinfo {pages} {885} (\bibinfo {year} {1998})},\ \Eprint
  {http://arxiv.org/abs/gr-qc/9708034} {arXiv:gr-qc/9708034} \BibitemShut
  {NoStop}%
\bibitem [{\citenamefont {Nagar}\ and\ \citenamefont
  {Shah}(2016)}]{Nagar:2016ayt}%
  \BibitemOpen
  \bibfield  {author} {\bibinfo {author} {\bibfnamefont {A.}~\bibnamefont
  {Nagar}}\ and\ \bibinfo {author} {\bibfnamefont {A.}~\bibnamefont {Shah}},\
  }\href {\doibase 10.1103/PhysRevD.94.104017} {\bibfield  {journal} {\bibinfo
  {journal} {Phys. Rev. D}\ }\textbf {\bibinfo {volume} {94}},\ \bibinfo
  {pages} {104017} (\bibinfo {year} {2016})},\ \Eprint
  {http://arxiv.org/abs/1606.00207} {arXiv:1606.00207 [gr-qc]} \BibitemShut
  {NoStop}%
\bibitem [{\citenamefont {Nagar}\ \emph {et~al.}(2025)\citenamefont {Nagar},
  \citenamefont {Chiaramello}, \citenamefont {Gamba}, \citenamefont {Albanesi},
  \citenamefont {Bernuzzi}, \citenamefont {Fantini}, \citenamefont {Panzeri},\
  and\ \citenamefont {Rettegno}}]{Nagar:2024oyk}%
  \BibitemOpen
  \bibfield  {author} {\bibinfo {author} {\bibfnamefont {A.}~\bibnamefont
  {Nagar}}, \bibinfo {author} {\bibfnamefont {D.}~\bibnamefont {Chiaramello}},
  \bibinfo {author} {\bibfnamefont {R.}~\bibnamefont {Gamba}}, \bibinfo
  {author} {\bibfnamefont {S.}~\bibnamefont {Albanesi}}, \bibinfo {author}
  {\bibfnamefont {S.}~\bibnamefont {Bernuzzi}}, \bibinfo {author}
  {\bibfnamefont {V.}~\bibnamefont {Fantini}}, \bibinfo {author} {\bibfnamefont
  {M.}~\bibnamefont {Panzeri}}, \ and\ \bibinfo {author} {\bibfnamefont
  {P.}~\bibnamefont {Rettegno}},\ }\href {\doibase 10.1103/PhysRevD.111.064050}
  {\bibfield  {journal} {\bibinfo  {journal} {Phys. Rev. D}\ }\textbf {\bibinfo
  {volume} {111}},\ \bibinfo {pages} {064050} (\bibinfo {year} {2025})},\
  \Eprint {http://arxiv.org/abs/2407.04762} {arXiv:2407.04762 [gr-qc]}
  \BibitemShut {NoStop}%
\bibitem [{\citenamefont {Bultheel}(1984)}]{Bultheel:1984aa}%
  \BibitemOpen
  \bibfield  {author} {\bibinfo {author} {\bibfnamefont {A.}~\bibnamefont
  {Bultheel}},\ }\href@noop {} {\emph {\bibinfo {title} {Laurent series and
  their Padé approximations}}}\ (\bibinfo  {publisher} {Birkh\"auser Verlag},\
  \bibinfo {address} {Basel},\ \bibinfo {year} {1984})\BibitemShut {NoStop}%
\bibitem [{\citenamefont {Fleischer}\ and\ \citenamefont
  {Tarasov}(1994)}]{Fleischer:1994ef}%
  \BibitemOpen
  \bibfield  {author} {\bibinfo {author} {\bibfnamefont {J.}~\bibnamefont
  {Fleischer}}\ and\ \bibinfo {author} {\bibfnamefont {O.~V.}\ \bibnamefont
  {Tarasov}},\ }\href {\doibase 10.1007/BF01560102} {\bibfield  {journal}
  {\bibinfo  {journal} {Z. Phys. C}\ }\textbf {\bibinfo {volume} {64}},\
  \bibinfo {pages} {413} (\bibinfo {year} {1994})},\ \Eprint
  {http://arxiv.org/abs/hep-ph/9403230} {arXiv:hep-ph/9403230} \BibitemShut
  {NoStop}%
\bibitem [{\citenamefont {Fleischer}\ \emph {et~al.}(1997)\citenamefont
  {Fleischer}, \citenamefont {Smirnov},\ and\ \citenamefont
  {Tarasov}}]{Fleischer:1996ju}%
  \BibitemOpen
  \bibfield  {author} {\bibinfo {author} {\bibfnamefont {J.}~\bibnamefont
  {Fleischer}}, \bibinfo {author} {\bibfnamefont {V.~A.}\ \bibnamefont
  {Smirnov}}, \ and\ \bibinfo {author} {\bibfnamefont {O.~V.}\ \bibnamefont
  {Tarasov}},\ }\href {\doibase 10.1007/s002880050400} {\bibfield  {journal}
  {\bibinfo  {journal} {Z. Phys. C}\ }\textbf {\bibinfo {volume} {74}},\
  \bibinfo {pages} {379} (\bibinfo {year} {1997})},\ \Eprint
  {http://arxiv.org/abs/hep-ph/9605392} {arXiv:hep-ph/9605392} \BibitemShut
  {NoStop}%
\bibitem [{\citenamefont {Rothstein}\ and\ \citenamefont
  {Saavedra}(2024)}]{Rothstein:2024nlq}%
  \BibitemOpen
  \bibfield  {author} {\bibinfo {author} {\bibfnamefont {I.~Z.}\ \bibnamefont
  {Rothstein}}\ and\ \bibinfo {author} {\bibfnamefont {M.}~\bibnamefont
  {Saavedra}},\ }\href@noop {} {\  (\bibinfo {year} {2024})},\ \Eprint
  {http://arxiv.org/abs/2412.04428} {arXiv:2412.04428 [hep-th]} \BibitemShut
  {NoStop}%
\bibitem [{\citenamefont {Barcaro}\ and\ \citenamefont
  {Del~Duca}(2025)}]{Barcaro:2025ifi}%
  \BibitemOpen
  \bibfield  {author} {\bibinfo {author} {\bibfnamefont {D.}~\bibnamefont
  {Barcaro}}\ and\ \bibinfo {author} {\bibfnamefont {V.}~\bibnamefont
  {Del~Duca}},\ }\href {\doibase 10.1007/JHEP09(2025)041} {\bibfield  {journal}
  {\bibinfo  {journal} {JHEP}\ }\textbf {\bibinfo {volume} {09}},\ \bibinfo
  {pages} {041} (\bibinfo {year} {2025})},\ \Eprint
  {http://arxiv.org/abs/2506.11822} {arXiv:2506.11822 [hep-th]} \BibitemShut
  {NoStop}%
\bibitem [{\citenamefont {Alessio}\ \emph {et~al.}(2025)\citenamefont
  {Alessio}, \citenamefont {Del~Duca}, \citenamefont {Gonzo}, \citenamefont
  {Rosi}, \citenamefont {Rothstein},\ and\ \citenamefont
  {Saavedra}}]{Alessio:2025isu}%
  \BibitemOpen
  \bibfield  {author} {\bibinfo {author} {\bibfnamefont {F.}~\bibnamefont
  {Alessio}}, \bibinfo {author} {\bibfnamefont {V.}~\bibnamefont {Del~Duca}},
  \bibinfo {author} {\bibfnamefont {R.}~\bibnamefont {Gonzo}}, \bibinfo
  {author} {\bibfnamefont {E.}~\bibnamefont {Rosi}}, \bibinfo {author}
  {\bibfnamefont {I.~Z.}\ \bibnamefont {Rothstein}}, \ and\ \bibinfo {author}
  {\bibfnamefont {M.}~\bibnamefont {Saavedra}},\ }\href@noop {} {\  (\bibinfo
  {year} {2025})},\ \Eprint {http://arxiv.org/abs/2511.11457} {arXiv:2511.11457
  [hep-th]} \BibitemShut {NoStop}%
\bibitem [{\citenamefont {Driesse}\ \emph {et~al.}(2026)\citenamefont
  {Driesse}, \citenamefont {Jakobsen}, \citenamefont {Mogull}, \citenamefont
  {Nega}, \citenamefont {Plefka}, \citenamefont {Sauer},\ and\ \citenamefont
  {Usovitsch}}]{Driesse:2026qiz}%
  \BibitemOpen
  \bibfield  {author} {\bibinfo {author} {\bibfnamefont {M.}~\bibnamefont
  {Driesse}}, \bibinfo {author} {\bibfnamefont {G.~U.}\ \bibnamefont
  {Jakobsen}}, \bibinfo {author} {\bibfnamefont {G.}~\bibnamefont {Mogull}},
  \bibinfo {author} {\bibfnamefont {C.}~\bibnamefont {Nega}}, \bibinfo {author}
  {\bibfnamefont {J.}~\bibnamefont {Plefka}}, \bibinfo {author} {\bibfnamefont
  {B.}~\bibnamefont {Sauer}}, \ and\ \bibinfo {author} {\bibfnamefont
  {J.}~\bibnamefont {Usovitsch}},\ }\href@noop {} {\  (\bibinfo {year}
  {2026})},\ \Eprint {http://arxiv.org/abs/2601.16256} {arXiv:2601.16256
  [hep-th]} \BibitemShut {NoStop}%
\bibitem [{\citenamefont {Wheeler}\ and\ \citenamefont
  {Feynman}(1949)}]{Wheeler:1949hn}%
  \BibitemOpen
  \bibfield  {author} {\bibinfo {author} {\bibfnamefont {J.~A.}\ \bibnamefont
  {Wheeler}}\ and\ \bibinfo {author} {\bibfnamefont {R.~P.}\ \bibnamefont
  {Feynman}},\ }\href {\doibase 10.1103/RevModPhys.21.425} {\bibfield
  {journal} {\bibinfo  {journal} {Rev. Mod. Phys.}\ }\textbf {\bibinfo {volume}
  {21}},\ \bibinfo {pages} {425} (\bibinfo {year} {1949})}\BibitemShut
  {NoStop}%
\bibitem [{\citenamefont {Damour}\ and\ \citenamefont
  {Esposito-Farese}(1996)}]{Damour:1995kt}%
  \BibitemOpen
  \bibfield  {author} {\bibinfo {author} {\bibfnamefont {T.}~\bibnamefont
  {Damour}}\ and\ \bibinfo {author} {\bibfnamefont {G.}~\bibnamefont
  {Esposito-Farese}},\ }\href {\doibase 10.1103/PhysRevD.53.5541} {\bibfield
  {journal} {\bibinfo  {journal} {Phys. Rev. D}\ }\textbf {\bibinfo {volume}
  {53}},\ \bibinfo {pages} {5541} (\bibinfo {year} {1996})},\ \Eprint
  {http://arxiv.org/abs/gr-qc/9506063} {arXiv:gr-qc/9506063} \BibitemShut
  {NoStop}%
\bibitem [{\citenamefont {Damour}(2016)}]{Damour:2016gwp}%
  \BibitemOpen
  \bibfield  {author} {\bibinfo {author} {\bibfnamefont {T.}~\bibnamefont
  {Damour}},\ }\href {\doibase 10.1103/PhysRevD.94.104015} {\bibfield
  {journal} {\bibinfo  {journal} {Phys. Rev. D}\ }\textbf {\bibinfo {volume}
  {94}},\ \bibinfo {pages} {104015} (\bibinfo {year} {2016})},\ \Eprint
  {http://arxiv.org/abs/1609.00354} {arXiv:1609.00354 [gr-qc]} \BibitemShut
  {NoStop}%
\bibitem [{\citenamefont {Buonanno}\ \emph {et~al.}(2024)\citenamefont
  {Buonanno}, \citenamefont {Mogull}, \citenamefont {Patil},\ and\
  \citenamefont {Pompili}}]{Buonanno:2024byg}%
  \BibitemOpen
  \bibfield  {author} {\bibinfo {author} {\bibfnamefont {A.}~\bibnamefont
  {Buonanno}}, \bibinfo {author} {\bibfnamefont {G.}~\bibnamefont {Mogull}},
  \bibinfo {author} {\bibfnamefont {R.}~\bibnamefont {Patil}}, \ and\ \bibinfo
  {author} {\bibfnamefont {L.}~\bibnamefont {Pompili}},\ }\href {\doibase
  10.1103/PhysRevLett.133.211402} {\bibfield  {journal} {\bibinfo  {journal}
  {Phys. Rev. Lett.}\ }\textbf {\bibinfo {volume} {133}},\ \bibinfo {pages}
  {211402} (\bibinfo {year} {2024})},\ \Eprint
  {http://arxiv.org/abs/2405.19181} {arXiv:2405.19181 [gr-qc]} \BibitemShut
  {NoStop}%
\bibitem [{\citenamefont {Damour}\ and\ \citenamefont
  {Rettegno}(2023)}]{Damour:2022ybd}%
  \BibitemOpen
  \bibfield  {author} {\bibinfo {author} {\bibfnamefont {T.}~\bibnamefont
  {Damour}}\ and\ \bibinfo {author} {\bibfnamefont {P.}~\bibnamefont
  {Rettegno}},\ }\href {\doibase 10.1103/PhysRevD.107.064051} {\bibfield
  {journal} {\bibinfo  {journal} {Phys. Rev. D}\ }\textbf {\bibinfo {volume}
  {107}},\ \bibinfo {pages} {064051} (\bibinfo {year} {2023})},\ \Eprint
  {http://arxiv.org/abs/2211.01399} {arXiv:2211.01399 [gr-qc]} \BibitemShut
  {NoStop}%
\bibitem [{\citenamefont {Long}\ \emph {et~al.}(2024)\citenamefont {Long},
  \citenamefont {Whittall},\ and\ \citenamefont {Barack}}]{Long:2024ltn}%
  \BibitemOpen
  \bibfield  {author} {\bibinfo {author} {\bibfnamefont {O.}~\bibnamefont
  {Long}}, \bibinfo {author} {\bibfnamefont {C.}~\bibnamefont {Whittall}}, \
  and\ \bibinfo {author} {\bibfnamefont {L.}~\bibnamefont {Barack}},\ }\href
  {\doibase 10.1103/PhysRevD.110.044039} {\bibfield  {journal} {\bibinfo
  {journal} {Phys. Rev. D}\ }\textbf {\bibinfo {volume} {110}},\ \bibinfo
  {pages} {044039} (\bibinfo {year} {2024})},\ \Eprint
  {http://arxiv.org/abs/2406.08363} {arXiv:2406.08363 [gr-qc]} \BibitemShut
  {NoStop}%
\bibitem [{\citenamefont {Bini}\ and\ \citenamefont
  {Damour}(2024)}]{Bini:2024tft}%
  \BibitemOpen
  \bibfield  {author} {\bibinfo {author} {\bibfnamefont {D.}~\bibnamefont
  {Bini}}\ and\ \bibinfo {author} {\bibfnamefont {T.}~\bibnamefont {Damour}},\
  }\href {\doibase 10.1103/PhysRevD.110.064005} {\bibfield  {journal} {\bibinfo
   {journal} {Phys. Rev. D}\ }\textbf {\bibinfo {volume} {110}},\ \bibinfo
  {pages} {064005} (\bibinfo {year} {2024})},\ \Eprint
  {http://arxiv.org/abs/2406.04878} {arXiv:2406.04878 [gr-qc]} \BibitemShut
  {NoStop}%
\bibitem [{\citenamefont {Dlapa}\ \emph {et~al.}(2024)\citenamefont {Dlapa},
  \citenamefont {K{\"a}lin}, \citenamefont {Liu},\ and\ \citenamefont
  {Porto}}]{Dlapa:2024cje}%
  \BibitemOpen
  \bibfield  {author} {\bibinfo {author} {\bibfnamefont {C.}~\bibnamefont
  {Dlapa}}, \bibinfo {author} {\bibfnamefont {G.}~\bibnamefont {K{\"a}lin}},
  \bibinfo {author} {\bibfnamefont {Z.}~\bibnamefont {Liu}}, \ and\ \bibinfo
  {author} {\bibfnamefont {R.~A.}\ \bibnamefont {Porto}},\ }\href {\doibase
  10.1103/PhysRevLett.132.221401} {\bibfield  {journal} {\bibinfo  {journal}
  {Phys. Rev. Lett.}\ }\textbf {\bibinfo {volume} {132}},\ \bibinfo {pages}
  {221401} (\bibinfo {year} {2024})},\ \Eprint
  {http://arxiv.org/abs/2403.04853} {arXiv:2403.04853 [hep-th]} \BibitemShut
  {NoStop}%
\bibitem [{\citenamefont {Dlapa}\ \emph {et~al.}(2025)\citenamefont {Dlapa},
  \citenamefont {K{\"a}lin}, \citenamefont {Liu},\ and\ \citenamefont
  {Porto}}]{Dlapa:2025biy}%
  \BibitemOpen
  \bibfield  {author} {\bibinfo {author} {\bibfnamefont {C.}~\bibnamefont
  {Dlapa}}, \bibinfo {author} {\bibfnamefont {G.}~\bibnamefont {K{\"a}lin}},
  \bibinfo {author} {\bibfnamefont {Z.}~\bibnamefont {Liu}}, \ and\ \bibinfo
  {author} {\bibfnamefont {R.~A.}\ \bibnamefont {Porto}},\ }\href@noop {} {\
  (\bibinfo {year} {2025})},\ \Eprint {http://arxiv.org/abs/2506.20665}
  {arXiv:2506.20665 [hep-th]} \BibitemShut {NoStop}%
\bibitem [{\citenamefont {Long}\ \emph {et~al.}(2025)\citenamefont {Long},
  \citenamefont {Pfeiffer}, \citenamefont {Kidder},\ and\ \citenamefont
  {Scheel}}]{Long:2025tvk}%
  \BibitemOpen
  \bibfield  {author} {\bibinfo {author} {\bibfnamefont {O.}~\bibnamefont
  {Long}}, \bibinfo {author} {\bibfnamefont {H.~P.}\ \bibnamefont {Pfeiffer}},
  \bibinfo {author} {\bibfnamefont {L.~E.}\ \bibnamefont {Kidder}}, \ and\
  \bibinfo {author} {\bibfnamefont {M.~A.}\ \bibnamefont {Scheel}},\ }\href
  {\doibase 10.1103/bdsb-sp9c} {\bibfield  {journal} {\bibinfo  {journal}
  {Phys. Rev. D}\ }\textbf {\bibinfo {volume} {112}},\ \bibinfo {pages}
  {124038} (\bibinfo {year} {2025})},\ \Eprint
  {http://arxiv.org/abs/2511.10196} {arXiv:2511.10196 [gr-qc]} \BibitemShut
  {NoStop}%
\bibitem [{\citenamefont {Barack}\ \emph {et~al.}(2023)\citenamefont {Barack}
  \emph {et~al.}}]{Barack:2023oqp}%
  \BibitemOpen
  \bibfield  {author} {\bibinfo {author} {\bibfnamefont {L.}~\bibnamefont
  {Barack}} \emph {et~al.},\ }\href {\doibase 10.1103/PhysRevD.108.024025}
  {\bibfield  {journal} {\bibinfo  {journal} {Phys. Rev. D}\ }\textbf {\bibinfo
  {volume} {108}},\ \bibinfo {pages} {024025} (\bibinfo {year} {2023})},\
  \Eprint {http://arxiv.org/abs/2304.09200} {arXiv:2304.09200 [hep-th]}
  \BibitemShut {NoStop}%
\bibitem [{\citenamefont {Tange}(2018)}]{tange2018gnu}%
  \BibitemOpen
  \bibfield  {author} {\bibinfo {author} {\bibfnamefont {O.}~\bibnamefont
  {Tange}},\ }\href@noop {} {\emph {\bibinfo {title} {GNU parallel 2018}}}\
  (\bibinfo  {publisher} {Lulu. com},\ \bibinfo {year} {2018})\BibitemShut
  {NoStop}%
\bibitem [{\citenamefont {Bezanson}\ \emph {et~al.}(2017)\citenamefont
  {Bezanson}, \citenamefont {Edelman}, \citenamefont {Karpinski},\ and\
  \citenamefont {Shah}}]{bezanson2017julia}%
  \BibitemOpen
  \bibfield  {author} {\bibinfo {author} {\bibfnamefont {J.}~\bibnamefont
  {Bezanson}}, \bibinfo {author} {\bibfnamefont {A.}~\bibnamefont {Edelman}},
  \bibinfo {author} {\bibfnamefont {S.}~\bibnamefont {Karpinski}}, \ and\
  \bibinfo {author} {\bibfnamefont {V.~B.}\ \bibnamefont {Shah}},\ }\href@noop
  {} {\bibfield  {journal} {\bibinfo  {journal} {SIAM review}\ }\textbf
  {\bibinfo {volume} {59}},\ \bibinfo {pages} {65} (\bibinfo {year}
  {2017})}\BibitemShut {NoStop}%
\end{thebibliography}%

\end{document}